\documentclass{aa}  

\usepackage{graphicx}
\usepackage{txfonts}
\usepackage{hyperref}
\hypersetup{citecolor=green}
\usepackage{makecell}
\usepackage{gensymb}
\usepackage{booktabs}
\usepackage{float}

\DeclareMathOperator{\sech}{sech}
\makeatletter

\begin{document}

   \title{The Fornax Deep Survey (FDS) with the VST}

   \subtitle{XI. The search for signs of preprocessing between the Fornax main cluster and Fornax A group}

   \author{A. H. Su\inst{1}
          \and
          H. Salo\inst{1}
          \and
          J. Janz\inst{1,2}
          \and
          E. Laurikainen\inst{1}
          \and
          A. Venhola\inst{1}
          \and
          R. F. Peletier\inst{3}
          \and
E.~Iodice\inst{4}
\and
M.~Hilker\inst{5}
\and
M.~Cantiello\inst{6}
\and
N.~Napolitano\inst{4}
\and
M.~Spavone\inst{4}
\and
M. A. Raj\inst{4}
\and
G.~van de Ven\inst{7}
\and
S.~Mieske\inst{8}
\and
M.~Paolillo\inst{9}
\and
M.~Capaccioli\inst{9}
\and
E.~A.~Valentijn\inst{3}
\and
A.~E.~Watkins\inst{10}
}

   \institute{Space Physics and Astronomy Research Unit, University of Oulu, Pentti Kaiteran katu 1, FI-90014, Finland\\
              \email{hung-shuo.su@oulu.fi}
              \and
              Finnish Centre of Astronomy with ESO (FINCA), University of Turku, V\"ais\"al\"antie 20, FI-21500 Piikki\"o, Finland
              \and
              Kapteyn Institute, University of Groningen, Landleven 12, 9747 AD Groningen, The Netherlands
              \and
              INAF, Osservatorio Astronomico di Capodimonte, Salita Moiariello 16,I-80131 Napoli, Italy
              \and
              European Southern Observatory, Karl-Schwarzschild-Strasse 2, 85748 Garching bei München, Germany
              \and
              INAF, Osservatorio Astronomico d’Abruzzo, Via Mentore Maggini Teramo,TE I-64100, Italy
              \and
              Department of Astrophysics, University of Vienna, Türkenschanzstrasse 17, 1180 Wien, Austria
              \and
              European Southern Observatory, Alonso de Cordova 3107, Vitacura, Santiago, Chile
              \and
              University of Naples Federico II, C.U. Monte Sant’Angelo, Via Cinthia, 80126 Naples, Italy
              \and
              Astrophysics Research Institute, Liverpool John Moores University, IC2, Liverpool Science Park, 146 Brownlow Hill, Liverpool L3 5RF, UK
             }

   \date{Received 9 October 2020 / Accepted 11 January 2021}

 
  \abstract
   {Galaxies either live in a cluster, a group, or in a field environment. In the hierarchical framework, the group environment bridges the field to the cluster environment, as field galaxies form groups before aggregating into clusters. In principle, environmental mechanisms, such as galaxy--galaxy interactions, can be more efficient in groups than in clusters due to lower velocity dispersion, which lead to changes in the properties of galaxies. This change in properties for group galaxies before entering the cluster environment is known as preprocessing. Whilst cluster and field galaxies are well studied, the extent to which galaxies become preprocessed in the group environment is unclear.}
   {We investigate the structural properties of cluster and group galaxies by studying the Fornax main cluster and the infalling Fornax A group, exploring the effects of galaxy preprocessing in this showcase example. Additionally, we compare the structural complexity of Fornax galaxies to those in the Virgo cluster and in the field.}
   {Our sample consists of 582 galaxies from the Fornax main cluster and Fornax A group. We quantified the light distributions of each galaxy based on a combination of aperture photometry, Sérsic+PSF (point spread function) and multi-component decompositions, and non-parametric measures of morphology. From these analyses, we derived the galaxy colours, structural parameters, non-parametric morphological indices (Concentration $C$; Asymmetry $A$, Clumpiness $S$; Gini $G$; second order moment of light $M_{20}$), and structural complexity based on multi-component decompositions. These quantities were then compared between the Fornax main cluster and Fornax A group. The structural complexity of Fornax galaxies were also compared to those in Virgo and in the field.}
   {We find significant (Kolmogorov-Smirnov test $p$-value $<\alpha =0.05$) differences in the distributions of quantities derived from Sérsic profiles ($g'-r'$, $r'-i'$, $R_e$, and $\bar{\mu}_{e,r'}$), and non-parametric indices ($A$ and $S$) between the Fornax main cluster and Fornax A group. Fornax A group galaxies are typically bluer, smaller, brighter, and more asymmetric and clumpy. Moreover, we find significant cluster-centric trends with $r'-i'$, $R_e$, and $\bar{\mu}_{e,r'}$, as well as $A$, $S$, $G$, and $M_{20}$ for galaxies in the Fornax main cluster. This implies that galaxies falling towards the centre of the Fornax main cluster become fainter, more extended, and generally smoother in their light distribution. Conversely, we do not find significant group-centric trends for Fornax A group galaxies. We find the structural complexity of galaxies (in terms of the number of components required to fit a galaxy) to increase as a function of the absolute $r'$-band magnitude (and stellar mass), with the largest change occurring between -14\,mag $\lesssim M_{r'}\lesssim-19$\,mag ($7.5 \lesssim \log_{10}(M_*/M_{\odot}) \lesssim 9.7$). This same trend was found in galaxy samples from the Virgo cluster and in the field, which suggests that the formation or maintenance of morphological structures (e.g. bulges, bar) are largely due to the stellar mass of the galaxies, rather than the environment they reside in. }
   {}


   \maketitle
%

\section{Introduction}

The shapes and sizes of galaxies we observe today are a product of their initial conditions in the early Universe and their evolution through cosmic time to the present day. The large variety of structures present in galaxies (e.g. bulges, bars, spirals) has led to sophisticated classification schemes \citep[e.g.][]{hubble1936,devaucouleurs1959,sandage1961,buta2015}, which qualitatively describe the light distribution of galaxies. However, a qualitative description alone does not provide enough information on how certain mechanisms and physical processes affect structures; a quantitative description is required. 

One method to quantify the structure of galaxies is through photometric decomposition, where the light distribution of galaxies is broken into individual components. This method has been widely used in the literature as it allows for a systematic analysis of the structures (e.g. bulges, disks, bars) within galaxies. A wide range of tools have been developed to implement 2D photometric decompositions (e.g. GIM2D, \citealt{simard1998}; GALFIT, \citealt{peng2002}; BDBAR, \citealt{laurikainen2004}; BUDDA, \citealt{desouza2004}; GASP2D, \citealt{mendezabreu2008}; IMFIT, \citealt{erwin2015}). A number of studies have conducted photometric decomposition on large galaxy samples \citep[e.g.][]{allen2006,simard2011,lackner2012}, although in some cases only one- or two-component models are fitted. Single Sérsic or Sérsic and nucleus decompositions can provide a useful global characterisation of the galaxy light distribution, particularly for low mass galaxies \citep[e.g.][]{graham2003}. When multiple (i.e. three or more) components are fitted simultaneously, such as for a galaxy hosting a bulge, a disk, and a bar, the number of free parameters makes the decomposition model heavily degenerate and the resulting solution, if the fitting procedure converges to a solution at all, easily becomes uninformative. However, with reasonable, physical initial conditions for each component as well as human supervision of the fitting procedure, multi-component decompositions are feasible and can yield insight into galaxy formation scenarios \citep[e.g.][]{graham2002,laurikainen2007,laurikainen2018,erwin2008,gadotti2009,salo2015,mendezabreu2017,kruk2018,spavone2020}.

For massive galaxies (e.g. $M_* \gtrsim 10^9 M_{\odot}$), multi-component decompositions are important in obtaining accurate parameters of structures. For example, several works \citep[e.g.][]{laurikainen2004,laurikainen2010,aguerri2005,gadotti2008} have shown that if the bar is not accounted for in the decomposition model, the bar flux can be erroneously attributed to the bulge flux. This was confirmed for a larger sample of galaxies in \citet{salo2015}. It has also been shown that in a nearly face-on view the vertically thick inner bar component is often erroneously attributed to the bulge flux \citep[see][]{laurikainen2014, athanassoula2015, salo2017}. 

It is known that the evolution of galaxies is dependent on internal processes, which are correlated with the stellar mass \citep{kauffmann2003,haines2007}, and external processes due to the environment \citep{dressler1980,jaffe2016}. Considerable work has also been done to address the environmental dependence of galaxy evolution (e.g. star formation quenching) throughout cosmic history \citep{baldry2006,peng2010_quench,nantais2016,nantais2017,nantais2020,old2020}. Several mechanisms have been proposed by which the environment can change the morphology of galaxies, such as mergers \citep{barnes1992}, high speed close encounters with neighbouring galaxies combined with tidal interactions with the cluster potential \citep[i.e. harassment,][]{moore1998,smith2015}, the removal of cool gas via ram pressure stripping \citep{gunn1972,boselli2014}, or halting of gas accretion \citep[i.e. strangulation or starvation,][]{larson1980}. These processes can also affect the star formation of the galaxies, such as merging galaxies often hosting signs of strong star formation at certain merger stages \citep[e.g.][]{sanders1988,dimatteo2008}, or the removal of gas making galaxies become quiescent \citep{bekki2002}. The change in star formation can be large enough to create a dichotomy in colour-magnitude diagrams, with the gas-poor early-type galaxies (ETG) predominantly residing along the red sequence (RS), and the blue cloud which consists of gas-rich late-type galaxies (LTG). 

Although the potential environmental mechanisms for transforming galaxies from the blue cloud to the red sequence are known \citep[e.g.][]{gomez2003}, it is not clear at which stage of group and cluster evolution this occurs. Due to the differences in mass between clusters and groups, the efficiencies of the aforementioned mechanisms vary. For example, the lower velocity dispersion in groups allows for more prolonged interactions between galaxies. This can lead to more efficient tidal stripping, causing galaxies to become more asymmetric. Such changes in the morphology of galaxies is known as preprocessing \citep{zabludoff1998,fujita2004}. With different mechanisms becoming more significant depending on environment, the signs of preprocessing and the cluster environment's impact on the evolution of galaxies should be encoded in the galaxies' structures and stellar populations.

The Fornax cluster is a great laboratory for studying the impact of galaxy environments, as it consists of both a main cluster and an infalling group. The main cluster is centred on NGC~1399  \citep[R.A.$=03h38m29.1s$, Dec.$=-35d27m02s$,][]{kim2013} and located at a distance of $20.0\pm0.3\pm1.4$\,Mpc \citep{blakeslee2009}. Fornax is a relatively low mass cluster \citep[$(7\pm2)\times 10^{13}M_{\odot}$,][]{drinkwater2001} compared to the most nearby galaxy cluster, Virgo \citep[$(4.2\pm 0.5) \times 10^{14} M_{\odot}$,][]{mclaughlin1999}. Despite its relatively low mass, Fornax appears to be at a more advanced evolutionay state than Virgo. For example, the Fornax core appears more dynamically relaxed and is dominated by quenched ETGs \citep{ferguson1989}. Despite this, the infalling Fornax A group, demonstrates the ongoing assembly of the Fornax cluster. 

As one of the latest wide-field observations of the Fornax cluster, the Fornax Deep Survey (FDS) provides an opportunity to probe the cluster in outstanding detail. With FDS, \citet{iodice2019} studied the surface brightness profiles of bright ($m_B \leq 15$\,mag) ETGs within the inner 9 square degrees ($<700$\,kpc) of the Fornax cluster. They found that there is a high density of ETGs and intracluster light (ICL) in the western region of the cluster ($\sim 0.3$\,Mpc from the core) and that many of the ETGs show signs of asymmetry in their outskirts. This was interpreted as evidence for a group accretion event during the formation of the cluster, and also suggests that the core is still virialising. More recently, \citet{spavone2020} found the luminosity of the ICL within the virial radius of the Fornax cluster to be $\sim 34\%$ of all cluster members. Additionally, the Next Generation Fornax Survey (NGFS), another deep, wide-field survey of the Fornax region, has also studied the dwarf galaxies population, in terms of structural parameters \citep{eigenthaler2018} and sub-structures in their spatial distribution \citep{ordenesbriceno2018}. They found that nucleated galaxies tend to be larger, brighter, and tend to be concentrated towards NGC~1399 than non-nucleated galaxies, which agree with the results of \citet{venhola2019}.

\citet{venhola2018} presented the FDS Dwarf Catalogue (FDSDC), a catalogue comprised of 564 likely cluster member dwarf galaxies reaching a 50\% completeness limit at $M_{r'}=-10.5$\,mag and $\bar{\mu}_{e,r'}=26$\,mag\,arcsec$^{-2}$. The catalogue spans an $r'$-band absolute magnitude of $-9$\,mag to $-18.5$\,mag. Using this catalogue, \citet{venhola2019} found that dwarf galaxies tend to become redder, smoother, and more extended with decreasing distance to the cluster centre. The fraction of early- and late-type dwarfs was observed to vary as a function of cluster-centric distance. Additionally, they found that early- and late-type dwarfs follow different scaling relations in absolute magnitude, effective radius, and Sérsic index. The observations are consistent with the idea that the star formation in dwarf galaxies is quenched via ram pressure stripping. Only the most massive dwarfs can manage to hold on to some cold gas in their cores. 

Taking advantage of the deep imaging of FDS, we aim to study the signs of preprocessing between two different environments: the Fornax main cluster and Fornax A group (for brevity, we refer to the main Fornax cluster as 'Fornax main', and the Fornax A group as 'Fornax group'). Specifically, we compare the galaxies between the two environments to see if environmental processes can explain any of the observed differences, and discuss which of the processes are driving such differences. In order to compare the two environments, we accurately measure and quantify the structures in the galaxies. Several works demonstrate a first look at this, using azimuthally averaged multi-component decompositions to quantify accreted vs. in-situ component of the brightest Fornax galaxies \citep{iodice2019, raj2019, spavone2020, raj2020}. To this end, one of the main focuses in this work is the 2D multi-component decompositions of galaxies in the FDS for both dwarfs and giants. Not only do such decompositions provide information on the structural complexity of the galaxies, but also provide opportunities to study the structures themselves. Furthermore, as we follow the same decomposition procedures used in the Spitzer Survey of Stellar Structure in Galaxies \citep[S$^4$G][]{salo2015, sheth2010}, this work serves to expand the catalogue of multi-component decompositions to even lower stellar masses. To complement the multi-component decompositions, we also present the Sérsic+PSF models and non-parametric measures (see Sect.~\ref{sect:nonparametric} for definitions) of member galaxies of the Fornax main and the Fornax group. 

This paper is structured in the following manner. In Sect.~\ref{sect:data} we discuss the FDS data and the sample selection. Sect.~\ref{sect:preprocessing} outlines the steps taken to prepare the data for analysis. In Sect.~\ref{sect:parameters} we outline the process of measuring aperture photometry, conducting structural decompositions, and the calculation of nonparametric indices. In Sect.~\ref{sect:compare_params} we compare the magnitudes derived from photometry and with FDSDC, as well as calculate the galaxy stellar masses. In Sect.~\ref{sect:results} we present the galaxy properties as a function of stellar mass and projected cluster-/group-centric distance (hereon referred to as halo-centric distance), and compare the differences between Fornax main cluster and Fornax group with stellar mass trends removed. In Sect.~\ref{sect:lit_comp} we compare our results with similar studies in the literature and in Sect.~\ref{sect:discussion} we discuss the potential mechanisms which led to the observed differences between these environments. Finally, in Sect.~\ref{sect:conclusion} we summarise and draw our conclusions for this work. Throughout this work a distance modulus of 31.5\,mag (equivalent to a distance of 20\,Mpc) \citep{blakeslee2009} was used. At this distance, 1\,arcsec corresponds to $\sim 0.097$\,kpc.

\section{Data}\label{sect:data}
\subsection{Observations}
This work uses data from the Fornax Deep Survey (FDS), which is a joint collaboration between two guaranteed time observation surveys: FOCUS (PI: R. Peletier) and VEGAS \citep[PI: E. Iodice,~][]{capaccioli2015}. FDS covers the Fornax main cluster and the infalling Fornax group using the 2.6\,m ESO VLT Survey Telescope (VST), located at Cerro Paranal, Chile. Imaging was completed using OmegaCAM \citep{kuijken2002}, which covers a field of view of $1\,\degree \times 1\,\degree$ and the pixel size is 0.21\,arcsec\,pix$^{-1}$. Observations were conducted in $u'$, $g'$, $r'$, and $i'$ bands under average seeing, with full width at half maximum (FWHM) of 1.2\,\arcsec, 1.1\,\arcsec, 1.0\,\arcsec, and 1.0\,\arcsec, respectively \citep{venhola2018}. For more details on the observing strategy, see \citet{iodice2016} and \citet{venhola2018}. The steps for the reduction and calibration of FDS mosaics covering 26\,deg$^2$ can be found in \citet{venhola2018}. The final mosaics were resampled to have a pixel scale of 0.20\,arcsec\,pix$^{-1}$. Given that the FDS coverage of the $u'$-band is limited to the main cluster, we exclude $u'$-band data from our analyses.

\subsection{The sample}\label{sect:sample}
\begin{figure}
\centering
\includegraphics[width=\hsize]{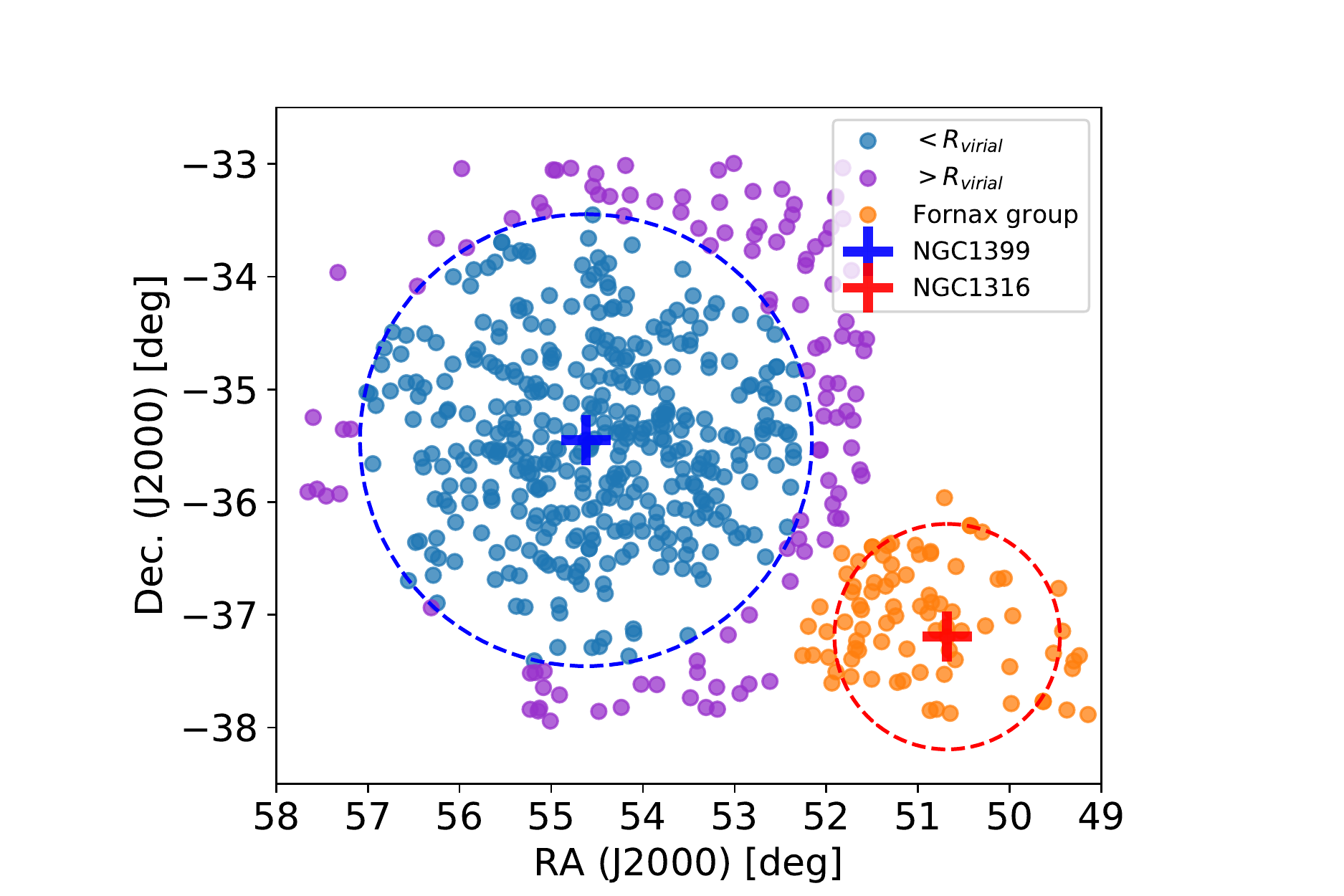}
  \caption{Positions of our sample of Fornax main member galaxies. The sample was split between the Fornax main cluster (\textit{blue} and \textit{purple}) and the Fornax group (\textit{orange}) based on the separation relative to NGC~1399 and NGC~1316. Using NGC~1399 and NGC~1316 as the centres of the Fornax main and Fornax group, respectively, the projected separation between the two centres is $\sim 3.6$\,$^{\circ}$. Galaxies which have a projected separation to NGC~1399 that is greater than twice the projected separation to NGC~1316 were classified as belonging to the Fornax group. The dashed blue and red lines denote 2\,$^{\circ}$ ($=700$\,kpc) and 1\,$^{\circ}$ from NGC~1399 and NGC~1316, respectively.}
     \label{fig:radec}
\end{figure}

The sample of galaxies used in this work is based on the catalogue of all FDS sources and cluster membership selection presented in \citet{venhola2018}. This consists of 594 galaxies, of which 564 overlap with FDSDC. The 30 additional galaxies not within FDSDC contain 29 bright ($M_{r'} <  -18.5$) galaxies, which include the samples from \citet{iodice2019}, \citet{raj2019,raj2020}, and one dwarf galaxy missing from FDSDC. From visual inspection of the images, we removed 8 FDSDC entries from our sample due to duplication (see Appendix~\ref{app:duplicate} for more detail). Due to the lack of $i'$-band coverage in the FDS33 field\footnote{For an overview of FDS fields, see \citet{venhola2018}, their Fig.~2}, the 4 galaxies\footnote{FDS33 galaxies: FDS33\_0081, FDS33\_0088, FDS33\_0121, FDS33\_0129} located in this field were excluded from the final sample, although decompositions were made in $g'$ and $r'$. Overall, our final sample consists of 582 galaxies. Figure~\ref{fig:radec} shows the positions of the galaxies and visualises how we define the Fornax main and Fornax group sub-samples. 

In terms of the assignment of galaxy IDs, we adapted the naming convention used in FDSDC to include both dwarfs and massive galaxies. Our format is FDSX\_YYYYz, where the prefix "FDS" is always present, X denotes the FDS field number (which can be up to two digits), Y denotes the identification number within the field and always contains four digits (filled with zeroes), and in some cases \citep[where nearby companions were detected during the catalogue creation of FDSDC][]{venhola2018} z is included as one letter as part of its identification number. We note that the majority of galaxies not contained in FDSDC already had identification numbers assigned during the catalogue creation process. For three galaxies\footnote{FDS7\_0736, FDS7\_0737, FDS31\_0607} which lacked identification numbers, their identifications were assigned as one plus the highest identification number within the FDS field (i.e. for FDS7 field the last identification number was 735, so the two galaxies not included in FDSDC were labelled as 736 and 737). This naming convention is used in the table of multi-component decompositions in Appendix~\ref{tab:multicomp_summary} as well as on the webpage\footnote{\url{https://www.oulu.fi/astronomy/FDS_DECOMP/main/index.html}} where all the decompositions are presented (see Appendix~\ref{app:website}). 


\section{Data processing}\label{sect:preprocessing}
To prepare the data for decompositions, we followed the procedures used in the S$^4$G decomposition pipeline \citep{salo2015}. The processing steps include determining the central pixel of each galaxy, creating masks with SExtractor \citep{bertin1996} which were modified (if necessary) after visual inspection, determining average sky values, and using IRAF \texttt{ellipse} to obtain isophotal position angle and ellipticity profiles. From this, initial guesses for decomposition parameters can be estimated. This reduces possible degeneracies in the model and the probability of $\chi^2$ minimisation becoming 'stuck' in erroneous local minima. Appendix~\ref{app:skysub} illustrates the various steps taken in the data processing procedure.

\subsection{Image preparation}\label{sect:centre}
First, the postage stamp images of each galaxy were cut out from the FDS mosaics reduced by \citet{venhola2018}. The postage stamp images cover at least five effective radii \citep[from][]{venhola2018}, with a lower limit of 100\,arcsec ($=500$\,pix) in both dimensions. 

Before creating the mask images, IRAF \citep{tody1986} \texttt{ellipse} and \texttt{bmodels} were used to provide model images of the galaxies. Residual images were constructed by subtracting the model images from the galaxy images, which were used as input for SExtractor. By using the residual images instead of the galaxy images directly, the binary segmentation maps cover any unwanted sources (e.g. stars) overlapping the galaxies. The binary segmentation maps were convolved with a Gaussian kernel with $\sigma=3$\,pix in order to extend the size of the resultant masked areas. The convolved segmentation maps were then converted back to binary mask images by using a threshold in the pixel values (in this case 0.03). Each mask image was inspected by eye and manually edited where necessary (e.g. when structures belonging to the galaxy are erroneously masked, or those with close bright companions). 

The central pixel coordinates of the galaxies were determined using a centroid fitting routine which finds the brightest pixel within a threshold radius from an input x and y coordinate. The initial guesses were chosen via a cursor on the galaxy images, which were typically chosen to be near the brightest pixels. The routine then computes the centroid based on a section of the image centred on the brightest pixel. The centroid is calculated as the coordinate where the change in pixel intensity with x and y become zero. This process is iterated six times, where each time the output coordinate is used as the new input coordinate, but with an increasingly lower radius. This maximises the chances of convergence to determine the brightest pixel, and allows for a larger margin of error in the initial guess. Nevertheless, the centroid routine may fail in some cases when the light distribution has no clear central peak (i.e. a flat light profile). In such cases, the central coordinates were selected by cursor after detailed inspection of the galaxy images based on both geometric considerations and the brightest pixel position. These typically agreed with those from FDSDC, which used isophotal coordinates\footnote{SExtractor galaxy coordinates were calculated as the first order moment of the detection image, according to SExtractor manual \url{https://sextractor.readthedocs.io/en/latest/Position.html\#pos-iso-def}.} from SExtractor for flat galaxies \citep{venhola2018}. 

\subsection{Sky subtraction} \label{sect:skysubtraction}
Although the FDS mosaics were first-order background subtracted during the data reduction process, the local level of sky background must be taken into account for each galaxy. We estimate the sky levels through three different steps. First, we placed square boxes with widths of 12\,arcsec (hereafter: skyboxes) via cursor around the masked postage stamp image. The locations of the skyboxes were chosen as regions devoid of obvious sources, external to the galaxy. On average 12 skyboxes were placed for each galaxy. For each skybox, the median and the standard deviation of the flux values enclosed (excluding masked pixels) were calculated. The sky level was then estimated as the mean of the median values and the root mean square (RMS) of the sky level as the mean of the standard deviation values. From this initial sky level estimate, we create a sky subtracted image and produce an ellipticity and position angle (PA) profile as a function of radius from the centre, via the IRAF \texttt{ellipse} routine. Through inspection of the sky subtracted image and the ellipticity and position angle profiles we calculate the mean ellipticity and position angle for the galaxy outskirts (see Fig.~\ref{fig:ellipse}).

Fixing the ellipticity and position angle to the mean values calculated, we construct elliptical annuli of increasing radii upon the postage stamp image. The width of the annuli was defined to increase by 2\% logarithmically (with a minimum width of 1\,pix) to increase the signal-to-noise ratio (S/N) in the galaxy outskirts. For each annulus we applied $3\sigma$ clipping of pixel intensities (where $\sigma$ is the RMS value within the annulus). Pixels which were rejected or masked away were replaced with the average pixel value within the annulus plus Gaussian RMS noise. Through this process, a 'cleaned' image of the galaxy is created, hereon referred to as cleanimage. 

From the cleanimage, an azimuthally averaged flux profile and a cumulative flux profile were constructed for each galaxy. The profiles allowed for clear visual inspection of the radius at which the galaxy flux is no longer significant compared to the background noise (we refer to this radius hereon as radgal, see Fig.\,\ref{fig:fsumplot1}). Beyond this conservative maximum galaxy radius, we chose an additional range in radius to create a sky annulus in the cleanimage. The second estimate of sky level and RMS value were calculated as the mean and standard deviation of the pixel values within this sky annulus, respectively. 

By default, a flat sky subtraction based on the sky level from the second approach was applied to the postage stamp images to create the data images used in GALFIT. In select cases (37 in total) where a strong gradient in the sky was observed (e.g. galaxies close to bright foreground stars) we also fit and subtracted a plane from the background sky, as in such cases we found that a flat sky subtraction significantly biased the galaxy's measured total magnitude. After the sky subtraction, the ellipticity and PA profiles were reiterated with the data images to recalculate the mean ellipticity and PA of the outer isophotes. These values were taken as initial values for the decompositions.

\subsection{Point spread function}\label{sect:psf}
In order to obtain accurate parameters in decompositions, the point spread function (PSF) must be taken into account. 
In principle there are variations in the PSF across different observations, so deriving the PSF on a galaxy-by-galaxy basis can be more accurate. However, there must be enough suitable stars close to the galaxy, else the uncertainty in these local PSFs will be skewed by the small sample sizes. As a result, we produced a separate PSF for each FDS field instead to apply to the galaxies residing in the corresponding field. To sample the PSFs we first use SExtractor to build a catalogue of objects in each FDS field. From the catalogue, a set of selection criteria was applied to select only point sources. Upper and lower limits were placed on the magnitude of the objects (typically $20 > MAG_{AUTO} > 15.5$ across the fields) to ensure that they are neither saturated nor have too low S/N. Then, to exclude extended objects, an upper limit on the parameter $FWHM_{IMAGE}$ was placed which varied from field to field (typically $\sim 1$\,arcsec). Furthermore, $1/ELONGATION > 0.95$ was used to exclude elongated objects, such as inclined galaxies. In addition, the quality flag produced in the catalogue was used so that objects with $FLAG \ge 1$ were excluded. From the selection cuts, a sample of point sources was made which typically consisted of a few hundred objects. 

From the sample of point sources, we first normalised the postage stamp images by its total flux (based on $MAG_{AUTO}$). Next, a 2D Gaussian was fitted to the image ($101 \times 101$\,pix) of each source in order to determine the central peak with sub-pixel accuracy. Using the new centres, radial flux profiles for all sources were made and 'stacked' to create a general profile for each field. The profile was median averaged in bins of 0.05\,arcsec within the inner region (<1.5\,arcsec), whereas the outer region (1 to 10\,arcsec) was median averaged in bins of 0.5\,arcsec. The smaller bin size in the inner region was applied to sample the most significant part of the PSF. The inner and outer profiles were combined and interpolated to create an azimuthally averaged 1D flux profile, with the slight overlap ensuring a smooth transition between the two regions. Based on the interpolated profile, an axisymmetric PSF was created which was oversampled by a factor of 2 compared to the image pixel size. A comparison to the PSFs used in \citet{venhola2018} can be found in Fig.\,\ref{fig:psf_rall}. 

\subsection{Sigma images}
We construct sigma images as one of the inputs for GALFIT decompositions. The sigma image quantifies the uncertainties in each pixel from the corresponding data image, in order to provide weighting in the fitting process of GALFIT. In principle there are two main sources of uncertainty: the uncertainty from the detectors and due to photon counts. Information on the noise due to instrumentation is contained in the weight images, which have pixel values calculated based on the standard deviation of the background noise in each frame and the number of overlapping frames for each pixel. In the weight images bad pixels were assigned a value of zero. The photon counts can be thought to follow a Poissonian distribution. Combining both sources of noise, the corresponding sigma values for each pixel are defined as 

\begin{equation}
    \text{sigma}=\left(\frac{0.20}{0.21}\right)^2 \sqrt{\frac{1}{W} + \frac{f}{g}},
\end{equation}
where $W$ is the weight image, $f$ is the flux value from sky subtracted science image (i.e. data image), $g$ is the conversion value between analog to digital units (ADUs) and electrons, also known as the gain (read from image headers), and the factor of $(0.20/0.21)^2$ accounts for the resampling of the science and weight images during calibration, changing from the instrumental pixel scale of 0.21\,arcsec to the final image pixel scale of 0.2\,arcsec. The sigma images were checked via inspection to ensure that their pixel values calculated in the sky region correspond with the measured sky RMS values.

\section{Photometric parameters}\label{sect:parameters}
Several techniques have been developed to extract information about objects from images. Given that different methods have distinct advantages and disadvantages, we employ a number of techniques to explore our data. We apply aperture photometry to measure the distribution of light and calculate parameters (e.g. integrated magnitudes, effective radii). We also employ photometric decompositions to quantify the light distribution of physically motivated components (e.g. bulge, disks, bars etc.). Additionally, we calculate non-parametric morphological indices in order to characterise our galaxies without explicitly imposing any model assumptions upon them.

\subsection{Aperture photometry}\label{sect:aper_phot}
In order to measure the light distribution of the galaxies, we utilised the azimuthally averaged and cumulative flux profiles made from cleanimages, as defined in Sect.~\ref{sect:skysubtraction}. To summarise, the cleanimages were constructed via iterative $3\sigma$ clipping of the masked postage stamp images within concentric elliptical annuli. The masked regions and pixels rejected during the iterations were filled with the average flux values at the given radii plus sky RMS noise, after which the sky level was subtracted. The elliptical annuli were constructed based on the mean outer isophotal position angle and axial ratio from IRAF \texttt{ellipse}. The aperture was then defined as the ellipse with semi-major axis equal to the maximum galaxy radius (defined in Sect\ref{sect:skysubtraction}). The aperture magnitude $m_{\text{aper}}$ was defined as
\begin{equation}
    m_{\text{aper}}=-2.5\log_{10}\left(\sum_{i} f_i\right), \label{eqn:mag_aper}
\end{equation}
where $f_i$ denotes the flux value of pixel $i$. The summation of $f_i$ applies only to pixels within the defined elliptical aperture\footnote{The FDS mosaics were calibrated such that no additional zeropoint term in Eqn. (\ref{eqn:mag_aper}) is necessary.}. From the aperture magnitude, the aperture effective radius was determined as the radius within which the cumulative flux is equal to half the total aperture flux\footnote{Each aperture effective radius was determined using fractional pixel flux.}. We derive the aperture quantities in the $g'$ and $i'$ band using the limiting galaxy radius. In case the cumulative flux profile in $g'$ or $i'$-band starts to decline before the limiting radius, the maximum cumulative flux was taken instead.

\subsection{Structural decompositions}\label{sect:decomp_theory}
Photometric decomposition of galaxies involves fitting parametric functions to their light profiles \citep[e.g. the de Vaucouleurs' profile for elliptical galaxies,][]{devaucouleurs1948}. By calculating parameters which best fit a galaxy, one can characterise the features of the galaxy with a simple set of values. This becomes particularly useful when structures are broken down into a linear combination of different functions. By decomposing a galaxy into components, not only can one study the different structures individually but also relative to the rest of the galaxy. 

\subsubsection{GALFIT}
For the galaxy decomposition we utilise GALFIDL \citep{salo2015}, an IDL interface which allows for batch processing of galaxies with GALFIT, ver. 3 \citep{peng2010_galfit} and visualisation of the output decomposition models. GALFIT is a galaxy fitting tool which fits parametric functions to 2D light profiles of galaxies and outputs the best fitting parameters. Minimisation is done using the Levenberg-Marquadt algorithm, with the goodness of fit $\chi^2$ defined as
\begin{equation}
    \chi^2=\sum_{x}\sum_{y}\frac{\left[O(x,y)-M(x,y)\right]^2}{\sigma(x,y)^2},
\end{equation}
where $O$ is the observation (data image), $M$ is the model image (parametric function convolved with PSF), $\sigma$ is the uncertainty in the observation (sigma image), and $x$ and $y$ denote the pixel index in the x and y axes of the images, respectively.

Additionally, GALFIT outputs the reduced $\chi^2$, defined as $\chi_{\nu}^2 = \chi^2 / \nu$, where $\nu$ is the number of degrees of freedom in the fit. In practice, $\nu$ is equivalent to the number of pixels used minus the number of free parameters in the model. $\chi_{\nu}^2$ is a better goodness-of-fit indicator than $\chi^2$ as the value is normalised to the size of the image. In the ideal case where the model matches the data within the given uncertainties, the expected value of $\chi_{\nu}^2=1$. $\chi_{\nu}^2<1$ implies that the uncertainties are likely to be overestimated, whereas $\chi_{\nu}^2>1$ indicates a difference between the model and the data (or underestimated uncertainties). In practice, $\chi_{\nu}^2$ typically exceeds 1, particularly for larger, more massive galaxies as the decomposition models do not completely account for the real structures in galaxies. For the same reason, the formal uncertainties calculated for the parameters are usually not very useful; the actual uncertainties relate more to model selection. 

In GALFIT, the model images have isophotes in the form of a generalised ellipse \citep{athanassoula1990}, which can be parameterised by the centre coordinates ($x_c, y_c$), axial ratio ($q=b/a$, the ratio of semi-minor to semi-major axis), position angle ($PA$, the orientation of the semi-major axis), and the shape parameter. The shape parameter $C_{shape}$, which can adjust the shape of the ellipse to be disky ($C_{shape}<2$) or boxy ($C_{shape}>2$), was not modified in our decompositions (i.e. we assumed simple elliptical shapes, $C_{shape}=2$, for all cases). In conjunction with the ellipses, there are several radial parametric functions to fit the 2D galaxy light distribution:
\begin{itemize}
\item The Sérsic function (\emph{sersic}) has the form
\begin{equation}
    \Sigma(r)=\Sigma_e \exp\left(-b_n\left[\left(\frac{r}{r_e}\right)^{1/n}-1\right]\right),
\end{equation}
where $r$ is the isophotal radius, $r_e$ is the effective radius (i.e. radius which contains half the total flux), $\Sigma_e$ is the surface brightness at effective radius (in sky-plane), $n$ is the Sérsic index, and $b_n$ is the normalisation factor dependent on the Sérsic index. 

\item The exponential function (\emph{expdisk}) has the form
\begin{equation}
    \Sigma(r)=\Sigma_0q^{-1}\exp\left(-\frac{r}{r_s}\right),
\end{equation}
where $r_s$ is the scale length (i.e. radius where the peak flux has fallen by $1/e$), $\Sigma_0$ is the central surface brightness (face-on), and $q$ is the axial ratio ($=b/a$). The combination of $\Sigma_0q^{-1}$ is equivalent to the central surface brightness in the sky-plane.

\item The edge-on disk function (\emph{edgedisk}) has the form
\begin{equation}
    \Sigma(r,h)=\Sigma_0\left[\frac{r}{r_s}\right]K_1\left(\frac{r}{r_s}\right) \sech^2\left(\frac{h}{h_s}\right),
\end{equation}
where $r_s$ is the scale radius, $h_s$ is the scale height, $\Sigma_0$ is the central surface brightness (face-on, same as in exponential function), and $K_1$ is the modified Bessel function. The function is derived from \citet{vanderkruit1981} (see their Eqn.~5), which has an exponential radial dependence. Therefore here $\Sigma_0$ is the same as denoted in the \emph{expdisk} function.

\item The Ferrers function (\emph{ferrer}) has the form
\begin{equation}
    \Sigma(r)=\Sigma_0\left[1-\left(\frac{r}{r_{out}}\right)^{2-\beta}\right]^\alpha,
\end{equation}
where $r_{out}$ is the outer truncation radius, $\Sigma_0$ is the central surface brightness (in sky-plane), $\alpha$ is the parameter which dictates the gradient of the outer truncation, and $\beta$ is the parameter which controls the gradient of the central slope. The Ferrers function is only evaluated within $r < r_{out}$.

\item If a galaxy contains an unresolved element (e.g. a nucleus), the PSF is used to model this component. 

\end{itemize}

\subsubsection{Decomposition strategy}
We apply two main types of decomposition models for our sample of galaxies: Sérsic+PSF and multi-component. The Sérsic+PSF models are useful as there is a significant number of galaxies that contain unresolved components within their light distributions (e.g. nucleus and a disk). 

Such galaxies are not well described by a single Sérsic function, as the unresolved component can cause misleadingly high Sérsic indices. For Sérsic+PSF models, all Sérsic parameters aside from the centre coordinates were free to vary. For the PSF function, which accounts for the unresolved nucleus, the same centre coordinate as for the Sérsic function was used and the PSF magnitude parameter was allowed to vary up to a limiting magnitude of 35. This constraint was applied to ensure that GALFIT does not crash due to unreasonably faint PSF magnitude values, such as in cases where the galaxy does not contain a nucleus (the model effectively turns to single-Sérsic). We note that the limit of 35 magnitudes is very conservative and in practice a PSF magnitude fainter than 30 can be regarded as non-nucleated, given that the contribution of flux from such a component becomes negligible. 

In order to fit multiple components in galaxies, one must first choose a parametric function. In principle there is no limitation on the choice of function for decompositions, but in practice some limitations are useful. For example, dwarf galaxies are generally well described by Sérsic functions. However, for more complex galaxies \citep[i.e. with distinct morphological structures, such as in][]{caon1994} we assign a specific set of functions in GALFIT to fit the physical components in a galaxy. This allows us to remain systematic in conducting decompositions. The list of functions used are: 
\begin{itemize}
    \item bulge: \emph{sersic},
    \item disk: \emph{expdisk}, or \emph{edgedisk} if edge-on,
    \item bar: \emph{ferrer2},
    \item nucleus: \emph{psf},
    \item barlens\footnote{By barlens we mean the vertically extended central part of the bar which, when viewed face-on, has a round appearance \citep[see][]{laurikainen2014,athanassoula2015,salo2017}.}: \emph{expdisk}.
\end{itemize}

For the multi-component decompositions, we employ the philosophy of beginning with a simple model and gradually building up complexity as required. As an initial step, visual inspection of the data images provided conservative estimates of the structures present within the galaxies. This identifies the most significant (in terms of flux contribution) structures and components (e.g. bulges, disks) which allow for the majority of the galaxy's flux to be modelled. Additionally, the model and residual images of the Sérsic+PSF decompositions were inspected for additional inner structures (or lack thereof). The ellipticity and position angle profiles also occasionally suggested bars were present, based on an increase in ellipticity but relatively constant position angle with increasing radii. 

For all multi-component models all component centres were fixed to the values determined in Sect.~\ref{sect:centre}. Furthermore, the axial ratio and position angle of the outermost component (i.e. the component with the largest effective radius/scale length, typically a disk component) were fixed to the outer isophote values measured in Sect.~\ref{sect:skysubtraction}. This reduced the degeneracy and increased GALFIT's fitting speed. After the initial decompositions, we inspected the residuals and iterated the decomposition models accordingly. Due to degeneracies from simultaneously fitting multiple components, the output model parameters from GALFIT which minimise $\chi^2$ are not always physically meaningful. This can occur with the Sérsic $n$ and the Ferrers $\alpha$ and $\beta$ parameters. 

\begin{figure*}
\centering
\resizebox{\hsize}{!}{\includegraphics[width=\hsize]{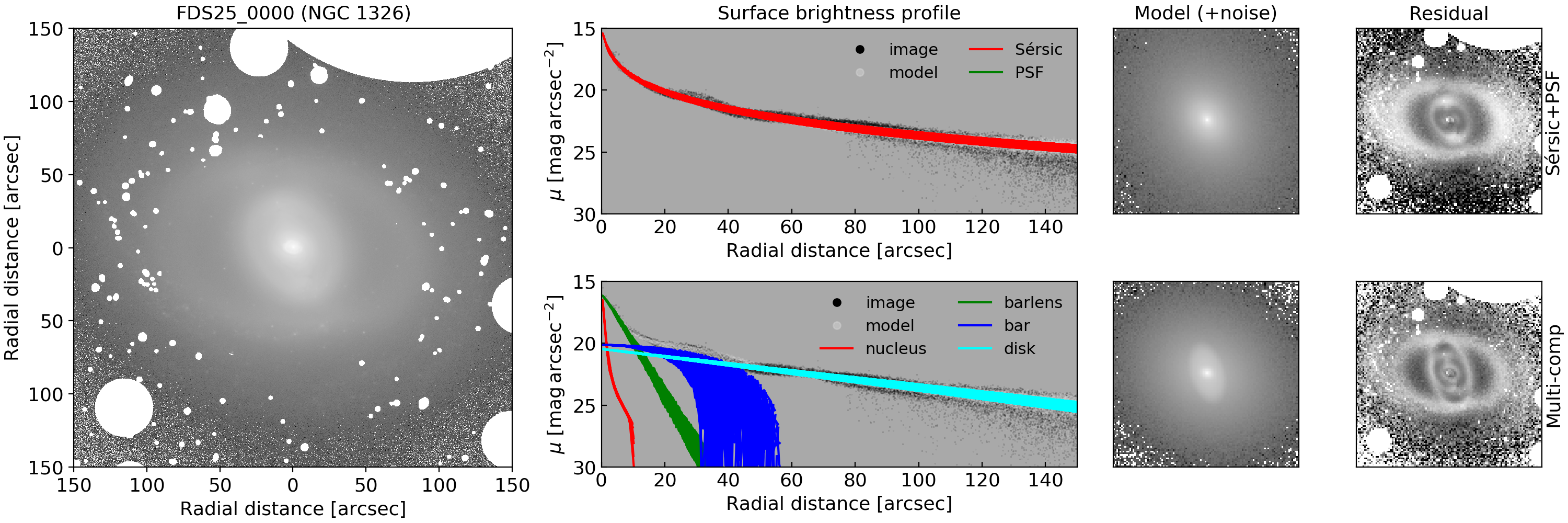}}
\caption{Overview of the Sérsic+PSF (\textit{upper row}) and multi-component (\textit{lower row}) decompositions for FDS25\_0000 (NGC~1326). In the \textit{first column from the left}, we show the masked galaxy image (within the inner 150\,arcsec region). The \textit{second column} shows the surface brightness profiles of the corresponding masked galaxy image as well as the decomposition models. The individual functions/components of the models are also shown, which highlight their contributions to the overall model. The \textit{third} and \textit{fourth columns} show the model and residual images, respectively, within the same region as the masked galaxy image. A range of 30 to 15\,mag\,arcsec$^{-2}$ was used to display the masked galaxy image, the surface brightness profiles, and the model images. For the residual images a range of -1 to 1 mag\,arcsec$^{-2}$ was used instead. }
\label{fig:decomp_summary}
\end{figure*}

As an example, Fig.~\ref{fig:decomp_summary} shows an overview of the Sérsic+PSF and multi-component decompositions for FDS25\_0000 (NGC~1326). In the surface brightness profiles we show the values from the masked galaxy data image as well as the overall decomposition models. We add Gaussian noise to the decomposition models (with standard deviation equal to the sky RMS measured in Sect.~\ref{sect:skysubtraction}) so that visual comparison with the masked image is possible. The differences in surface brightness between the (masked) galaxy and the models are shown in the residual images, which in this case amounts to prominent galactic ring(s). In Appendix~\ref{app:multicomp_table} we list the type of multi-component models for the 50 most massive galaxies (via stellar mass) in our sample. A full overview of the decompositions and tables of decompositions values can be found on our complementary website (see Appendix~\ref{app:website}). 

\subsection{Non-parametric measures} \label{sect:nonparametric}
To further quantify our sample of galaxies, we also compute some non-parametric morphological measures for each galaxy. From the literature, there are several non-parametric indices which measure morphological features. The most popular indices have been based on \citet{conselice2003} and \citet{lotz2004}, although the exact definitions can vary from study to study. Therefore, here we introduce and define the specific parameters we use in this work. All measures were calculated using elliptical apertures with position angles and ellipticities defined in Sect.~\ref{sect:skysubtraction}. When referring to radius we mean the semi-major axis of the elliptical aperture.

\subsubsection{Concentration ($C$)}
The concentration index, following \citet{conselice2003}, is defined as 
\begin{equation}
    C=5\log_{10}(R_{80}/R_{20}),
\end{equation}
where $R_{80}$ and $R_{20}$ are the radii which enclose $80\%$ and $20\%$ of the Petrosian flux, respectively. The Petrosian flux is determined as the total flux enclosed within 1.5 times the Petrosian radius ($r_{\text{petro}}$), which in turn is defined as the radius where the flux is equal to 0.2 times the average flux within the same radius. The concentration index provides insight into how much of the flux of a galaxy is distributed towards the centre. The more centrally concentrated the flux, the higher the value. For a single Sérsic model, $n=1$ corresponds to $C=2.80$, $n=2$ to $C=3.80$, and $n=4$ to $C=5.27$. 

\subsubsection{Asymmetry ($A$)}
The asymmetry index is defined as 
\begin{equation}
    A=\text{min}\left(\frac{\sum|I_0 - I_{180}|}{\sum |I_0|}\right) - A_{\text{background}},\label{eqn:asym}
\end{equation}
where $I_0$ and $I_{180}$ are the original and 180$^\circ$ rotated galaxy images, respectively, $A_{\text{background}}$ is a correction term to account for the contribution to $A$ from the background noise and is defined as 
\begin{equation}
    A_{\text{background}}=\frac{\sum|B_0 - B_{180}|}{\sum |I_0|},
\end{equation}
where similarly $B_0$ and $B_{180}$ are the original and 180$^\circ$ rotated area of the background sky, respectively. The centre of rotation (i.e. asymmetry centre) is determined as the coordinate which minimises the first term of Eqn.~(\ref{eqn:asym}). Furthermore, only pixels within $1.5\times r_{\text{petro}}$ are used in this minimisation term. We calculated $A_{\text{background}}$ using the same asymmetry centre for rotation, unlike what was defined in \citet{conselice2003} where $A_{\text{background}}$ was minimised separately. As the number of pixels in $I_0$ and $B_0$ may differ for each galaxy, we multiply $A_{\text{background}}$ with a scale factor based on the ratio of pixels in $I_0$ and $B_0$ (i.e. $n_{\text{pix},I}/n_{\text{pix},B}$). The asymmetry index can range from 0--which implies a symmetric galaxy--to 2--which implies a highly asymmetric galaxy. 

\subsubsection{Asymmetry profile}
To probe the outskirts of galaxies, particularly for potential signs of galaxy--galaxy interactions/tidal disruptions, we measure the asymmetry parameter as a function of isophotal radius. The centre is fixed to the brightness centre (determined from Sect.~\ref{sect:centre}) and $A$ is calculated within elliptical annuli. The brightness centre was used instead of the asymmetry centre as the profile becomes more sensitive to asymmetric features in the galaxy outskirts, rather than at the centre for which the overall asymmetry is minimised. We used ellipse semi-major axes ranging from 0 to $1.5r_{\text{petro}}$ in steps of $0.25r_{\text{petro}}$

\subsubsection{Clumpiness ($S$)}
The clumpiness (sometimes referred to as the smoothness) index is defined following \citet{rodriguezgomez2019}:
\begin{equation}
    S=\frac{\sum (I_0 - I_{\sigma})}{\sum I_0} - S_{\text{background}},\label{eqn:smoothness}
\end{equation}
where $I_0$ and $I_{\sigma}$ are the original and convolved galaxy images, respectively, and $S_{background}$ is the clumpiness of the background region, defined as 
\begin{equation}
    S_{\text{background}}=\frac{\sum (B_0 - B_{\sigma})}{\sum I_0},
\end{equation}
where $B_0$ and $B_{\sigma}$ are the original and convolved regions of the background sky, respectively. A higher $S$ value means that the galaxy is clumpier (e.g. due to star formation). The convolved images are created by using a Gaussian kernel with $\sigma=0.25\times r_{\text{petro}}$ on the original images. Again, a scale factor of $n_{\text{pix},I}/n_{\text{pix},B}$ must be applied to $S_{\text{background}}$. Pixels within $1.5\times r_{\text{petro}}$ are used to calculate $S$, although there are two caveats: i) the central $0.25\times r_{\text{petro}}$ region of the galaxy is excluded when calculating $S$ as it can be highly concentrated, and ii) only positive residual pixel values are used in calculating $S$, as the 'clumpy' features should be brighter than the convolved counterpart. This applies to the summation of the galaxy images and the background sky region in Eqn.~(\ref{eqn:smoothness}). 

\subsubsection{Gini ($G$)}
The Gini coefficient measures the distribution of flux values amongst the pixels. We define the Gini coefficient as in \citet{lotz2004}
\begin{equation}
    G=\frac{1}{\langle \lvert f \rvert \rangle n_{\text{pix}}(n_{\text{pix}}-1)}\sum_i^{n_{\text{pix}}} (2i-n_{\text{pix}}-1)\lvert f_i \rvert, \label{eqn:gini}
\end{equation}
where $\lvert f_i \rvert$ is the absolute flux value of pixel $i$, $\langle \lvert f \rvert \rangle$ is the mean of the absolute flux values of the image region used, $n_{\text{pix}}$ is the total number of pixels, and where the summation is evaluated over pixels ranked in ascending order of absolute flux. To evaluate $G$ we create the Gini segmentation map, where first a smoothed image was created by convolving the original galaxy image with a Gaussian kernel of $\sigma=0.2\times r_{\text{petro}}$. From the smoothed image, the mean flux at $r_{\text{petro}}$ was calculated and set as a threshold, so the segmentation map consists of pixels with flux greater than this threshold. If all pixels have equal flux value, then $G=0$. Conversely, if all the flux is concentrated to one pixel, $G=1$. The Gini values for galaxies tend to correlate with $C$, as galaxies tend to be brightest at their centres. However, $G$ does not depend on the spatial location of the pixels, meaning galaxies which have highly concentrated light that is not necessarily at the centre of the galaxy can still have high $G$ values. 

\subsubsection{$M_{20}$}
The normalised second order moment of the brightest $20\%$ of light, $M_{20}$, traces the spatial extent of the brightest regions in a galaxy. To begin, the total second order moment of light is defined as 
\begin{equation}
    M_{\text{tot}}=\sum_i^{n_{\text{pix}}} M_i = \sum_i^{n_{\text{pix}}} f_i\left((x_i-x_c)^2+(y_i-y_c)^2\right), \label{eqn:mi}
\end{equation}
where $M_i$ is the second order moment of pixel $i$, $n_{\text{pix}}$ is the total number of pixels, $f_i$ is the flux value of pixel $i$, and $x_i$, $y_i$ and $x_c$, $y_c$ denote the $x$ and $y$ coordinates of pixel $i$ and the centre, respectively. The centre ($x_c$, $y_c$) is defined as the coordinate which minimises $M_{tot}$. $M_{20}$ is then defined as 
\begin{equation}
    M_{20}=\log_{10}\left(\frac{\sum_i^{n_{\text{pix},20}} M_i}{M_{\text{tot}}}\right), \label{eqn:m20}
\end{equation}
where $n_{\text{pix},20}$ denotes the brightest 20\% of pixels. In practice, the summations are evaluated over pixels set by the Gini segmentation map, which defines $n_{\text{pix}}$. Additionally, the pixels in the summations are sorted by flux in descending order to ensure that $n_{\text{pix},20}$ contains 20\% of the total flux from $n_{\text{pix}}$. 

\subsubsection{Colour difference}
In addition to the integrated colours (see Sect.~\ref{sect:aper_phot}), we also compare the radial colour distributions of the galaxies within elliptical apertures. The radial changes in colour can provide insight into the star formation history within a galaxy, hence environmental effects can potentially be imprinted. We define the colour difference as
\begin{equation}
    \Delta \text{colour}=\text{colour}(1R_e < r < 2R_e) - \text{colour}(r < 0.5R_e), \label{eqn:col_grad_aper}
\end{equation}
where $r$ is the isophotal radius, $\text{colour}(1R_e < r < 2R_e)$ denotes the colour (e.g. $g'-r'$) based on the total fluxes calculated between one and two effective radii of the galaxy, and similarly $\text{colour}(r < 0.5R_e)$ denotes the colour calculated within half an effective radius of the galaxy. We calculate the colours within elliptical annuli with position angles and ellipticities measured as explained in Sect.~\ref{sect:skysubtraction}. A positive $\Delta \text{colour}$ implies a bluer inner region and redder outer region, and vice versa for negative $\Delta \text{colour}$. 

\section{Comparison of parameters}\label{sect:compare_params}
\begin{figure}
\centering
\includegraphics[width=\hsize]{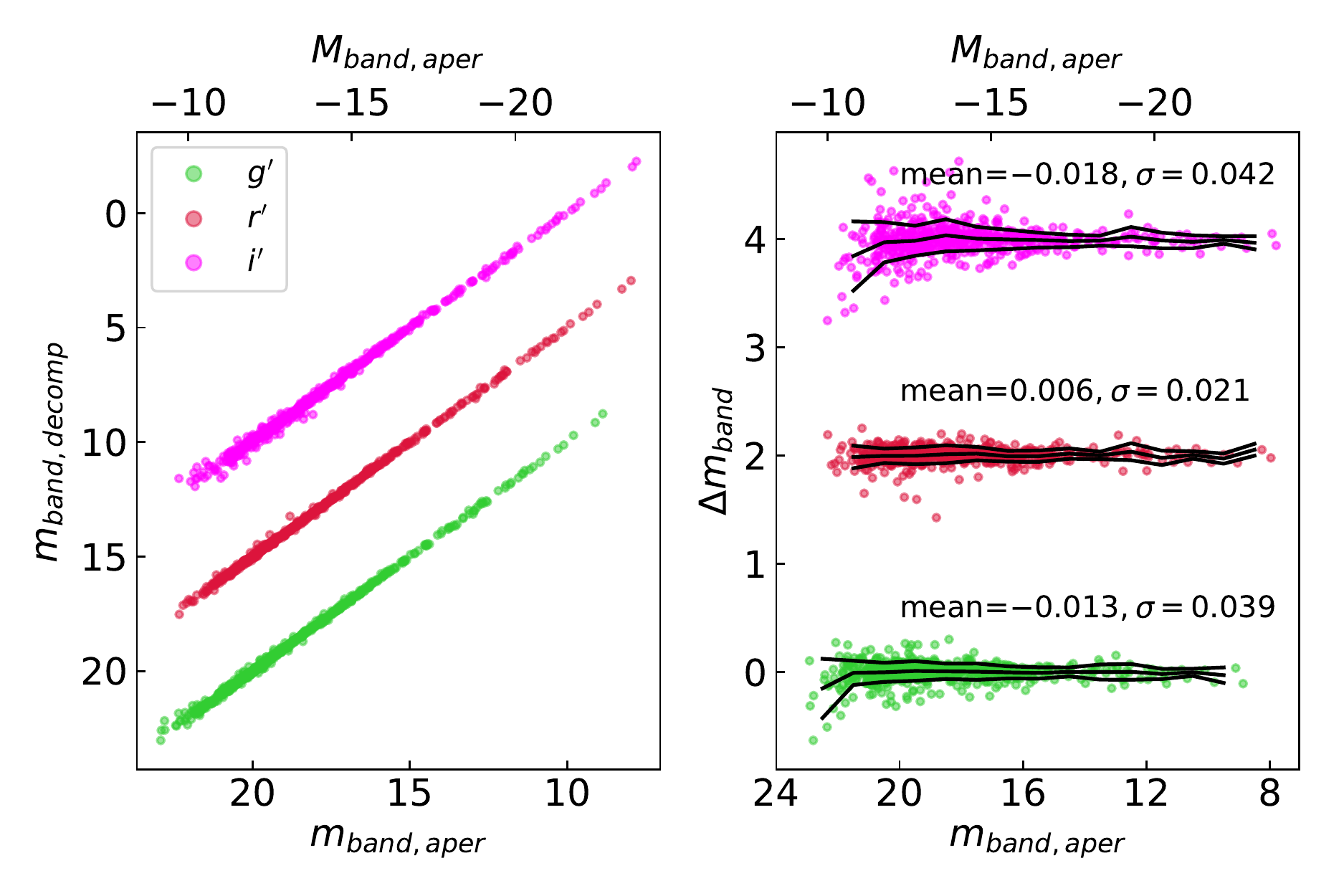}
  \caption{\textit{Left panel}: Total magnitudes calculated from multi-component decompositions as a function of aperture magnitudes, in $g'$ (\textit{green}), $r'$ (\textit{red}), and $i'$-band (\textit{fuchsia}). \textit{Right panel}: Difference in magnitudes $\Delta m_{\text{band}}$ (i.e. $m_{\text{band,decomp}} - m_{\text{band,aper}}$) as a function of aperture magnitudes. The black lines denote the mean and $\pm$RMS in bins of 1\,mag. The annotations show the mean and standard deviation of the means of $\Delta m_{\text{band}}$. For better visibility, each magnitude relation is offset by 5\,mag (\textit{left}) and 2\,mag (\textit{right}) relative to each other, with $g'$-band as the reference relation.}
     \label{fig:compare_mag1}
\end{figure}

We compare the parameters from aperture photometry against those from structural decompositions, both from this work as well as those from FDSDC. This allows us to estimate the uncertainty on the magnitudes, stellar masses, and decomposition parameters. Figure.~\ref{fig:compare_mag1} shows the total magnitudes calculated in this work via aperture photometry (see Sect.~\ref{sect:aper_phot}) and multi-component decompositions (see Sect.~\ref{sect:decomp_theory}). The right panel shows the difference between aperture and decomposition magnitudes for each galaxy. Overall, the magnitudes agree quite well in the three bands, with the largest scatter occurring at the faintest magnitudes. The scatter at the faint magnitudes are likely due to the lower S/N. The uncertainties in the magnitudes for each band are tabulated in left of Table~\ref{tab:mag_errors}. 

\begin{table}
    \caption{Uncertainties in $g'$, $r'$, and $i'$ magnitudes and stellar mass.}
    \centering
    \begin{tabular}{rccc|rc}
\hline
  $m_{\text{band,aper}}$ & RMS$_{g'}$ & RMS$_{r'}$ & RMS$_{i'}$ & $\log_{10} (\frac{M_*}{M_{\odot}})$ & RMS$_{M_*}$ \\
\hline
 8.50 &  - & 0.05 & 0.06 & - & -\\
 9.50 &  0.07 & 0.05 & 0.03 & 11.25 & 0.06 \\
10.50 &  0.03 & 0.03 & 0.06 & 10.75 & 0.05 \\
11.50 &  0.05 & 0.07 & 0.07 & 10.25 & 0.07 \\
12.50 &  0.07 & 0.08 & 0.09 &  9.75 & 0.10 \\
13.50 &  0.07 & 0.03 & 0.05 &  9.25 & 0.09 \\
14.50 &  0.04 & 0.05 & 0.06 &  8.75 & 0.07 \\
15.50 &  0.05 & 0.05 & 0.07 &  8.25 & 0.09 \\
16.50 &  0.06 & 0.05 & 0.09 &  7.75 & 0.11 \\
17.50 &  0.08 & 0.06 & 0.11 &  7.25 & 0.14 \\
18.50 &  0.07 & 0.08 & 0.15 &  6.75 & 0.17 \\
19.50 &  0.09 & 0.08 & 0.14 &  6.25 & 0.20 \\
20.50 &  0.09 & 0.07 & 0.19 &  5.75 & 0.29 \\
21.50 &  0.11 & 0.10 & 0.32 &  5.25 & 0.37 \\
22.50 &  0.27 & - & - & - & -\\
\hline
\end{tabular}

    \tablefoot{\textit{Left}: The RMS of the difference in aperture and decomposition magnitudes for a range of aperture magnitude bins (black lines in Fig.~\ref{fig:compare_mag1}). These can be used as the estimated uncertainties of the calculated magnitudes. \textit{Right}: The mean RMS for stellar mass within bins of $\log_{10}(M_*/M_{\odot})$, which estimates the uncertainty in stellar mass due to uncertainties in the total magnitudes (i.e. values from the \textit{left}). The RMS were calculated based on 1000 Monte-Carlo realisations of stellar mass for each galaxy in our sample (see Eqn.~\ref{eqn:mstar}, and the accompanying text in Sect.~\ref{sect:compare_params}).}
    \label{tab:mag_errors}
\end{table}

For further comparisons, stellar masses were calculated to characterise the galaxies. To estimate the stellar mass of our sample galaxies we adopt the empirical relation between colours and stellar mass to light ratio ($M_*/L$) from \citet{taylor2011}, adapted for SDSS bands \citep[i.e.][their Eqn.~2]{venhola2019}:
\begin{equation}
\log_{10}\left( \frac{M_*}{M_{\odot}} \right)=1.15+0.70(g'-i')-0.4M_{r'}+0.4(r'-i') \label{eqn:mstar}
\end{equation}
where $M_*$ is the stellar mass, $M_{\odot}$ is the solar mass, $M_{r'}$ is the absolute $r'$-band magnitude, and $g'$, $r'$, $i'$ denote the total magnitudes in their respective bands. 
The stellar mass estimates have a reported uncertainty of 0.1\,dex within $1\sigma$ accuracy using only $g'$ and $i'$ bands \citep{taylor2011}. However, galaxies with $\log_{10}(M_*/M_{\odot})<7.5$ were not included in deriving Eqn.~(\ref{eqn:mstar}), so the relation must be extrapolated for lower masses. \citet{venhola2019} used an independent stellar mass estimate based on \citet{bell2001} to test the uncertainty in the low mass region. They reported that both methods gave consistent results within $\sim 10\%$ error. On top of the uncertainty due to differences in the methods used, we also calculate the contribution to stellar mass uncertainties from the uncertainty in the magnitudes. This was done by conducting 1000 Monte-Carlo simulations for the stellar mass for each galaxy in our sample, using the total magnitudes and corresponding uncertainties from the left of Table~\ref{tab:mag_errors}. We then calculate the RMS and mean stellar mass for each galaxy and use the average RMS within bins of stellar mass as the uncertainty for a given stellar mass. This is shown in the right of Table~\ref{tab:mag_errors}. 

\begin{figure}
\centering
\includegraphics[width=\hsize]{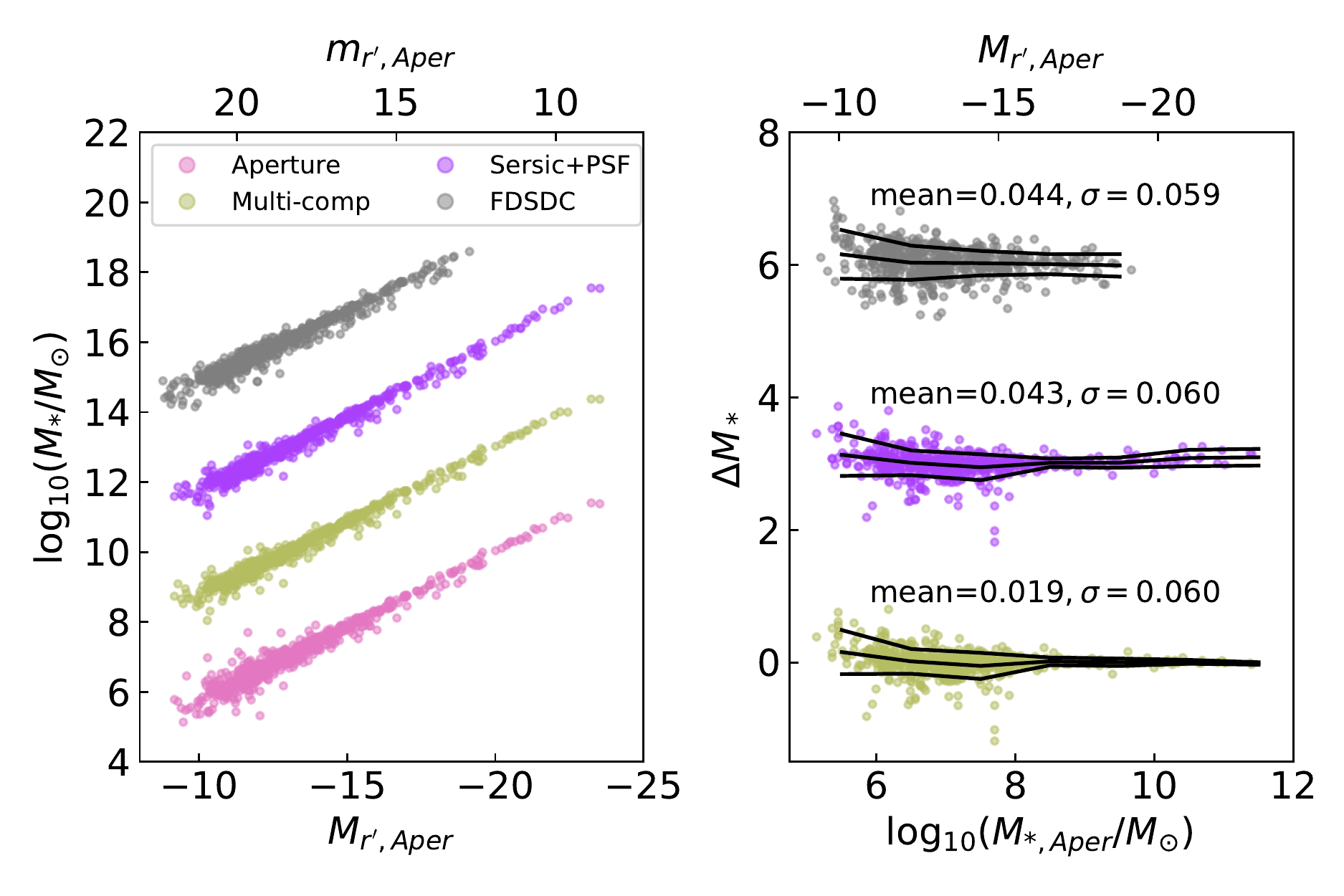}
  \caption{Comparison of stellar mass calculated via different methods of determining total magnitudes, as a function of $r'$-band total magnitudes (\textit{left}) and stellar mass from aperture photometry (\textit{right}). Stellar masses for FDSDC were estimated by multiplying the fluxes in each band by a factor of two, as the aperture used in \citet{venhola2018} only extended to one effective radius. The black lines denote the mean and $\pm$RMS in bins of 1\,dex. The annotations show the mean and standard deviation of the means of $\Delta M_*$, where $\Delta M_*$ is defined as the difference in exponents (i.e. $\log_{10}(M_{*,\text{Aper}}) - \log_{10}(M_{*,\text{other}})$). For better visibility, each relation is offset by 3\,dex relative to each other, with "Aperture" (pink) as the reference relation on the left plot whilst "Multi-comp" (green) as the reference relation on the right plot.}
     \label{fig:compare_mag2}
\end{figure}

Figure~\ref{fig:compare_mag2} also compares the stellar masses we derive from different magnitude estimates, plotted as a function of aperture $r'$-band magnitude. The stellar masses were calculated from magnitudes obtained by Sérsic+PSF and multi-component decompositions, as well as from FDSDC. Overall, there is good agreement regardless of method. As with magnitude estimates, the scatter in stellar masses is larger for the lower mass galaxies. The multi-component stellar masses give the smallest mean difference with the aperture values ($\sim 0.02$\,dex). For $\log_{10}(M_*/M_{\odot}) > 8$, the RMS of the mean $\Delta M_{*,Multi}$ values is $~0.01$. Henceforth, we refer to the stellar mass of galaxies as the values calculated via Eqn.~(\ref{eqn:mstar}) using magnitudes from multi-component decompositions. 

\begin{figure}
\centering
\includegraphics[width=\hsize]{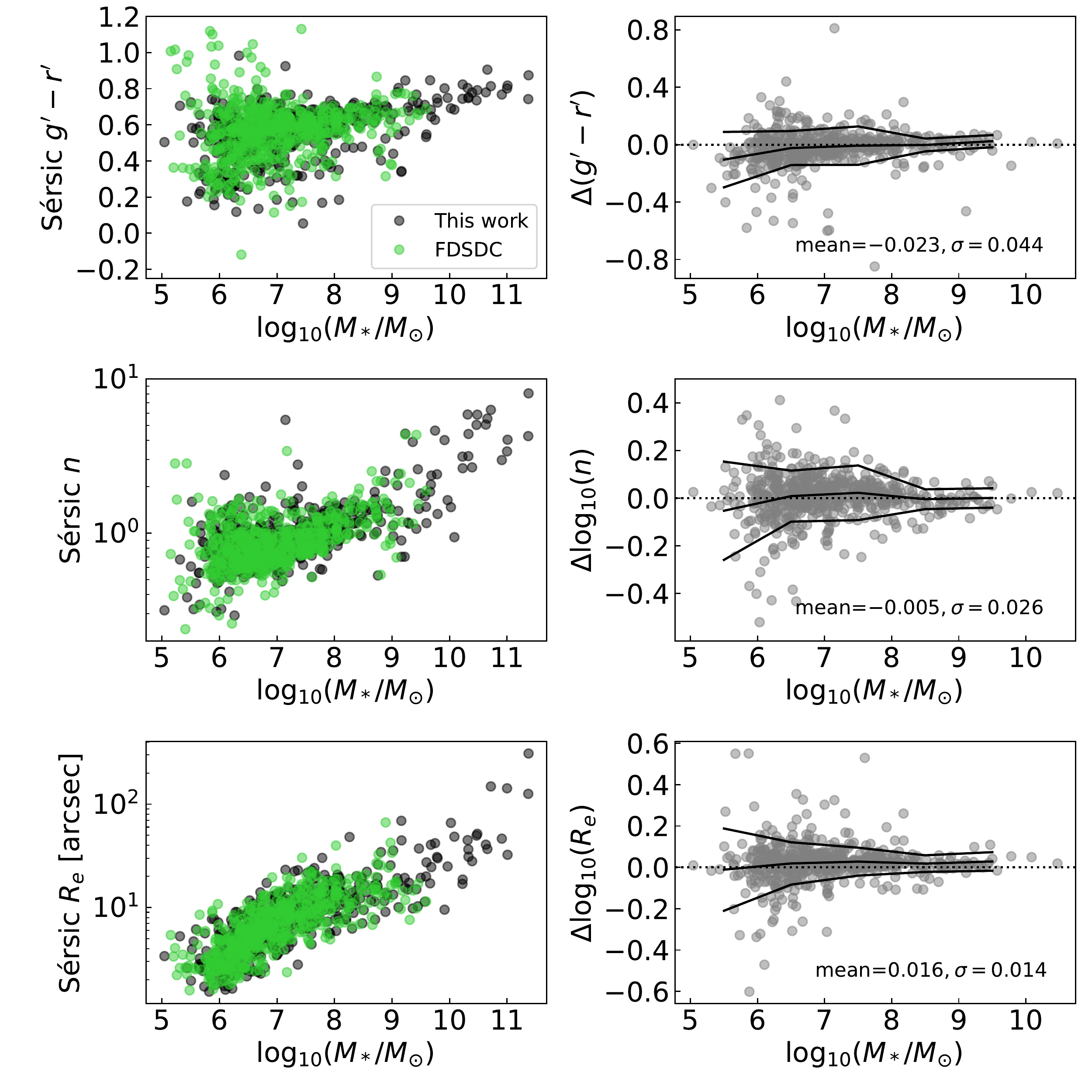}
  \caption{Comparison of the $r'$-band Sérsic-derived quantities from Sérsic+PSF decompositions between those made in this work and in FDSDC\protect \footnotemark. FDSDC sample ranges from $-9>M_{r'}>-18.5$, corresponding to $10^5 M_{\odot} \lesssim M_* \lesssim 10^9 M_{\odot}$. \textit{Left panels}: The relations of Sérsic-derived quantities as a function of stellar mass. \textit{Right panels}: The difference in Sérsic-derived quantities for each galaxy (where the samples overlap) as a function of the parameters determined in this work. Here $\Delta$ is defined as the values from this work minus the values from FDSDC. The black lines denote the mean and $\pm$RMS in bins of 1\,dex. The annotations show the mean and standard deviation of the means of $\Delta$.}
     \label{fig:compare_sersic}
\end{figure}

In Fig.~\ref{fig:compare_sersic} we compare the quantities derived from the Sérsic+PSF models ($g'-r'$, Sérsic $n$, and $R_e$) made in this work with those from FDSDC. The Sérsic $n$ and $R_e$ were calculated based on Sérsic+PSF decompositions on $r'$-band images. For the $g'-r'$ colour, the $g'$-band magnitudes were calculated using Sérsic+PSF models calculated from $r'$-band (i.e. model with $n$, $R_e$, $q$, and $PA$ fixed to values found from $r'$-band decomposition for the Sérsic component, but the magnitude from Sérsic and PSF were free parameters to be fitted). For brevity, we denote these quantities derived from the Sérsic component of Sérsic+PSF models as 'Sérsic-derived quantities'. Regarding $g'-r'$ colours, FDSDC reports a redder colour for a handful of galaxies at the low mass range ($5.5 < \log_{10}\left(M_*/M_{\odot}\right) < 6.5$), but is otherwise consistent with our colours. Values of Sérsic $n$ also agree quite well. We note, however, that some galaxies (in both studies) show $n<0.5$. When converting to 3D, this would imply that there is a depression or 'dent' in the light distribution of the galaxy. In our cases it is more likely due to low S/N. In general the distributions of Sérsic-derived quantities agree between the two measures, with remarkably little scatter in $R_e$.  

\footnotetext{FDS10\_0143 and FDS5\_0010 fall outside the Sérsic $n$ and $\Delta n$ plots, respectively, due to extreme values from this work compared to FDSDC. For FDS10\_0143 this is likely due to what we think is a foreground star overlapping with the galaxy, which the PSF component does not fully account for. For FDS5\_0010 the cause for the abnormally low Sérsic $n$ is uncertain, although its $r'$-band surface brightness profile does appear remarkably flat.}

\section{Fornax main versus Fornax group}\label{sect:results}
\begin{figure*}
\centering
\resizebox{\hsize}{!}{\includegraphics[width=\hsize]{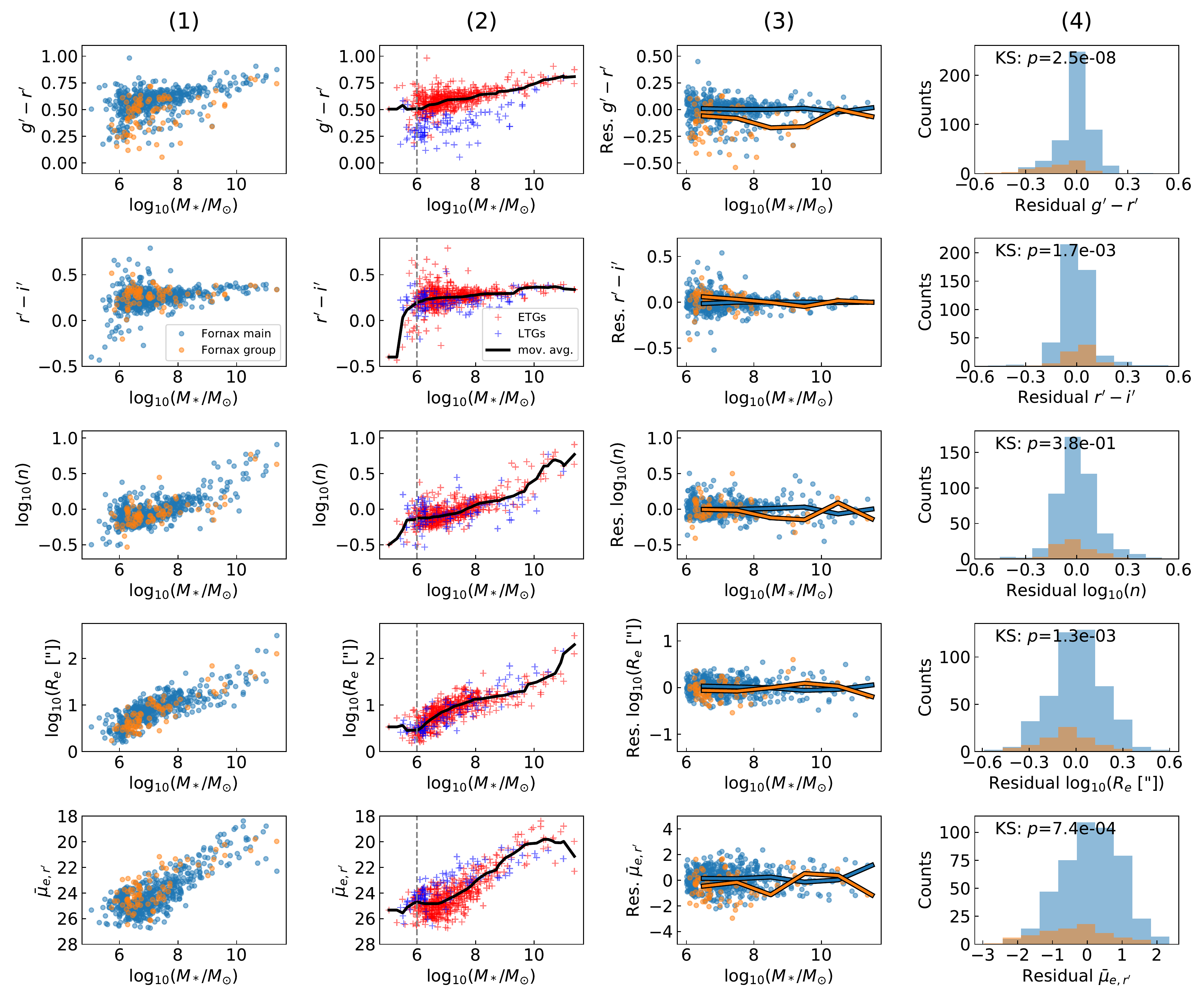}}
  \caption{Scaling relations for $r'$-band Sérsic-derived quantities from Sérsic+PSF decompositions. \textit{Column 1} $g'-r'$ colour, $r'-i'$ colour, Sérsic index, effective radius (in arcsec), and mean effective surface brightness (in mag\,arcsec$^{-2}$) as a function of stellar mass, between Fornax main (\textit{blue}) and Fornax group (\textit{orange}) galaxies. \textit{Column 2} Similar to \textit{column 1}, but split between ETGs (\textit{red crosses}) and LTGs (\textit{dark blue crosses}). The \textit{solid black} lines denote the moving averages (median) based on moving a fixed-size bin through increasing stellar mass to find the mass dependence in each parameter. The vertical \textit{grey dashed} line denotes $\log_{10}(M_*/M_{\odot})=6$. \textit{Column 3} Residual parameters (i.e. parameters minus derived mass-trend from moving averages) as a function of mass. The solid lines denote the median residual value within stellar mass bins of width 1\,dex for the Fornax main (\textit{blue}) and Fornax group (\textit{orange}) sub-samples, respectively. Galaxies with $\log_{10}(M_*/M_{\odot})<6$ were excluded due to incompleteness in the sample. \textit{Column 4} Distribution of the residual parameters for the Fornax main (\textit{blue}) and Fornax group (\textit{orange}) sub-samples. The $p$-value from null hypothesis tests (i.e. both sub-samples are drawn from the same distribution) are annotated in the plot, indicating the probability that both distributions are drawn from the same sample.}
     \label{fig:scale_rel_nomasstrend}
\end{figure*}

In order to investigate potential effects of the cluster environment compared to the group environment, we split our galaxy sample into sub-samples (see Fig.~\ref{fig:radec}): the Fornax main galaxies (centred on NGC~1399/FDS11\_0003) and the Fornax group galaxies (centred on NGC~1316/FDS26\_0001). From \citet{drinkwater2001} the main cluster has a virial radius of 2\,deg, whereas the Fornax group does not have a well defined virial radius as it is in the act of falling into the main cluster. Nevertheless, we can estimate the group's size using a $2\sigma$ limit in number density \citep{drinkwater2001}, which covers a region of $\sim$1\,deg in radius \citep[see also][Fig. 3]{venhola2019}. Using these regions, we assigned to the Fornax group sub-sample all galaxies with an angular separation to NGC~1399 that is greater than twice the angular separation compared to NGC~1316. We exclude the two aforementioned galaxies with anomalous Sérsic $n$ from further analyses of Sérsic-derived quantities. The sub-sample sizes for Fornax main and Fornax group are 497 and 83 galaxies, respectively. Additionally, we utilise the ETG/LTG classification scheme of \citet{venhola2018} for the dwarfs and extend it to the bright galaxies via visual inspections. 

\subsection{Scaling relations}\label{sect:scale_rel}
\subsubsection{Sérsic-derived quantities}
Based on the Sérsic component of $r'$-band Sérsic+PSF decompositions, we investigate trends in the Sérsic index ($n$), Sérsic effective radius ($R_e$), and mean effective surface brightness ($\bar{\mu}_{e,r'}$), as well as $g'-r'$ and $r'-i'$ colours as functions of stellar mass. The mean effective surface brightness is defined as
\begin{equation}
    \bar{\mu}_{e,r'}=-2.5\log_{10}\left(\frac{\frac{1}{2}f_{\rm tot}}{q\pi R_e^2}\right),\label{eqn:mue}
\end{equation}
where $f_{\rm tot}$ is the total flux, $q$ is the axial ratio, and $R_e$ is the effective radius (in arcseconds).

The first column in Fig.~\ref{fig:scale_rel_nomasstrend} shows the scaling relations of each Sérsic-derived quantity. We find redder $g'-r'$ and $r'-i'$ colour, increasing Sérsic $n$, larger $R_e$, and brighter $\bar{\mu}_{e,r'}$ with higher stellar mass for our sample. The most distinct difference between Fornax main galaxies and Fornax group galaxies occur in the $g'-r'$ colour, where Fornax group clearly contains a higher number fraction of blue galaxies. This is in agreement with what was observed in \citet{raj2020}, \citet{iodice2019}, and \citet{spavone2020}. It is worth mentioning that the trends between $R_e$ and brighter $\bar{\mu}_{e,r'}$ with stellar mass are a form of the Kormendy relations \citep{kormendy1977, kormendy2009}. We also note an apparent steepening in the $\log_{10}(R_e) - \log_{10}(M_*/M_{\odot})$ relation, for galaxies with $\log_{10}(M_*/M_{\odot})<8$ (see also \citealt{munozmateos2015}, their Fig.~15; \citealt{venhola2019}, their Fig.~12). In the second column of Fig.~\ref{fig:scale_rel_nomasstrend} we show the same scaling relations as the first column, but we highlight ETGs and LTGs. For the $g'-r'$ colour, there is a clear separation between ETGs--which dominate the red sequence--and LTGs--forming the blue cloud. In contrast, the $r'-i'$ colour does not show such distinctions between ETGs and LTGs, with the majority of galaxies lying along the red sequence. It is possible that because $g'-r'$ can reliably estimate the mean stellar age of galaxies \citep{eminian2008,cerulo2019}, $g'-r'$ can better separate red and blue galaxies compared to $r'-i'$. As such, we primarily use $g'-r'$ when discussing galaxy colours.

The aforementioned trends we find with stellar mass, however, are general across galaxy populations, so in order to isolate the effects of the environment, we must remove the stellar mass trends for each parameter. To this end, we calculated the moving average for each quantity using the whole sample (i.e Fornax main and Fornax group galaxies)\footnote{We also tried using only galaxies within the virial radius of the Fornax main cluster to calculate the moving averages. This, in principle, could better highlight the differences in parameters compared to the outskirts of Fornax main and to the Fornax group. However, the results did not differ significantly, due to the fact that our sample galaxies are dominated by galaxies in the inner region of the Fornax main cluster.} and use them as the stellar mass trends/scaling relations. A bin of fixed width at 1\,dex was moved through the values of stellar mass in ascending order, with a step size of 0.2\,dex. The median of the bin was calculated at each step to create moving average profiles for each quantity. The mass trends were subtracted from all galaxies in each sub-sample. We exclude residual parameters which have $\log_{10}(M_*/M_{\odot})<6$ as FDSDC is not complete at low stellar masses (Venhola et al. in prep.). This excludes 32 galaxies from our sample. Figure~\ref{fig:scale_rel_nomasstrend} shows the averages as well as the distributions of residual parameters. 

Overall, we observe a clear separation in the median values of the residual $g'-r'$ colour between Fornax main and Fornax group (see column 3 of Fig.~\ref{fig:scale_rel_nomasstrend}). There are no obvious differences in the residual Sérsic $n$ for galaxies with $\log_{10}(M_*/M_{\odot})\lesssim 8$, beyond which Fornax group galaxies appear to have marginally lower Sérsic $n$. Fornax group galaxies tend to have smaller effective radii at $\log_{10}(M_*/M_{\odot})\lesssim 8$, but are larger at higher masses. A similar feature can be seen in $\bar{\mu}_{e,r'}$ (likely due to the $R_e^2$ term in Eqn.~\ref{eqn:mue}), where around $\log_{10}(M_*/M_{\odot})\lesssim 8$ Fornax group galaxies tend to have brighter $\bar{\mu}_{e,r'}$ than Fornax main galaxies. 

To quantitatively differentiate the distributions of residual parameters between Fornax main and Fornax group, two sample Kolmogorov-Smirnov (KS) tests were carried out. KS tests compare the difference in distributions from two samples by measuring the separation between their cumulative distributions. It tests the null hypothesis that both samples are drawn from the same distribution. For each comparison, we calculate the probability that the observed KS statistic can occur assuming the null hypothesis (i.e. $p$-value, see column 4 of Fig.~\ref{fig:scale_rel_nomasstrend}). Based on the $p$-values (using significance level $\alpha=0.05$), the $g'-r'$\footnote{Whilst the bluer $g'-r'$ colour and significant $p$-value from KS testing can indicate that Fornax group galaxies are generally younger than their Fornax main counterparts, we note that the colour difference is small, such that the difference in age (if any) are likely small.}  and $r'-i'$ colours, $R_e$ and $\bar{\mu}_{e,r'}$ distributions appear to be very unlikely to be drawn from the same population, whereas in terms of Sérsic $n$ we cannot exclude the possibility that they are drawn from the same population.

\begin{figure}
\centering
\includegraphics[width=\hsize]{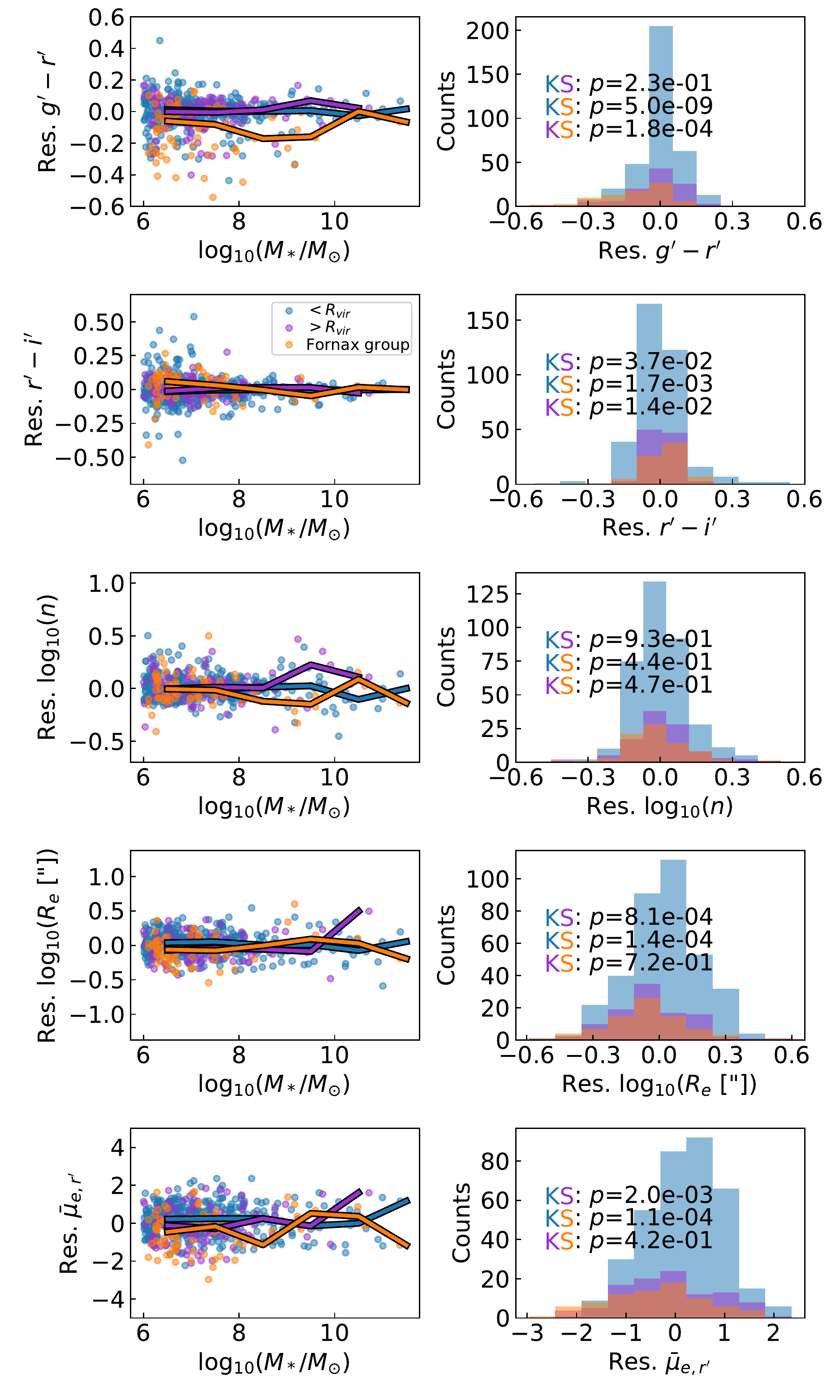}
  \caption{Same as columns 3 and 4 of Fig.~\ref{fig:scale_rel_nomasstrend}, but with three sub-samples: Fornax main - within $R_{\rm vir}$ (\textit{blue}), Fornax main - beyond $R_{\rm vir}$ (\textit{purple}), and Fornax group (\textit{orange}).}
     \label{fig:scale_rel_nomasstrend_3sub}
\end{figure}

\begin{figure*}
\centering
\resizebox{\hsize}{!}{\includegraphics[width=\hsize]{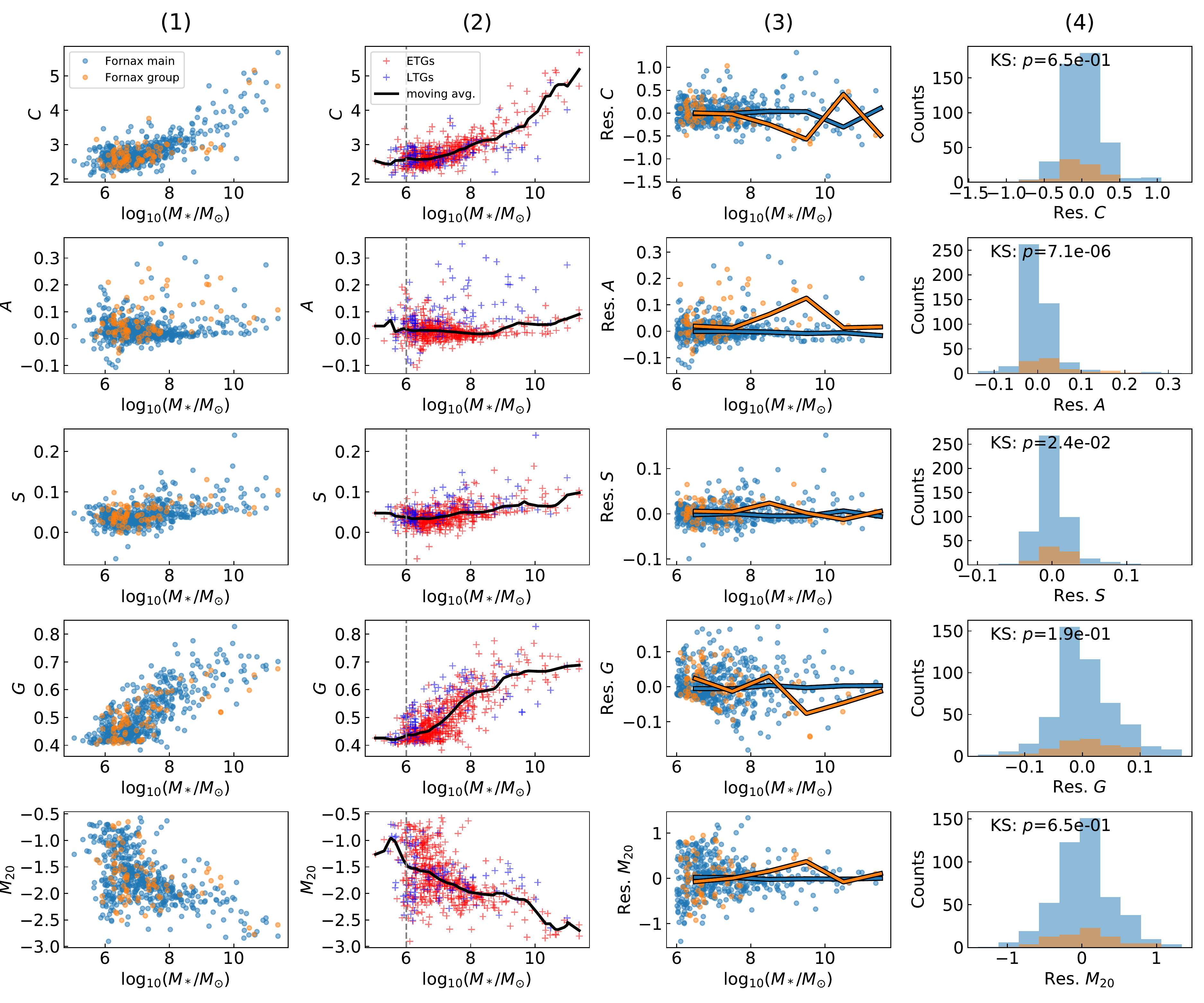}}
  \caption{Same as Fig.~\ref{fig:scale_rel_nomasstrend}, but for non-parametric morphological indices defined in Sect.~\ref{sect:nonparametric}.}
     \label{fig:nonparametric_scalerel_full}
\end{figure*}

In order to investigate any potential differences between the core and outskirts of the Fornax main cluster, we further split it at the virial radius ($700\,$kpc, see Fig.~\ref{fig:radec}) and re-examine the scaling relations. We also include to the comparison the Fornax group. The scaling relations of all three sub-samples are shown in Fig.~\ref{fig:scale_rel_nomasstrend_3sub}. We find that the residual $g'-r'$ colours for galaxies beyond the virial radius do not differ much from those within the virial radius, but both remain significantly different from the Fornax group sub-sample. Regarding Sérsic $n$ we do not find significant differences between the sub-samples. For $R_e$ and $\bar{\mu}_{e,r'}$, however, galaxies beyond the virial radius appear to be more similarly distributed to Fornax group than those from within the virial radius. The $p$-values reflect this, implying that it is highly unlikely that $R_e$ and $\bar{\mu}_{e,r'}$ for galaxies within and beyond the virial radius are drawn from the same distribution.

\subsubsection{Non-parametric morphological indices}
Besides Sérsic-derived quantities, we also compare the non-parametric morphological measures as a function of the galaxies' stellar mass. The non-parametric measures were calculated using $r'$-band images. Figure~\ref{fig:nonparametric_scalerel_full} shows, for each non-parametric index, the measured and residual (i.e. mass-trend subtracted) values as a function of stellar mass, following the format of Fig.~\ref{fig:scale_rel_nomasstrend}. 

Overall, the median residual values of $C$ (within mass bins) for galaxies within $6<\log_{10}(M_*/M_{\odot})<7.5$ appear remarkably similar between Fornax main and Fornax group. For $\log_{10}(M_*/M_{\odot})>8$ the scatter becomes larger and the median values appear to differ between Fornax main and Fornax group. Based on the two-sample KS test the $p$-value suggests that we cannot confidently reject the possibility that both distributions are drawn from the same sample. This is similar to what we observed for Sérsic $n$ (see Fig.~\ref{fig:scale_rel_nomasstrend}), which is expected given the correlation between the two quantities. In fact, \citet{trujillo2001} showed that for a pure Sérsic model, $C$ is a monotonically increasing function of Sérsic $n$ (see also \citealt{graham2005}, their Eqn.~21; \citealt{venhola2018}, their Fig.~13).

\begin{figure}
\centering
\includegraphics[width=\hsize]{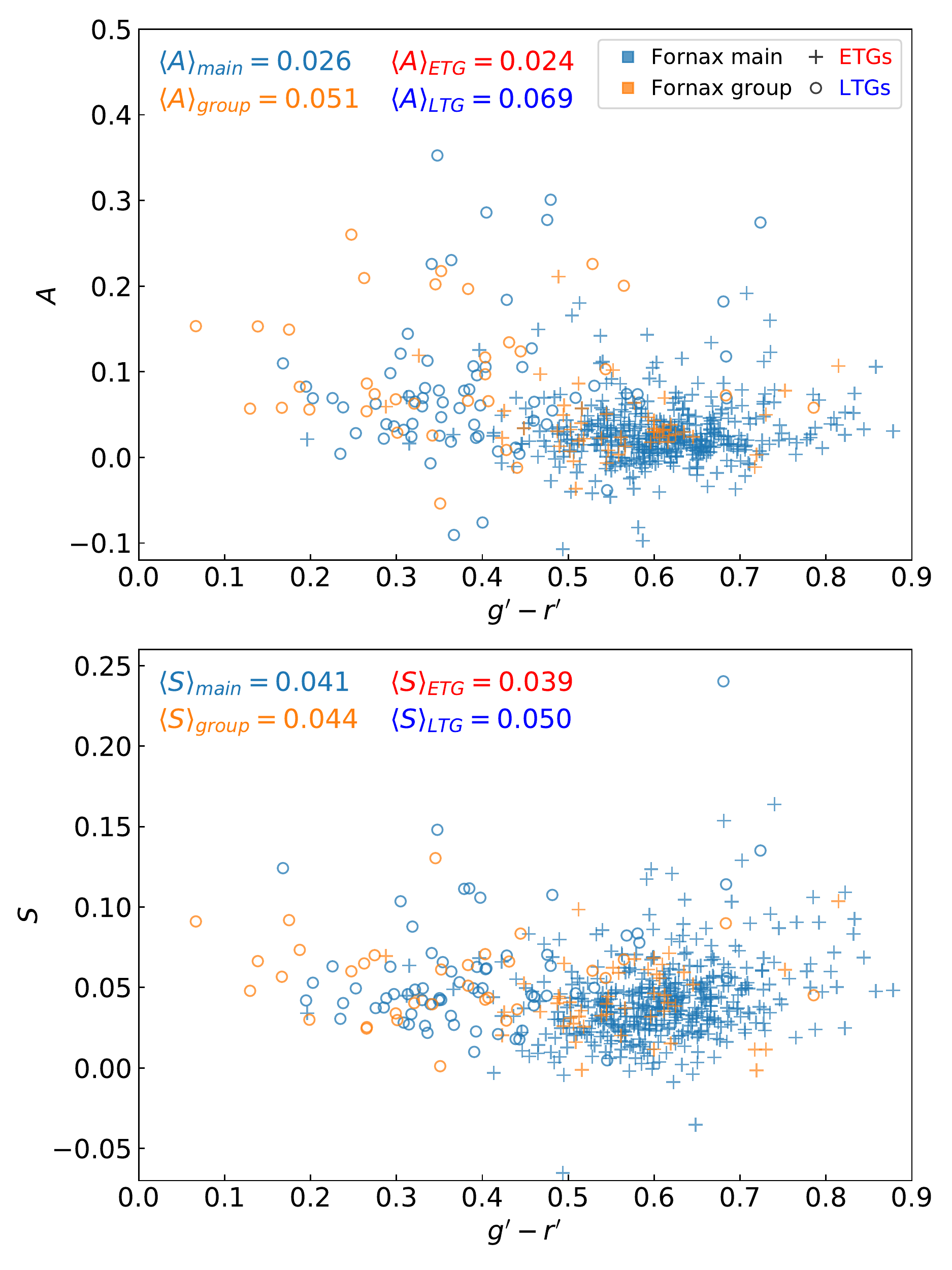}
  \caption{Asymmetry (\textit{upper panel}) and clumpiness (\textit{lower panel}) index as a function of $g'-r'$ colour, split between Fornax main (\textit{blue}) and Fornax group (\textit{orange}), as well as ETGs (\textit{plus}) and LTGs (\textit{circle}). Here $\langle A \rangle$ and $\langle S \rangle$ denote the median values of asymmetry and clumpiness, respectively, for each sub-sample.}
     \label{fig:clump_asym_vs_colour}
\end{figure}

The residual asymmetry index $A$ generally appears to be higher for Fornax group galaxies compared to Fornax main galaxies, based on the binned median values. The difference between Fornax group and Fornax main is also reflected by the low $p$-value of the KS test (well below the significance level of 0.05). Figure~\ref{fig:clump_asym_vs_colour} shows a clear difference in $\langle A \rangle$ between ETGs and LTGs, which suggests that the difference between Fornax group and Fornax main can potentially be explained by the different fractions of ETGs and LTGs.

Regarding the residual clumpiness index $S$, Fornax group galaxies are clumpier than their Fornax main counterparts. The difference is reflected in the low $p$-value from the KS test, which suggests that the distributions are likely drawn from different samples. This is also likely due to the higher fraction of LTGs in Fornax group (see Fig.~\ref{fig:clump_asym_vs_colour}), which tend to have more sub-structures in star forming regions than ETGs. 

The median residual $G$ and $M_{20}$ values for Fornax group galaxies appear similar to their Fornax main counterparts up to $\log_{10}(M_*/M_{\odot}) \approx 9$. This suggests that the overall distribution of flux, as well as the spatial distribution of the brightest regions within each galaxy is comparable between the two environments. 

To summarise the scaling relations of non-parametric indices, the moving averages show that $C$ and $G$ have clear positive correlations with stellar mass, whereas $M_{20}$ shows a clear negative correlation. In contrast, for $A$ and $S$ the moving averages appear relatively flat with stellar mass. Calculating Spearman's $\rho$ for these two quantities we find that $S$ has a significant positive correlation ($\rho_{S} \approx 0.4$), whereas for $A$ there is no significant correlation ($\rho_{A} \approx 0.001$).

From Fig.~\ref{fig:clump_asym_vs_colour} we find the median asymmetry values for Fornax main galaxies are consistent with what \citet{conselice2003} (see their Table~6) found for bright and dwarf ellipticals ($\langle A \rangle_{\rm elliptical} \sim 0.02 \pm 0.02$ and $\langle A \rangle_{\rm dE} \sim 0.02 \pm 0.03$, respectively). For the late-types and dwarf irregulars the medium values ($\langle A \rangle_{\rm late-type\text{ }spiral} \sim 0.15 \pm 0.06$ and $\langle A \rangle_{\rm irregular} \sim 0.17 \pm 0.10$, respectively) are higher than what we find for our LTGs, although differences in the datasets (e.g. shallower depth and more distant cluster) could play a factor.

\subsection{Halo-centric relations}\label{sect:pars_vs_dist}
\begin{figure}
\centering
\includegraphics[width=\hsize]{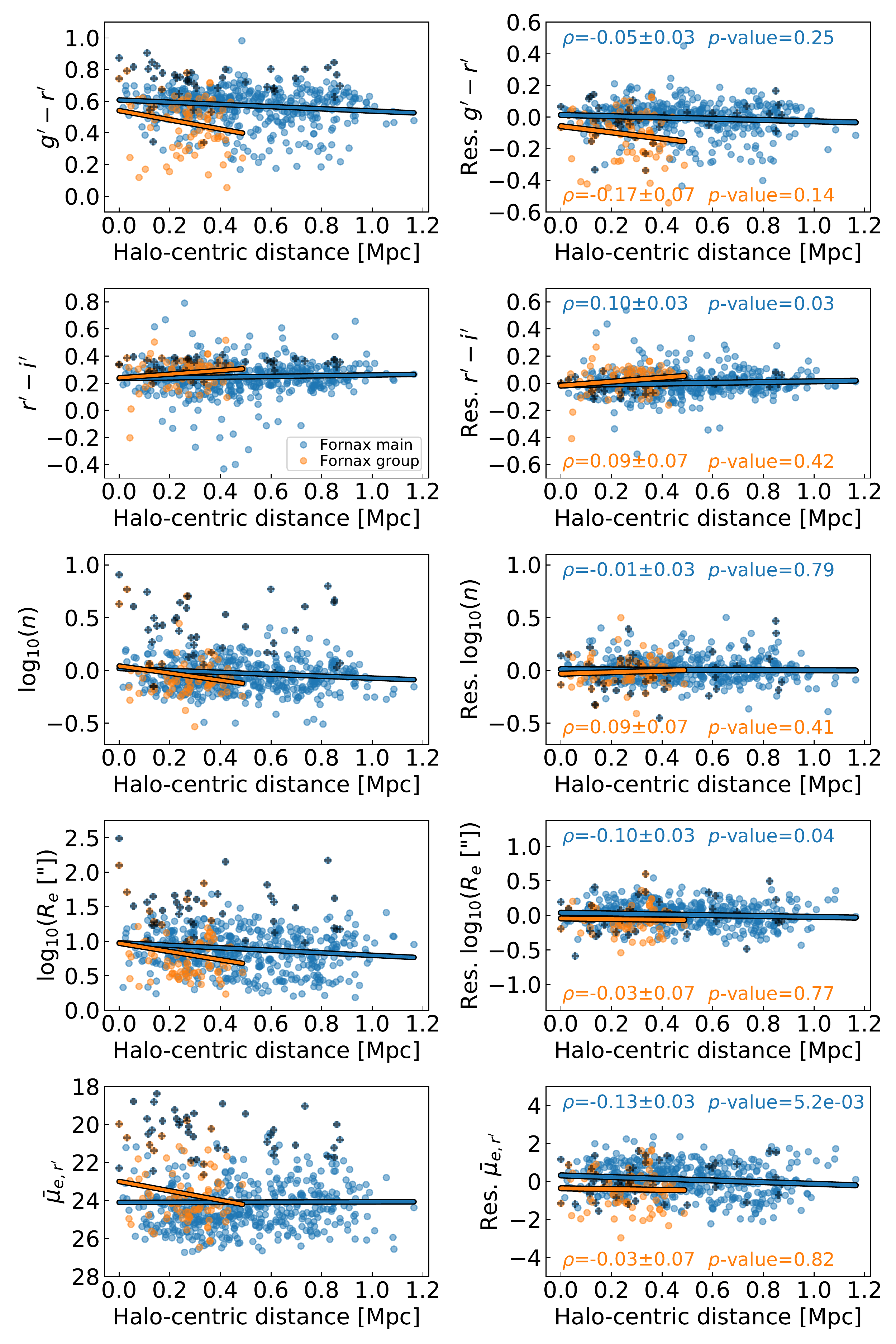}
  \caption{Measured (\textit{left}) and residual (\textit{right}) Sérsic-derived quantities as a function of projected halo-centric distance. The coloured lines denote the linear fits to the sub-samples. The residual parameters are defined as the mass-trend subtracted parameter values. Galaxies with $\log_{10}(M_*/M_{\odot}) >9$ are also marked with a black cross. The Spearman's $\rho$ rank coefficient and the $p$-value are annotated in the subplots. We use the same y-axes range between the corresponding measured and residual quantities to highlight the reduction in scatter.}
     \label{fig:cluster_dist}
\end{figure}

Echoing the quantities derived from Sérsic profiles, we test for trends as a function of the projected halo-centric distances (which, as a reminder, refers to the projected cluster- and group-centric distances collectively). Figure~\ref{fig:cluster_dist} shows both these measured and residual (i.e. mass-trend subtracted) trends\footnote{We show in Appendix~\ref{app:dist_vs_mass} that there is no significant correlation between the stellar mass and the projected halo-centric distance. Nevertheless, even if there were, any stellar mass trends would have been removed in calculating the residual quantities.}. We find that the measured and residual $g'-r'$ colour, Sérsic $n$, and $R_e$ decrease with increasing projected distance in both Fornax main and Fornax group, whilst $r'-i'$ appears to increase\footnote{In principle $r'-i'$ and $g'-r'$ should behave similarly. Given that the $r'-i'$ trend appears to be weak (and only marginally significant) and its ability to separate ETGs and LTGs to be poor, we do not discuss it further.}. The trends in Sérsic $n$, $R_e$, and $\bar{\mu}_{e,r'}$ show large scatter, which is mainly contributed by the massive galaxies (crosses in Fig.~\ref{fig:cluster_dist}). As such, mass-normalised trends with halo-centric distance show less scatter (second column in Fig.~\ref{fig:cluster_dist}). 

Based on the rank correlation coefficients, Spearman's $\rho$, and the corresponding $p$-value\footnote{Based on the null hypothesis that Spearman's $\rho$ is zero (i.e. no correlations with halo-centric distance). Here the $p$-value is the two-sided value, which represents the probability that we obtain values at least as extreme as the $\rho$ found for our samples}, the residual parameters imply that galaxies located at the outskirts of the Fornax main cluster tend to have smaller effective radius and brighter effective surface brightness than those closer to the cluster centre. In contrast, Sérsic $n$ and $g'-r'$ colour for galaxies does not appear to vary much as a function of halo-centric distance, with rank correlations similar to its uncertainties. In comparison, Fornax group galaxies have higher absolute values of Spearman's $\rho$ for the residual $g'-r'$ colour with increasing halo-centric distance, but lower absolute values ($\sim 0$) for $R_e$ and $\bar{\mu}_{e,r'}$. Their $p$-values ($>\alpha=0.05$) suggest we cannot exclude the probability that the quantities are uncorrelated with projected halo-centric distance. We note that the trends in $R_e$ and $\bar{\mu}_{e,r'}$ are consistent with what was found in Fig.~\ref{fig:scale_rel_nomasstrend_3sub}, where galaxies beyond the virial radius have significantly different distributions in $R_e$ and $\bar{\mu}_{e,r'}$ than those within the virial radius.

\begin{figure}
\centering
\includegraphics[width=\hsize]{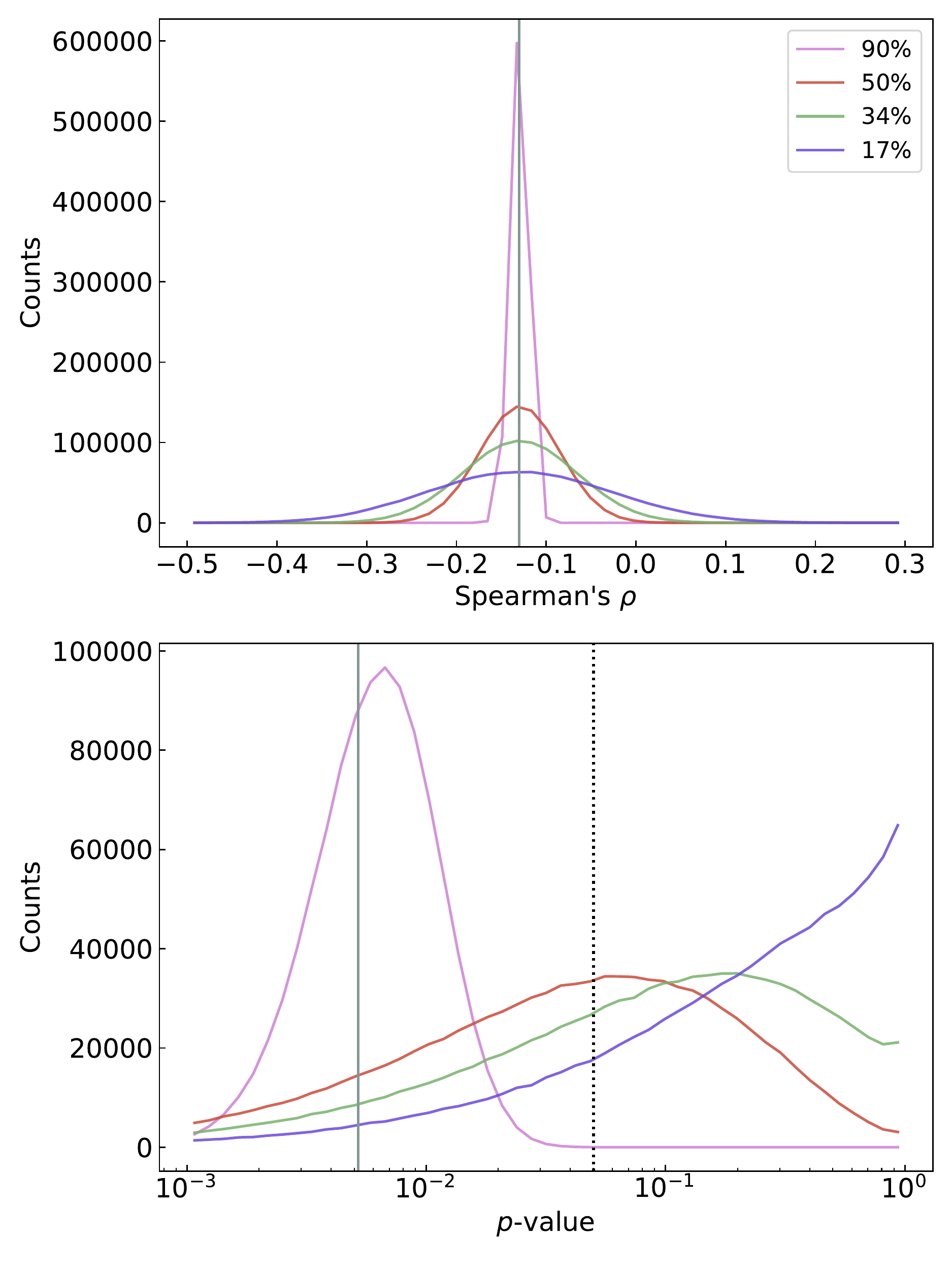}
  \caption{Distributions of Spearman's $\rho$ (\textit{upper}) and the corresponding $p$-values (\textit{lower}) from residual $\bar{\mu}_{e,r'}$ and cluster-centric distance, based on 90\% (\textit{lilac}), 50\% (\textit{orangered}), 34\% (\textit{pale green}), and 17\% (\textit{indigo}) of the Fornax main sample. The vertical \textit{grey} lines denote the Spearman's $\rho$ and $p$-value from the whole Fornax main sample (i.e. from Fig.~\ref{fig:cluster_dist}). The dotted \textit{black} line shows the $\alpha=0.05$ significance level. }
     \label{fig:test_samplesize}
\end{figure}

\begin{figure}
\centering
\includegraphics[width=\hsize]{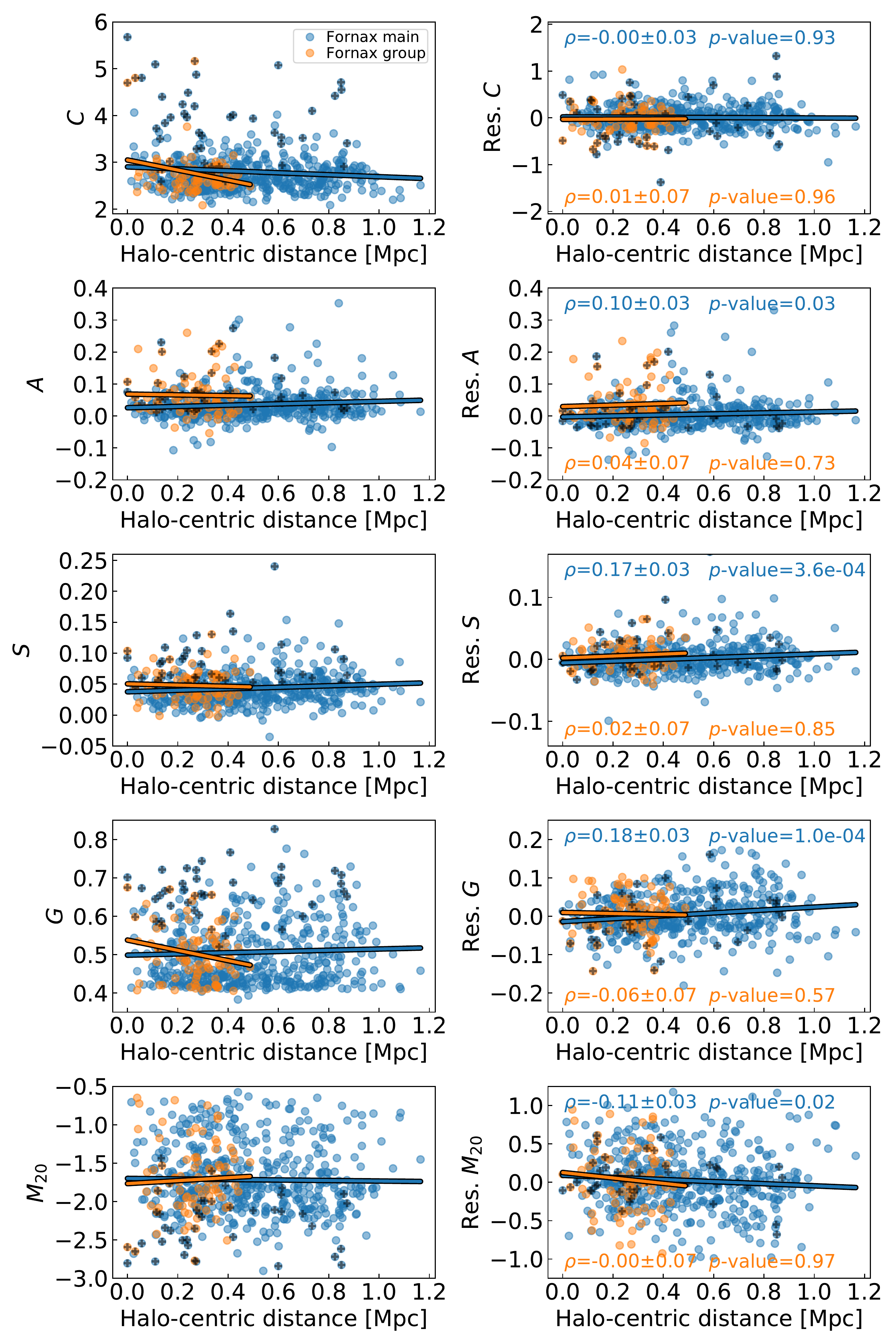}
  \caption{Measured and residual (i.e. with mass-trends removed) non-parametric indices as a function of projected halo-centric distance. The black crosses denote galaxies with $\log_{10}(M_*/M_{\odot})>9$. Spearman's $\rho$ is annotated for the Fornax main (\textit{blue}) and Fornax group (\textit{orange}), as well as the $p$-value of the null hypothesis that $\rho$ is zero.}
     \label{fig:nonparametric_dist}
\end{figure}

Naively, one could attribute the lack of significant group-centric trends to the lower sample size of the Fornax group compared to the Fornax main. This begs the question: If the sample size of Fornax main was decreased to that of the Fornax group, would the cluster-centric trends shown in Fig.~\ref{fig:cluster_dist} remain significant? To explore this, we use the residual $\bar{\mu}_{e,r'}$ and clustercentric distance from Fornax main as the reference sample, since $\bar{\mu}_{e,r'}$ has a clear trend with $p$-value $\ll \alpha$. We use bootstrapping to create sub-samples with 90\%, 50\%, 34\%, and 17\% the size of the Fornax main sample. The 17\% sample size is approximately the sample size of the Fornax group. For each sub-sample we calculate Spearman's $\rho$ and corresponding $p$-value, with the distributions based on 1000000 sub-samples shown in Fig.~\ref{fig:test_samplesize}. We find that the distributions of Spearman's $\rho$ are centred around the same value, regardless of sample size, with the peaks located at the Spearman's $\rho$ found in Fig.~\ref{fig:cluster_dist}. Additionally, the spread in Spearman's $\rho$ values increases with decreasing sample size. However, the distributions of $p$-values differ greatly with sample size, with a 50\% sample size already displaying a peak $p$-value greater than the significance level $\alpha = 0.05$. For the $17\%$ sample size, we find the $p$-values are typically well above the significance level. Overall, this suggests that if Fornax main has a sample size similar to Fornax group, it would have a more or less similar Spearman's $\rho$ value, but a higher corresponding $p$-value which would likely be deemed insignificant. 

Figure~\ref{fig:nonparametric_dist} shows the measured and residual non-parametric morphological indices for Fornax main and Fornax group galaxies. Overall, residual $A$, $S$, and $G$ show significant positive rank correlations in Fornax main, which suggests that galaxies with high halo-centric distance tend to be more asymmetric and generally have less smooth light distributions. With respect to residual $M_{20}$, Fornax main galaxies follow a significant negative correlation i.e. lower $M_{20}$ with increasing distance. We find that $C$, similar to Sérsic $n$, does not show a clear correlation with halo-centric distance, which, as previously mentioned, is expected due to the relation between the two quantities. For all residual parameters from Fornax group, the high $p$-values mean we cannot reject the possibility that there are no correlations with halo-centric distance. 

\begin{figure}
\centering
\includegraphics[width=\hsize]{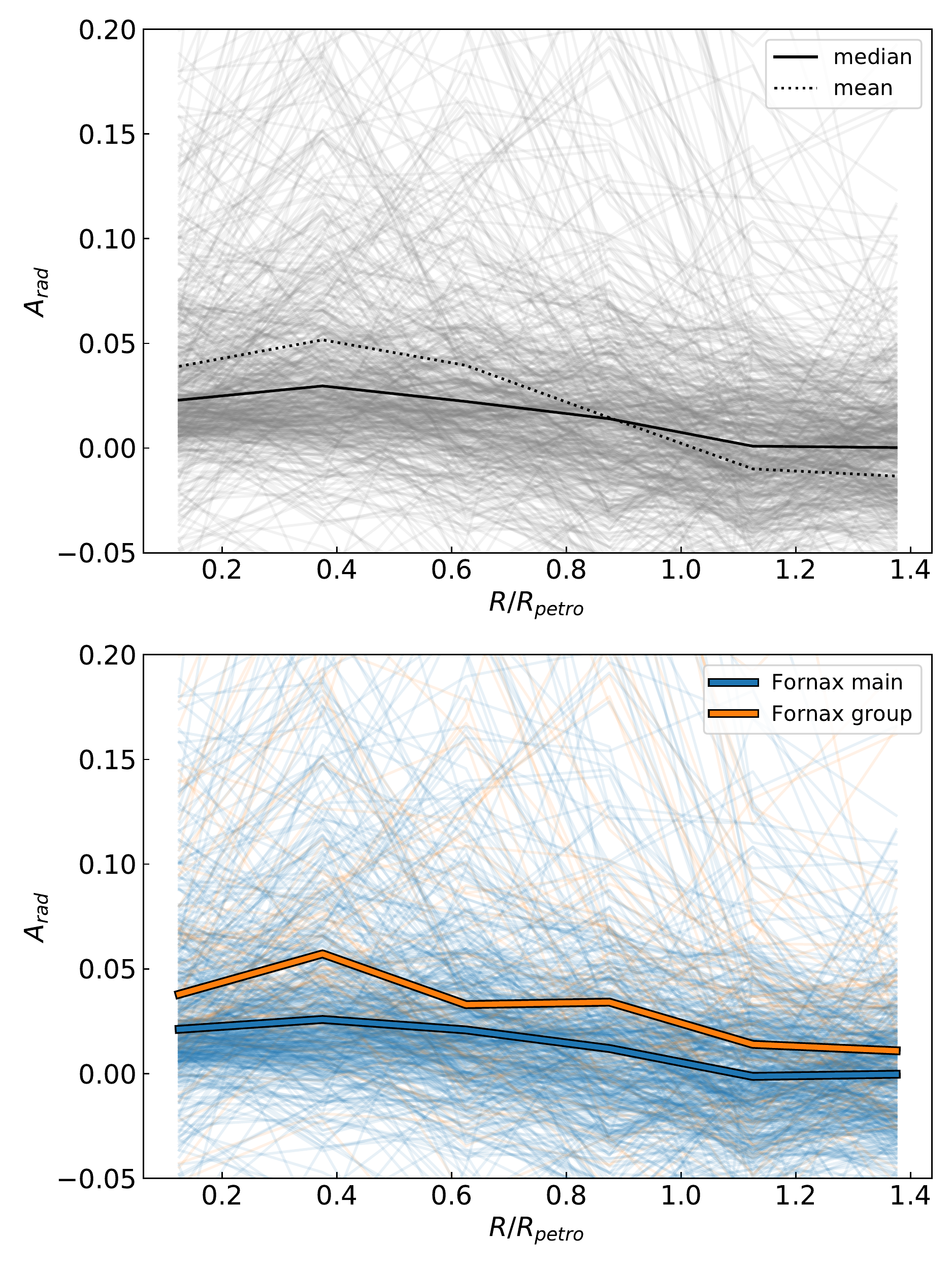}
  \caption{Asymmetry index as a function of radius for all galaxies ($A_{rad}$). \textit{Upper panel}: The asymmetry profiles for all galaxies (grey). The black solid and dashed lines indicate the median and mean values, respectively, of asymmetry at fixed factors of Petrosian radius. \textit{Lower panel}: The asymmetry profiles split between Fornax main (blue) and Fornax group (orange). The coloured solid lines show the median values of asymmetry at a given factor of the Petrosian radius.}
     \label{fig:asym_prof}
\end{figure}

\begin{figure}
\centering
\includegraphics[width=\hsize]{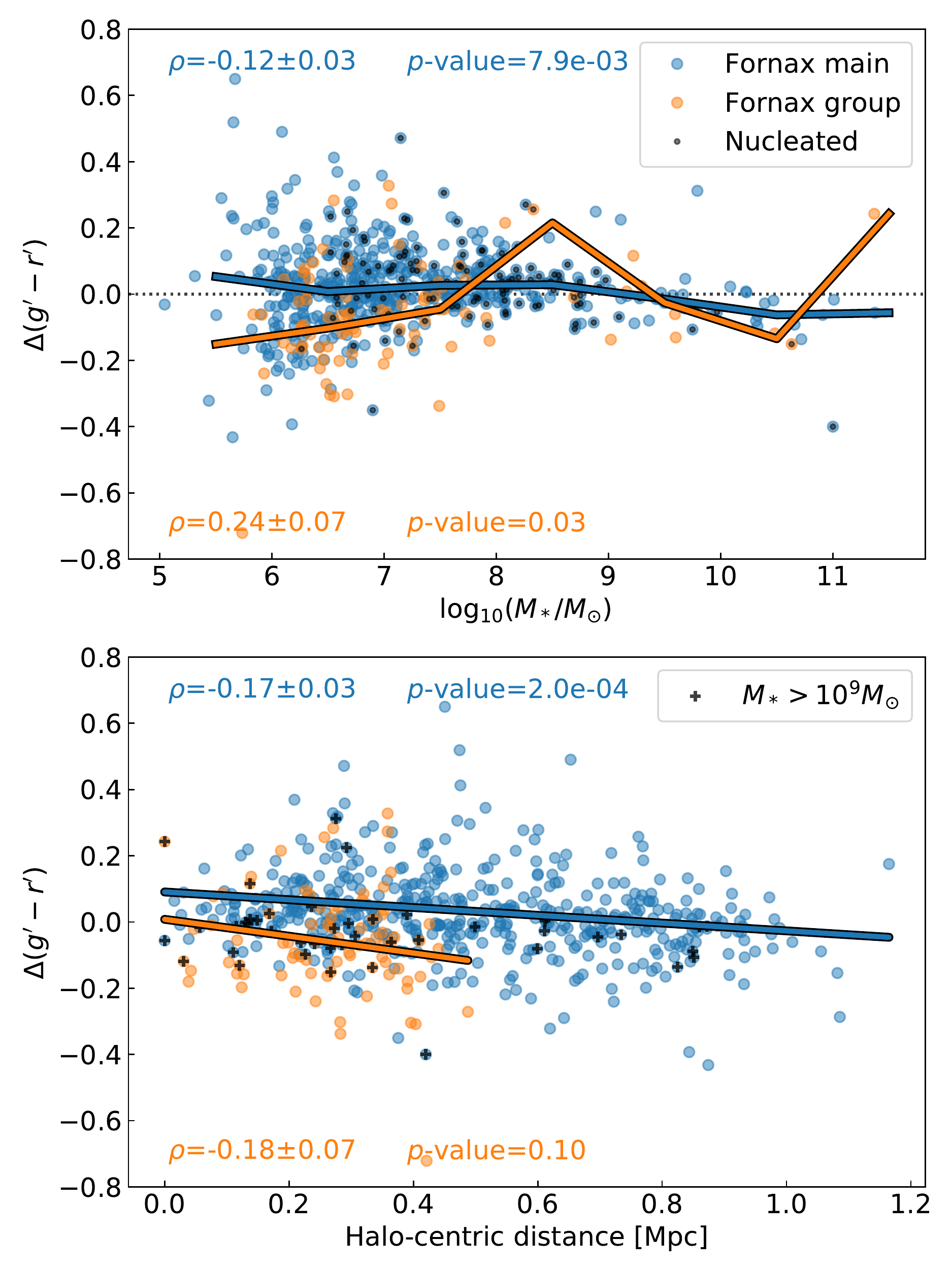}
  \caption{$g'-r'$ colour differences (defined as the colour of the 1--2$R_e$ region minus the inner 0.5$R_e$ region, see Eqn.~\ref{eqn:col_grad_aper}) as a function of stellar mass (\textit{upper}) and halo-centric distance (\textit{lower}), split between Fornax main (\textit{blue}) and Fornax group (\textit{orange}). $\Delta(g'-r')>0$ implies a redder outer region, whereas $\Delta(g'-r')<0$ implies a bluer outer region. The solid coloured lines in the upper plot denote the median values calculated within bins of 1\,dex in stellar mass, whereas those in the lower plot show the linear fits. The dotted line denotes $\Delta(g'-r')=0$. The Spearman's $\rho$ and corresponding $p$-values for Fornax main and Fornax group are annotated in the plots. The KS-test for $\Delta(g'-r')$ between Fornax main and Fornax group provided a $p$-value$=6.7\times 10^{-10}$.}
     \label{fig:colour_grad}
\end{figure}

\subsection{Asymmetry profiles}\label{sect:asym_prof}
Figure~\ref{fig:asym_prof} shows the radial asymmetry profiles of each galaxy in our sample, as well as the mean and median profiles based on all galaxies in our sample. In general, the median asymmetry values appear to be larger in the inner regions than the outskirts, peaking at $R/R_{\text{petro}} \sim 0.4$. This is consistent with the idea that most prominent structures reside in the inner regions of galaxies. Within Fornax itself, the median values are consistently higher for Fornax group galaxies compared to Fornax main at all radii. This is potentially caused by clumps or regions of star formation in Fornax group galaxies. Indeed, Fig.~\ref{fig:clump_asym_vs_colour} shows that the median asymmetry $\langle A \rangle$ for LTGs is higher than for ETGs. Therefore, the higher fraction of LTGs in Fornax group is likely the main source of disparity in $A$ between the two populations.

\subsection{Colour differences}\label{sect:col_diff}
The distribution of the $g'-r'$ colour difference (between the outer and inner portions of a galaxy), $\Delta (g'-r')$, with stellar mass and halo-centric distance is shown in Fig.\,\ref{fig:colour_grad}. For the lowest stellar mass, there appears to be a spread around $-0.3 \lesssim \Delta (g'-r') \lesssim +0.3$. However, taking into account the uncertainties in the magnitudes (from Table~\ref{tab:mag_errors}) and propagating\footnote{We propagate the uncertainty in $g'-r'$ colour as $\sigma_{(g'-r')} = \sqrt{\sigma_{g'}^2 + \sigma_{r'}^2}$. The uncertainty in $\Delta (g'-r')$ is then estimated from the propagation of uncertainties in the inner and outer colours.}, we find the uncertainty in $\Delta (g'-r')$ is $\sim 0.3$ (at 1 standard deviation level) for the faintest galaxies. On average, Fornax main galaxies have $\Delta (g'-r') \approx 0$, whereas there is a large portion of lower mass ($\log_{10}(M_*/M_{\odot})<8$) galaxies with negative colour differences (i.e. redder inner part) in Fornax group. We note that there appears to be a weak but significant trend between $\Delta (g'-r')$ and stellar mass for Fornax main, where the most massive galaxies tend to have negative colour differences whereas less massive galaxies typically have positive colour differences. In contrast, for Fornax group galaxies there is a positive correlation with stellar mass, as reflected by the significant, positive Spearman's $\rho$ value. A positive colour difference implies the galaxies tend to be bluer in their inner regions and redder in the outskirts, whereas the opposite is true for a negative colour difference. We find negative correlations between $\Delta(g'-r')$ and halo-centric distance for both sub-samples, but they are only significant for the Fornax main sample. This suggests that galaxies have larger colour differences (i.e. tendency for relatively bluer inner galaxy regions) towards the centre of the cluster.

\subsection{Multi-component decompositions}\label{sect:decomp_multicomp}
\begin{figure}
\centering
\includegraphics[width=\hsize]{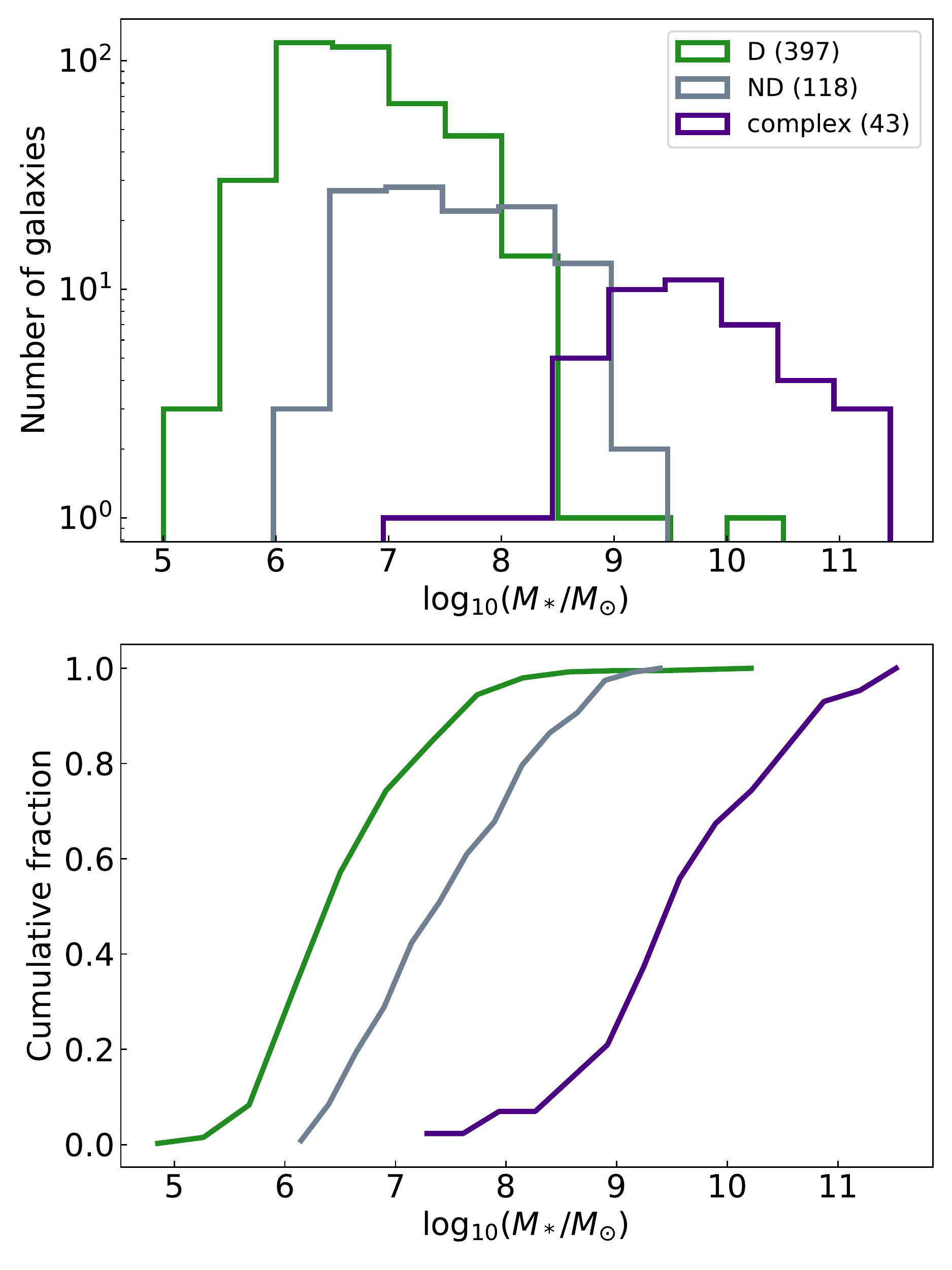}
  \caption{Overview of the multi-component decomposition models of our sample galaxies, split between purely disk (D, \textit{green}), nucleus+disk (ND, \textit{grey}), and galaxies with more complex models (complex, \textit{purple}). The \textit{upper} and \textit{lower} panels show the histograms and relative cumulative distributions, respectively, of the three sub-samples. For better readability, we shifted the ND and complex histograms by 0.025 and 0.05, respectively, to the left. No such shifts were applied to the cumulative distributions. A bin size of 0.5\,dex was used. }
     \label{fig:multicomp_hist}
\end{figure}

Before analysing the multi-component decompositions, we inspected all the models and removed 21 and 3 galaxies from Fornax main and Fornax group, respectively, due to uncertain decomposition parameters (see Sect.~\ref{app:uncertain}). Figure~\ref{fig:multicomp_hist} shows the distribution for galaxies best-fit through our decompositions with three different classes of models: purely disk (D), nucleus+disk (ND), and more complex models containing bulges or bars. Overall, the majority of our galaxies can be well described by the D and ND models. These galaxies tend to be almost featureless low mass dwarf galaxies. Conversely, high-mass galaxies more often require more complex models. Overall, there is a segregation in stellar mass between the simpler (D and ND) models and the complex models. The point of segregation occurs at $\log_{10}(M_*/M_{\odot}) \sim 9$.

\subsubsection{Nucleation}\label{sect:nucleation}
\begin{figure}
\centering
\includegraphics[width=\hsize]{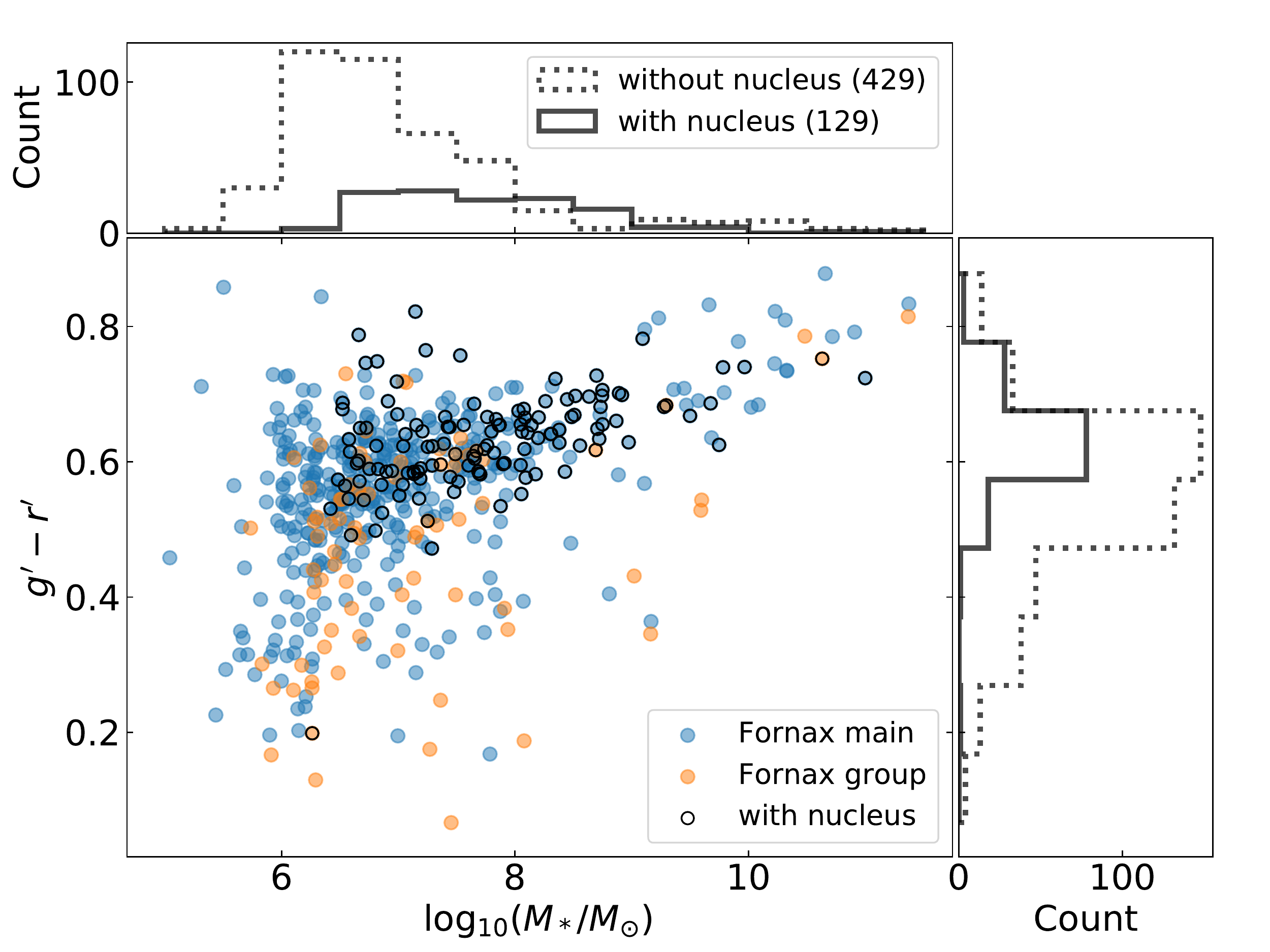}
  \caption{$g'-r'$ colour as a function of stellar mass for Fornax main (\textit{blue}) and Fornax group (\textit{orange}). Galaxies with a nucleus are highlighted with a \textit{black} circle. The \textit{upper} and \textit{right panels} show the distributions of stellar mass and $g'-r'$ colour, respectively, for galaxies with (\textit{bold}) and without (\textit{dashed}) a nucleus. Bins of 0.5\,dex and 0.1\,mag were used for the stellar mass and $g'-r'$ colour, respectively.}
     \label{fig:multicomp_nuc}
\end{figure}

Our sample contained of 129 galaxies where a nucleus was identified and a nucleus component fitted in the final model. Excluding any uncertain decomposition models, the Fornax main sample included 121 nucleated galaxies, whereas only 8 nucleated galaxies were found in Fornax group. We also observe that the Fornax main sample has a much higher nucleation fraction compared to the Fornax group sample (25\% against 10\%). Figure~\ref{fig:multicomp_nuc} shows the $g'-r'$ colour as a function of stellar mass, with nucleated galaxies indicated using black outlines. As a whole, the nucleated galaxies appear to span a tighter range in $g'-r'$ colour, as well as marginally redder $g'-r'$ colours on average compared to their non-nucleated counterparts. The redder colours reflect the nucleated galaxies' location along the red sequence and the general lack of bluer nucleated galaxies. The difference in the range of colours occupied by nucleated and non-nucleated galaxies suggests a formation mechanism for nucleation which goes beyond colour change. In other words, simply changing the colour of a blue galaxy (e.g. via ram pressure stripping) does not guarantee the formation of a nucleus, as there are also many non-nucleated galaxies which overlap with nucleated galaxies in the colour-stellar mass space. Given that the nucleation of galaxies in dense environments are of particular interest, we will present more detailed analyses in a forthcoming paper. 

\subsubsection{Bulge component}
\begin{figure}
\centering
\includegraphics[width=\hsize]{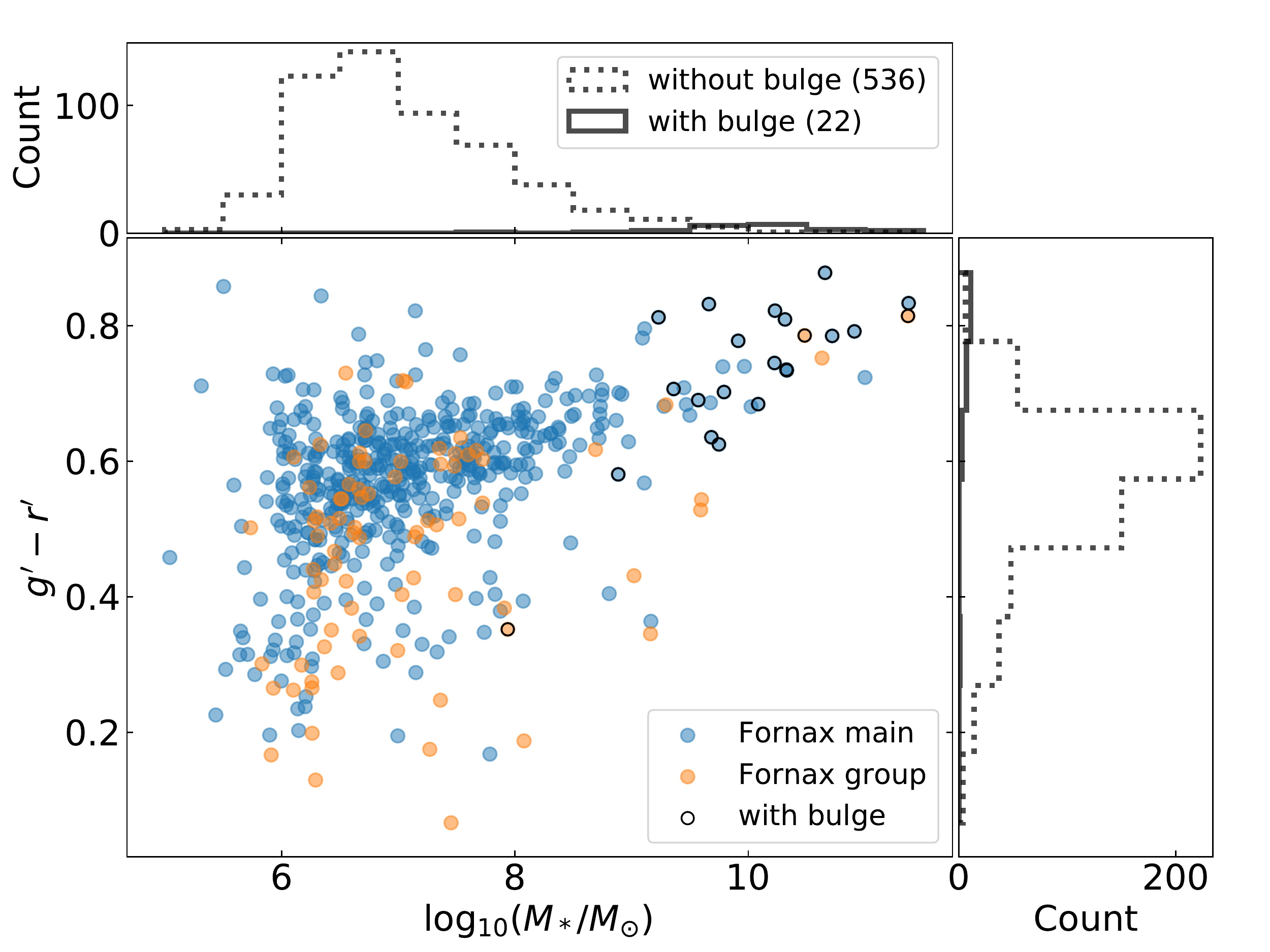}
  \caption{Same as Fig.~\ref{fig:multicomp_nuc}, but for bulges.}
     \label{fig:multicomp_bulge}
\end{figure}

In total, the sample contains 22 galaxies which have a bulge component fitted in the multi-component decomposition model. Figure~\ref{fig:multicomp_bulge} shows the CMD of $g'-r'$ colour, split between galaxies with and without a bulge component fitted. On average, galaxies with bulge components have redder colour than those without. This is likely linked to their stellar masses, which are amongst the highest of our sample. 

\subsubsection{Bar component}
\begin{figure}
\centering
\includegraphics[width=\hsize]{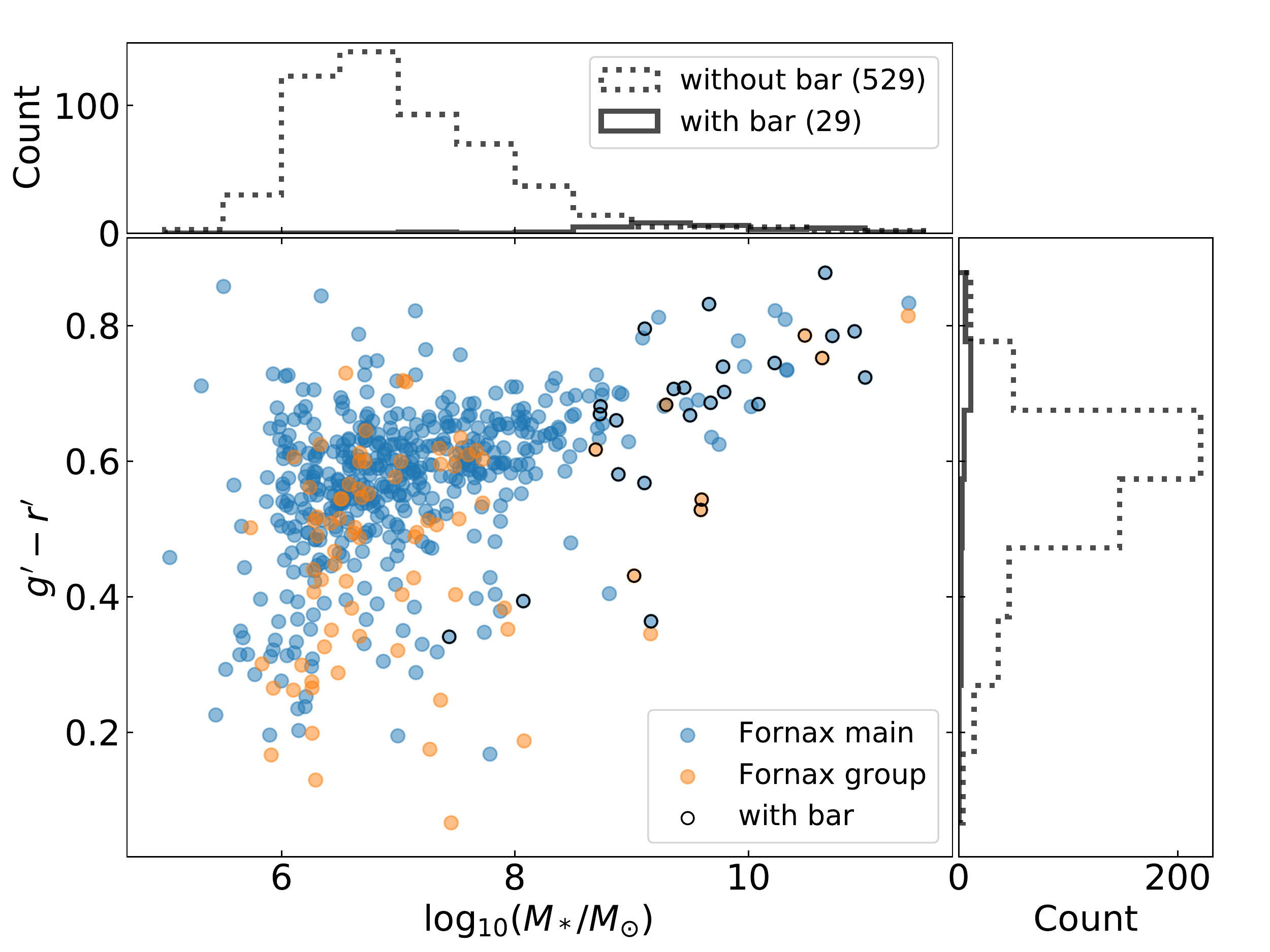}
  \caption{Same as Fig.~\ref{fig:multicomp_nuc}, but for bars.}
     \label{fig:multicomp_bar}
\end{figure}

Of our sample of galaxies, 29 galaxies were fitted with a bar component. Figure~\ref{fig:multicomp_bar} shows the $g'-r'$ colour as a function of stellar mass between galaxies with and without a bar component. The distribution of barred galaxies is skewed towards the brightest galaxies. As with the case of the bulge sample, the redder colours for barred galaxies are likely predominately due to the higher mass. The distribution of barred galaxies with stellar mass reflects what was found in S$^4$G \citep[see][their Fig.~19]{diazgarcia2016}, where the barred fraction decreases to zero at stellar masses of $\log_{10}(M_*/M_{\odot}) \sim 8$. Additionally, 7 barred galaxies have a barlens component fitted, spanning a mass range of $9.7 \lesssim \log_{10}(M_*/M_{\odot}) \lesssim 11.0$. Using the same stellar mass range of $9.7 < \log_{10}(M_*/M_{\odot}) < 11.4$ as the samples of \citet{laurikainen2014,laurikainen2018}, we find 7 out of 10 barred galaxies have a barlens component in our sample, compared to $\sim 50\%$ from \citet{laurikainen2014,laurikainen2018}.  

\subsubsection{Model complexity} \label{sect:fds_model_complexity}
From our sample of multi-component decomposition models, we investigate any trends regarding the complexity of the models fitted. We quantify this complexity using the number of components. The models were split into two classes: simple and complex. Models which have only one component, regardless of whether a nucleus component was fitted, were classified as simple (e.g. D and ND galaxies). Conversely, models with more than one component (apart from the nucleus) were classified as complex. A limit on the axial ratio of $b/a \le 0.423$ was applied to the decomposition models, which excluded 40 galaxies ($\sim 7\%$ of the sample). This effectively excluded models which exceed inclinations of $65\deg$ or are edge-on, where much of the galaxies' structures are obscured/lost. 

\begin{figure}
\centering
\includegraphics[width=\hsize]{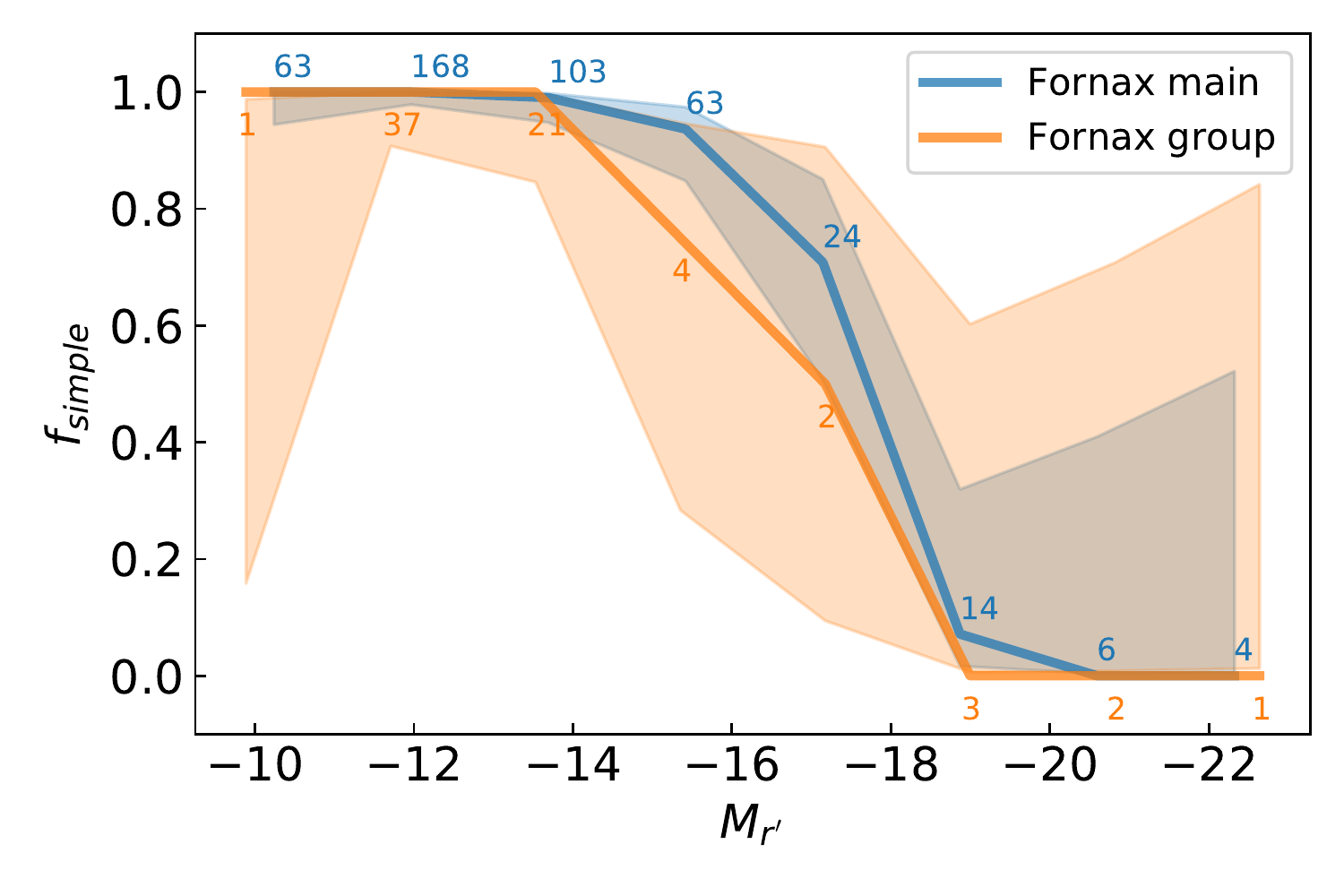}
  \caption{Fraction of galaxies with simple final decomposition models as a function of multi-component $r'$-band magnitudes. Here 'simple' refers to models which only consist of one component, excluding the nucleus. A bin size of 2\,mag was used for both samples. The numbers denote the number of galaxies within each bin. The shaded regions denote the $95\%$ confidence interval based on the methodology of \citet{cameron2011}. For reference, $M_{r'}=-14 \sim 10^{7.5}M_{\odot}$ and $M_{r'}=-19 \sim 10^{9.7}M_{\odot}$.}
     \label{fig:simple_fraction_fornax}
\end{figure}

\begin{figure}
\centering
\includegraphics[width=\hsize]{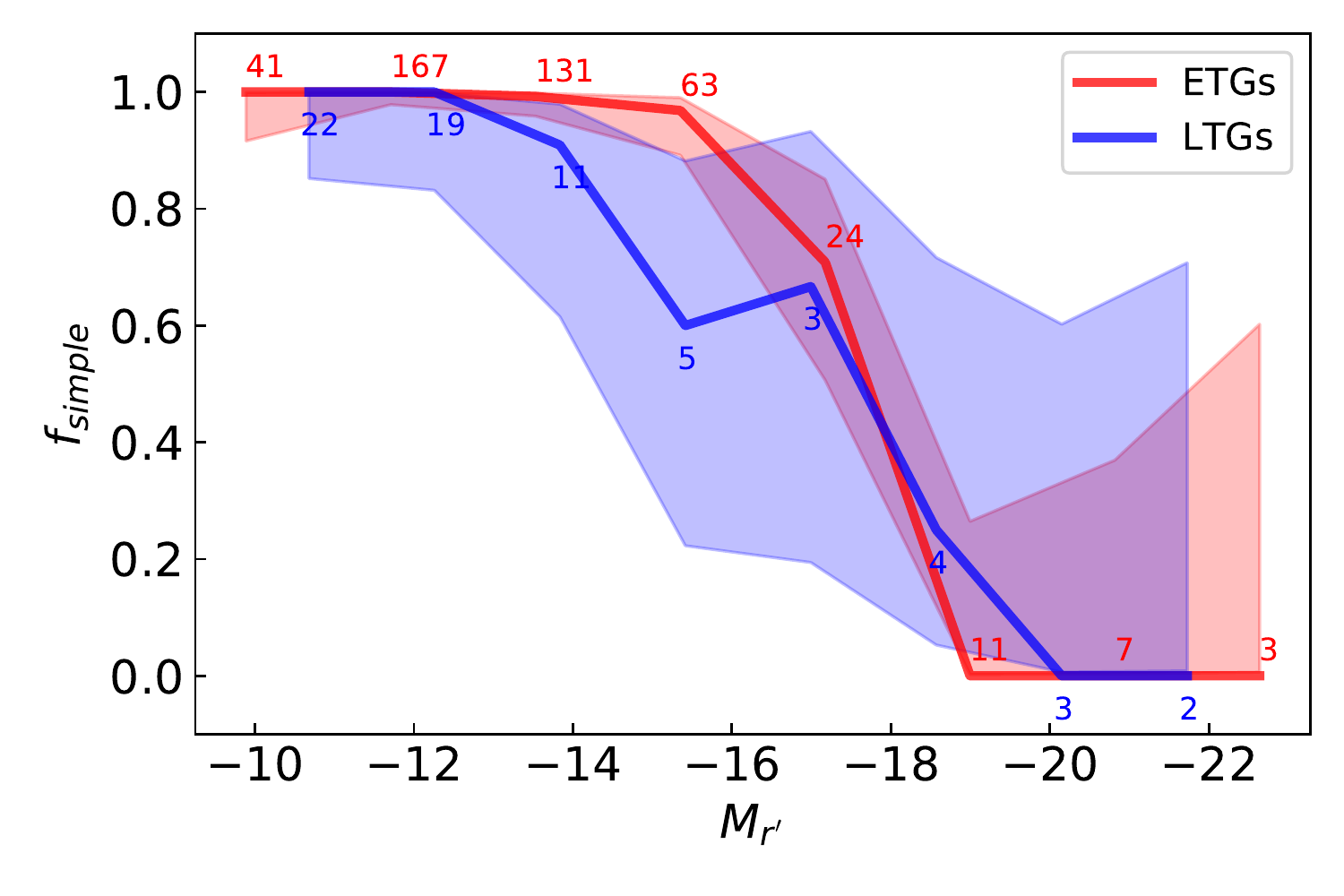}
  \caption{Same as Fig.~\ref{fig:simple_fraction_fornax}, but split between ETGs (\textit{red}) and LTGs (\textit{dark blue}).}
     \label{fig:simple_fraction_class}
\end{figure}

Figure \ref{fig:simple_fraction_fornax} shows the fraction of simple decompositions as a function of absolute $r'$-band magnitude\footnote{Magnitudes based on multi-component decompositions.}. Overall, there is a decrease in the simplicity fraction with brighter magnitudes. The simplicity fraction decreases rapidly from 1 to 0 between $M_{r'} \sim -14$ to $M_{r'} \sim -19$. Averaging the stellar masses within $\pm 0.1$ of $M_{r'}=-14$ and $M_{r'}=-19$, this corresponds to $\sim 10^{7.5}M_{\odot}$ and $\sim 10^{9.7}M_{\odot}$, respectively. To estimate the uncertainty in $f_{\text{simple}}$ the $95\%$ binomial confidence interval was calculated\footnote{The confidence intervals were calculated following the beta distribution quantile technique as described by \citet{cameron2011}.}. Within uncertainties, the distributions between Fornax group and Fornax main are similar. We note that the large uncertainties for the Fornax group sample is due to the small sample size, particularly towards the brighter (lower) magnitudes end where the bins only contain a few ($<5$) galaxies. 

Figure \ref{fig:simple_fraction_class} shows the simplicity fraction between ETGs and LTGs in Fornax main and Fornax group. Both types of galaxies follow the trend of decreasing simplicity fractions with brighter magnitudes. The mean values of $f_{\text{simple}}$ for LTGs appear to be lower than that of ETGs, especially in the range $M_{r'}>-18$. This implies that generally LTGs tend to contain more (morphological) structures than ETGs at comparable $M_{r'}$. However, the low sample sizes of LTGs in each bin lead to large uncertainties in $f_{\text{simple}}$.

\section{Comparisons with literature} \label{sect:lit_comp}

\begin{figure}
\centering
\includegraphics[width=\hsize]{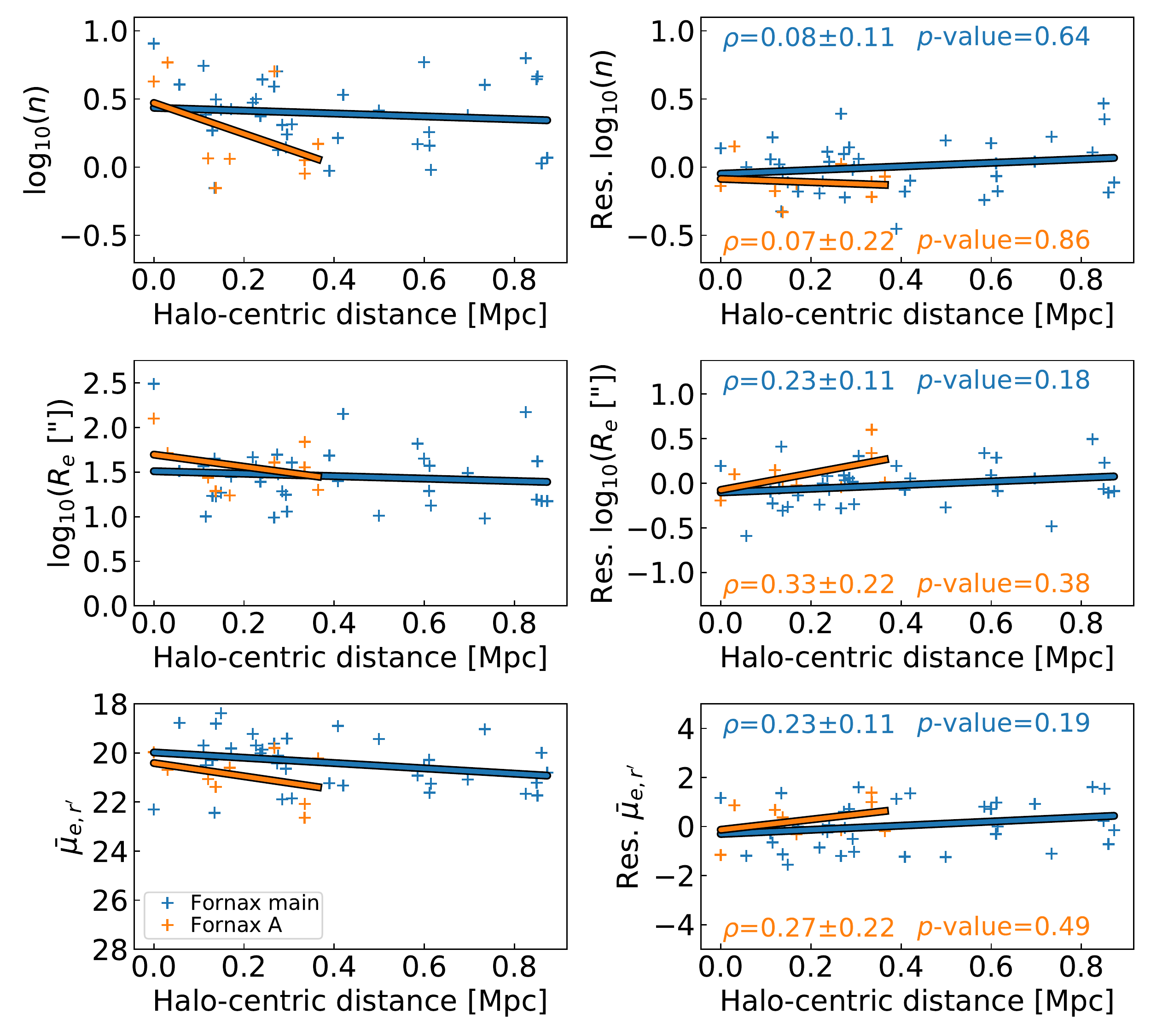}
  \caption{Same as Fig.~\ref{fig:cluster_dist}, but for galaxies with $\log_{10}(M_*/M_{\odot})>9$ (i.e. the black crosses from Fig.~\ref{fig:cluster_dist}).}
     \label{fig:cluster_dist_bigonly}
\end{figure}

We found significant differences in the properties of galaxies between Fornax main and Fornax group. This applies to the distributions of residual (i.e. mass-trend removed) properties (Sect.~\ref{sect:scale_rel}), as well as their correlations with halo-centric distance (Sect.~\ref{sect:pars_vs_dist}). In this section we compare our results to studies in the literature.

Given that the FDS mosaics and galaxy sample selection were taken from \citet{venhola2018}, it is prudent to compare our results with \citet{venhola2019}, who also studied the optical properties of dwarfs in Fornax via Sérsic+PSF decompositions. We stress that while both our work and \citet{venhola2019} use the reduced FDS mosaics from \citet{venhola2018}, the preparation and analysis steps were conducted independently from each other. Overall, our results agree with what \citet{venhola2019} found for fainter (i.e. $M_{r'}>-16.5$) dwarf galaxies, where they tend to become redder ($g'-r'$), larger and fainter with decreasing halo-centric distance, although we cannot say the $g'-r'$ trend is significant ($p$-value $<\alpha$). Additionally, they found that the Residual Flux Fraction\footnote{The RFF measures the difference in flux between a galaxy image and its model (i.e. the residual image) which cannot be attributed to noise. Typically, RFF is used as an indicator of the amount of additional components/structures in the galaxy image which are not accounted for by the model.} \citep[RFF, see][]{blakeslee2006, hoyos2011} decreases with decreasing halo-centric distance, which is consistent with the trends of $A$ and $S$. For the brightest (i.e. $M_{r'}<-16.5$, or $\log_{10}(M_*/M_{\odot}) \gtrsim 8.5$) dwarfs, \citet{venhola2019} found that they tend to become less centrally concentrated, smaller, and fainter with decreasing halo-centric distance, which, at a glance, appears to differ from our observed trends. However, by only consider galaxies with $\log_{10}(M_*/M_{\odot})>9$, Fig.~\ref{fig:cluster_dist_bigonly} shows the same trends \citet{venhola2019} found, although they are not significant. The apparent difference is likely due to the higher number of lower mass galaxies dominating over the massive galaxies (see e.g. Fig.~\ref{fig:cluster_dist}). 

\begin{figure}
\centering
\includegraphics[width=\hsize]{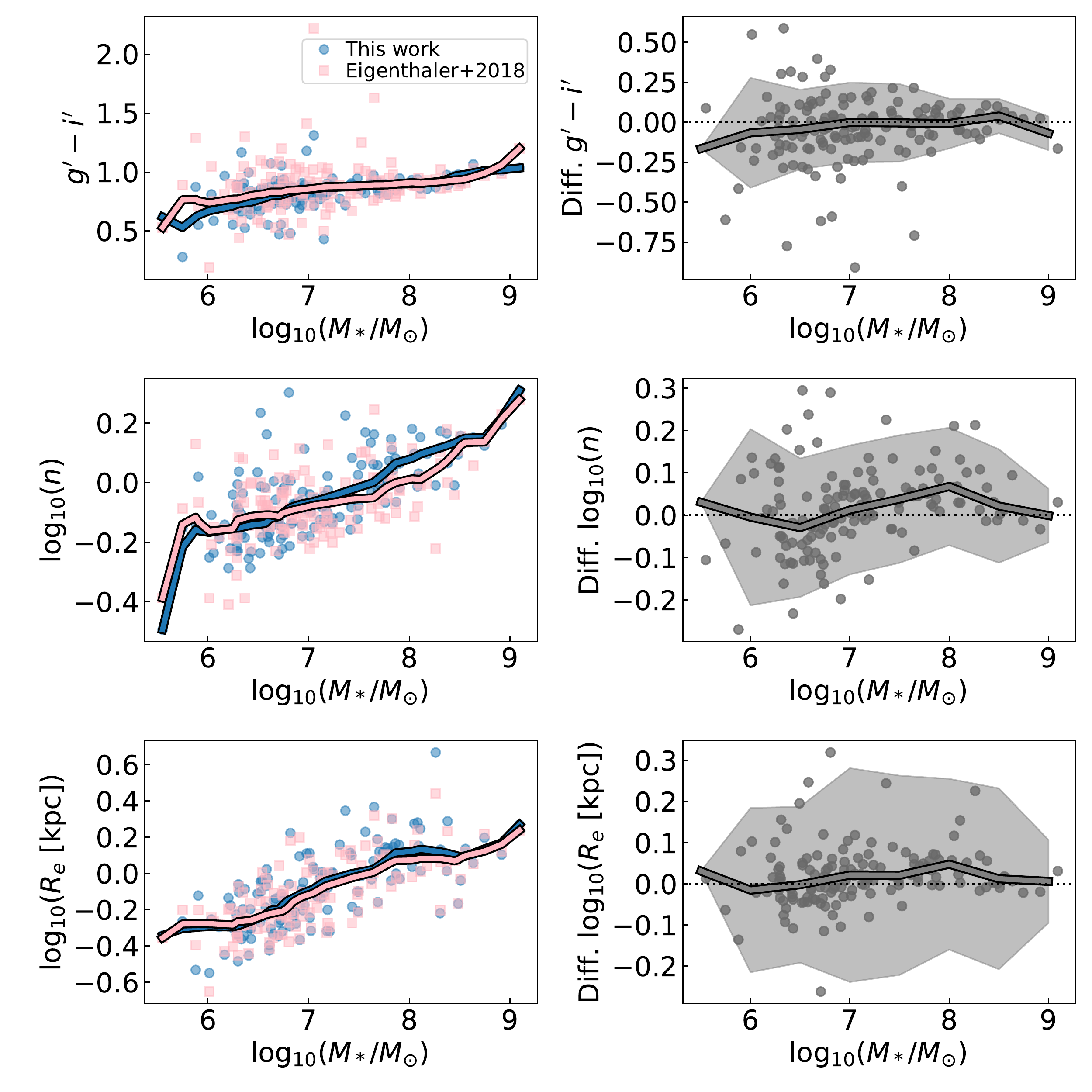}
  \caption{Comparison of $g'-i'$, Sérsic $n$, and $R_e$ as a function of the stellar mass calculated from this work for galaxies cross-matched between \citet{eigenthaler2018} (NGFS, \textit{pink}) and this work (\textit{blue}). The \textit{left panels} show the quantities based on Sérsic decompositions, whilst the \textit{right panels} show the difference (i.e. this work $-$ NGFS) in quantities for each matched galaxy (\textit{grey}). The solid lines show the moving averages calculated with a bin size of 2\,mag and bin step size of 0.2\,mag. The shaded grey regions denote the uncertainties propagated from the RMS values within the bins of the moving averages of both works. The dotted lines denote zero difference in parameters. }
     \label{fig:ngfs_comparison}
\end{figure}

Another study which observed the Fornax region in detail is the Next Generation Fornax Survey \citep[NGFS][]{munoz2015}. In particular, \citet{eigenthaler2018} presented Sérsic-derived quantities for dwarfs in the inner part ($\sim 0.4$\,Mpc) of the Fornax main cluster. For a direct comparison we cross-match the sources between \citet{eigenthaler2018} and our work, using a separation limit of 5\,arcsec. This resulted in 142 matched galaxies. In Fig.~\ref{fig:ngfs_comparison} we compare the $g'-i'$ colour, Sérsic $n$, and $R_e$ of the cross-matched sample. In general, the parameters from both samples appear to follow the same scaling relations with stellar mass. For $g'-i'$ colour, we find large differences between some matched galaxies, but very similar moving averages. The scatter in $g'-i'$ can be attributed to the large uncertainties in calculating the difference in colours (such as in Sect.~\ref{sect:col_diff}). The similarity in moving averages is due to the systematic differences in the $g'$- and $i'$-band magnitudes between FDS and \citet{eigenthaler2018} cancelling each other out, as reported by \citet{venhola2018} (see Appendix~\ref{app:ngfs_mag_offset}). Another possible source of difference comes from the difference in the band used for the Sérsic decomposition, where the Sérsic $n$ and $R_e$ used in our $g'$-band decompositions were fixed based on values from the $r'$-band decompositions. Therefore, the differences in the depth and seeing of each band can contribute to the difference in parameters. 

\begin{figure}
\centering
\includegraphics[width=\hsize]{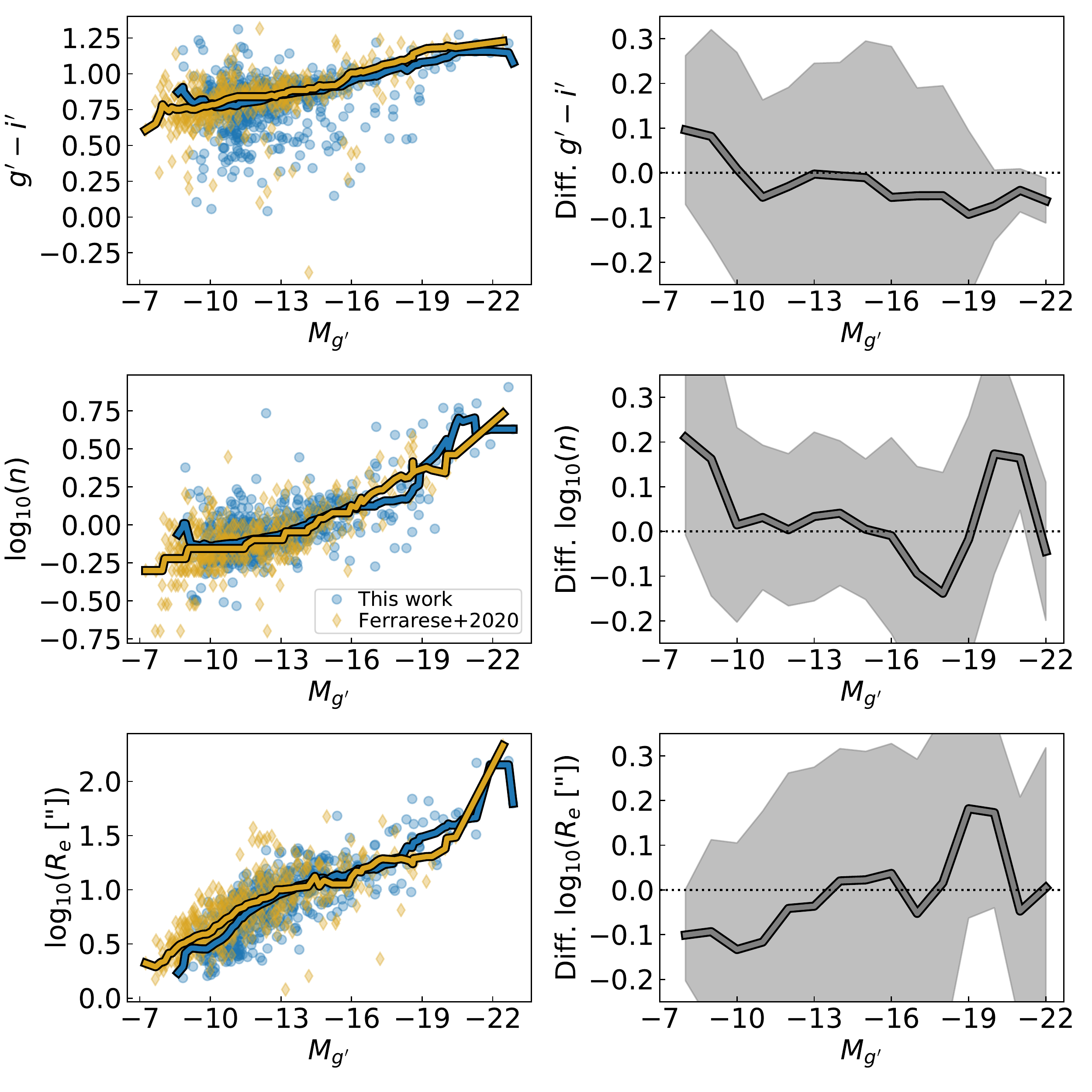}
  \caption{Comparison of parameters based on Sérsic decompositions as a function of $g'$-band magnitude for galaxies in \citet{ferrarese2020} (NGVS, \textit{gold}) and this work (\textit{blue}). \textit{Left panels}: The scaling relations of $g'-i'$, Sérsic $n$, and $R_e$. The coloured solid lines denote the moving averages based on a bin size of 2\,mag and bin step size of 0.2\,mag. \textit{Right panels}: the difference (i.e. this work $-$ NGVS) in the moving averages (\textit{solid grey lines}), shown from $-22<M_{g'}<-8$ in intervals of 1\,mag. The shaded grey regions denote the uncertainties propagated from the RMS values within the bins of the moving averages of both works. The dotted lines denote zero difference in parameters.}
     \label{fig:ngvs_comparison}
\end{figure}

Additionally, we compare our results from Fornax (i.e. Fornax main and Fornax group) to the Virgo cluster. For this purpose, we utilise the results of the Next Generation Virgo Survey \citep[NGVS][]{ferrarese2020}, which presented Sérsic-derived quantities based on MegaCam $g$-band data and colours determined via a curve-of-growth technique for the inner 3.71\,deg$^2$ (0.3\,Mpc$^2$) region of the Virgo cluster. In order to compare the magnitudes and colours from NGVS to FDS, we first transform the MegaCAM bands into SDSS bands\footnote{We use the transformations from \url{http://www.cadc-ccda.hia-iha.nrc-cnrc.gc.ca/en/megapipe/docs/filtold.html} to transform MegaCam bands to SDSS bands. Specifically, we use $g' = g_{\rm Mega} +0.103\times (g_{\rm Mega} - i_{\rm Mega})$ and $i' = i_{\rm Mega} - 0.003\times (g_{\rm Mega} - i_{\rm Mega})$.}. From Fig.~\ref{fig:ngvs_comparison} we find that galaxies from Fornax and Virgo appear to follow similar trends in $g'-i'$, $n$ and $R_e$, although the low number statistics at the brightest magnitudes make meaningful comparisons unreliable. In particular, the $g'-i'$ moving averages appear to well trace the red sequences of Virgo and Fornax. In the Virgo cluster, one could expect ram pressure stripping to be more efficient than in Fornax, which could lead to a generally redder galaxy population. Indeed, observations find higher central X-ray gas density and velocity dispersion for Virgo compared to Fornax, which point to more efficient ram pressure stripping \citep[see][Sect. 5.5]{hamraz2019}. However, in terms of the morphology of galaxies, Fornax contains a higher fraction of ETGs to LTGs compared to Virgo \citep[e.g.][]{ferguson1988}. 

\section{Discussion}\label{sect:discussion}

\subsection{Environmental effects}
In this section we explore the effects of environmental processes and whether they can explain the observed differences in the structures of galaxies between Fornax main and Fornax group. To summarise, our main results are:
\begin{itemize}
    \item We find significant (KS test $p$-value $<\alpha=0.05$) differences in the distributions of residual (i.e. stellar mass-trend removed) Sérsic-derived quantities ($g'-r'$, $r'-i'$, $R_e$, $\bar{\mu}_{e,r'}$), residual non-parametric morphological indices ($A$, $S$), and $\Delta(g'-r')$ between Fornax main and Fornax group. 
    \item We find significant halo-centric trends (based on $p$-values from Spearman's $\rho$) in residual Sérsic-derived quantities ($r'-i'$, $R_e$, $\bar{\mu}_{e,r'}$), residual non-parametric indices ($A$, $S$, $G$, $M_{20}$), and $\Delta(g'-r')$ for Fornax main galaxies. We do not find significant halo-centric trends for Fornax group. 
    \item From multi-component decomposition models, we find the fraction of simple galaxies ($f_{\text{simple}}$) drops from 1 (all simple) to 0 (all complex) within $-14 \lesssim M_{r'} \lesssim -19$ for both Fornax main and Fornax group galaxies.
\end{itemize}
The implications of the results can be interpreted as follows: 
\begin{itemize}
    \item Overall, galaxies in the Fornax group exhibit i) different stellar populations, and are ii) smaller, iii) brighter (in terms of mean effective surface brightness), and iv) less smooth (due to star formation and/or tidal perturbations), compared to galaxies in Fornax main.
    \item Galaxies in the Fornax main cluster become more extended, fainter, and have smoother light distribution closer to the cluster centre. The lack of significant halo-centric trends in Fornax group may be due to the lower sample size. 
    \item Overall, the structural complexity of galaxies appears to be mainly driven by (stellar) mass, and does not appear to be heavily influenced by the environment which they reside in (between the Fornax main cluster and Fornax group).
\end{itemize}
To explain the observed differences, we focus on the effects of ram pressure stripping and tidal interactions on galaxy structures, which are most efficient in the cluster and group environments. 

\subsubsection{Ram pressure stripping} 
The process of ram pressure stripping acts on the gas in galaxies, typically by removing it whilst leaving the stellar component intact. This process hampers the galaxies' star formation abilities, turning them redder over time \citep[see ][and references therein]{boselli2014}. Simulations have shown that ram pressure stripping can act on short timescales, removing the atomic gas within a few hundred Myr \citep{boselli2008}, and stripping the entire gas reservoir of galaxies in timescales of $\sim 1-1.5$\,Gyr \citep{tonnesen2007, roediger2007}. It is thought that ram pressure stripping affects the galaxies from outside-in, with the possibility of some gas remaining bound in the central regions of galaxies \citep{mori2000,vollmer2009}.

Given the effects of ram pressure stripping, it is likely that gas stripping played a role in the significant differences found in the distributions of (residual) $g'-r'$ colour between Fornax main and Fornax group (see Fig.~\ref{fig:scale_rel_nomasstrend}). We compare these trends to those found in simulations; \citet{steinhauser2016}, for example, studied the effects of ram pressure stripping on galaxies with $\log_{10}(M_*/M_{\odot}) \sim 10$ entering clusters with different combinations of initial parameters. By considering stellar particles as simple stellar populations, they found that the colours of their model galaxies can change significantly within a few hundred Myr to one Gyr, particularly at shorter wavelengths. Given that the majority of the galaxies in our sample are much less massive than $\log_{10}(M_*/M_{\odot}) < 10$, their shallower potential wells are more susceptible to the effects of ram pressure stripping \citep{boselli2008}. Furthermore, calculations by \citet{venhola2019} showed that ram pressure is able to strip away all of the gas from galaxies with masses $\log_{10}(M_*/M_{\odot}) < 7.5\text{--}8$ which enter within $\sim 0.5R_{\text{vir}}$ of the Fornax cluster centre. Additionally, the differences in distributions of residual $A$ and $S$, which we attributed to the disparity in galaxy morphological classes of galaxies (i.e. ETG or LTG), is potentially due to ram pressure stripping quenching star formation. \citet{derijke2010} reached a similar conclusion for dwarfs in Fornax.

In the Fornax group, it appears that ram pressure stripping is not as efficient as in the Fornax main cluster. X-ray observations show the hot gas density to be higher in NGC~1399 \citep{paolillo2002} than in NGC~1316 \citep{kim1998}, and the high density gas extends further out in the case of NGC~1399. To get an order-of-magnitude estimate of the differences in ram pressure between Fornax main and Fornax group, we adapt the expression from \citet{gunn1972}:
\begin{equation}
    P_{\rm ram} \approx \rho_{\rm ICM} v^{2},
\end{equation}
where $\rho_{\rm ICM}$ is the density of the ICM, and $v$ is the velocity of the galaxy travelling through the ICM. By substituting $\rho_{\rm ICM}$ with the hot gas density from \citet{paolillo2002} and \citet{kim1998}, and $v$ for the velocity dispersion from \citet{maddox2019} (318\,kms$^{-1}$ and 204\,kms$^{-1}$ for Fornax main and Fornax group, respectively), we find the ratio of $P_{\rm ram}$ to be close to unity, if the outermost measured density was used ($\sim 4\times 10^{-4}$\,cm$^{-3}$ at $\sim 100$\,kpc for Fornax main; $\sim 1\times 10^{-3}$\,cm$^{-3}$ at $\sim 10$\,kpc for Fornax group). However, if the density at $\sim 60$\,kpc (region where the gas density is dominated by NGC~1399) was used for the Fornax main instead, $P_{\rm ram, main}/P_{\rm ram, group} \approx 1.5$. Likewise, if the density at the same radial distance ($\sim 10$\,kpc) was used, $P_{\rm ram, main}/P_{\rm ram, group} \approx 10$. 

Although the calculations support stronger ram pressure stripping in Fornax main, we cannot rule out the possibility that galaxies in Fornax group also experience gas stripping, albeit at a lower efficiency. In fact, there is evidence to suggest that gas stripping does occur in galaxy groups \citep[e.g.][]{brown2017}. More recently, \citet{steyrleithner2020} performed simulations of ram pressure stripping on dwarf galaxies ($6 \lesssim \log_{10}(M_*/M_{\odot}) \lesssim 8$) by varying the wind velocity of the intracluster medium (ICM) to match cluster and group environments. They found that a dwarf galaxy, subject to winds analogous to those of cluster environments, follows the same trends in star formation rate (SFR) and gas fraction with time as a dwarf galaxy subject to winds typical of group environments. However, there is one crucial difference: the stripping timescale is longer in the group environment by several hundreds Myr \citep[see][Fig. 1 and 2]{steyrleithner2020}. Furthermore, they found that ram pressure stripping in the group environment does not appear to be able to completely remove the gas. As such, the lower density group environment of Fornax group may be to blame for the relatively milder gas stripping compared to Fornax main. The relative inefficiency of gas stripping in the Fornax group also translates to a longer timescale for colour change, hence the difference in the distribution of $g'-r'$ colour between Fornax main and Fornax group.

\citet{donnari2020}, who focused on the quenching\footnote{\citet{donnari2020} defined as quenched galaxies with SFR 1\,dex below the star forming main sequence.} of galaxies across environment in IllustrisTNG, found that the majority ($\sim 75\%$) of satellite galaxies become quenched within their final host halo (of similar mass to Fornax main), whereas just over 20\% of the galaxies have been pre-quenched (i.e. quenched before infall into their current host halo). Although the fraction of pre-quenched galaxies is low, \citet{donnari2020} also noted that up to $\sim 50\%$ of satellite galaxies quenched by their final host have been in groups before, which may have helped to lower star formation in the galaxies before they are defined as quenched (see their Sect. 4.4 for more detail). This supports the aforementioned scenario that ram pressure stripping occurs in the group environment, but over a longer timescale. Indeed, this could be the case for the Fornax group, where the environmental effects are not efficient enough to show significant trends (e.g. $g'-r'$ colour with halo-centric distance in Fig.~\ref{fig:cluster_dist}), but still have mild effects on the galaxies. Furthermore, the fraction of pre-quenched galaxies appears to increase with decreasing stellar mass ($\sim 30\%$ of $\log_{10}(M_*/M_{\odot})\approx 9$ galaxies compared to a minimum of $\sim 15\%$ of $\log_{10}(M_*/M_{\odot})\approx 10$). Therefore, it is not impossible that for $\log_{10}(M_*/M_{\odot})<9$ galaxies the fraction of pre-quenched galaxies increases. 

In terms of $\Delta (g'-r')$ (see Fig.~\ref{fig:colour_grad}), our results imply that Fornax group galaxies tend to have redder inner regions than their outskirts, whereas Fornax main contains a higher fraction of galaxies with bluer inner regions than their outskirts. It is worth noting that several works have reported a bluer inner region in cluster galaxies (e.g. Virgo, \citealt{lisker2007,urich2017}; Coma, \citealt{denbrok2011}; Fornax, \citealt{hamraz2019}). In particular, using radial colour profiles derived from Hubble Space Telescope data, \citet{hamraz2019} found that ETGs in Fornax overwhelmingly have a blue core\footnote{These cores typically cover a much smaller central area (radius within 1\,arcsec) than the inner regions we defined, but are comparable to the nuclei we find in this work.}. Whilst it is tempting to directly attribute the blue cores to the nucleation of galaxies, where nuclei can represent nuclear star clusters (NSC) or even active galactic nuceli (AGN), \citet{hamraz2019} found that the sizes of nuclei tend to be smaller than the blue cores (see their Fig.~14).

Additionally, given that ram pressure stripping works outside-in, we would expect the outer regions of a galaxy experiencing ram pressure stripping to become more gas deficient and hence host less star formation than its inner region. Gas retained in the innermost regions can continue to form stars, hence generating a bluer inner region/redder outer region (i.e. $\Delta (g'-r')>0$). Therefore, again, this effect may be attributable to the different efficiencies of ram pressure stripping between Fornax main and Fornax group. Lastly, it is worth keeping in mind that as star formation becomes quenched, there can be a 'final' burst of star formation within the galaxy due to compression of the interstellar medium (ISM) from ram pressure stripping \citep{bekki2014,steinhauser2016,steyrleithner2020}. For galaxies in the Fornax group, where ram pressure stripping is not as efficient, inside-out formation occurs due to the presence of gas throughout the galaxies. This leads to a more evolved central region, which translates to negative $\Delta (g'-r')$.

\subsubsection{Tidal interactions} \label{sect:harassment}
Tidal interactions of galaxies can occur due to nearby encounters between cluster galaxies (at high relative speeds this becomes galaxy--galaxy harassment), and between a galaxy and the cluster potential itself. This mechanism can be very efficient in unbinding dark matter halo mass from galaxies which can lead to changes in effective radii \citep[see][and references therein]{recchi2014}. However, \citet{smith2010} found that tidal interactions are not efficient in transforming recently accreted late-type dwarf galaxies into dwarf ellipticals. In general, the effects of tidal interactions are sensitive to the initial conditions of the galaxy as it falls into the cluster. For example, \citet{smith2015} found that the fraction of mass bound to the galaxy, as well as the change in effective radii of the stellar component, are dependent on the number of pericentre passages about the cluster potential. Furthermore, the inclination of the galaxy with respect to its orbital plane can affect the amount of mass loss \citep{bialas2015}. It is worth noting, however, that the galaxy models from \citet{smith2015} and \citet{bialas2015} are of the order $\log_{10}(M_*/M_{\odot})\sim 9$, which is at the high mass end of the dwarfs in our sample. 

From simulations, we would expect harassed galaxies with multiple pericentre passages to be close to the cluster centre \citep[see][their Fig.~2]{smith2015}. However, given that we only have the projected halo-centric distances, there is significant contamination in using our halo-centric distances as a proxy for the galaxy's time spent in the cluster, as well as for the number of pericentre passages. Nevertheless, a general trend in the time spent within the halo with halo-centric distance should still be possible to observe \citep[e.g.][their Fig.~4]{lisker2013}. 

From Fig.~\ref{fig:cluster_dist} we find a negative trend in residual $R_e$ with increasing halo-centric distance, meaning that galaxies closer to the cluster centre have larger $R_e$ compared to the counterparts in the cluster outskirts. This may have occurred because, as mentioned, $R_e$ can increase as dark matter is stripped from galaxies interacting with the cluster potential. As the galaxies 'relax' into their new equilibriums, the stellar components become more loosely bound (due to the mass loss), hence increasing $R_e$. This mechanism is consistent with what has been found in comological simulations \citep{fattahi2018, sales2020}. In particular, \citet{fattahi2018} studied dwarf galaxies from Local Group analogs via hydrodynamical simulations from APOSTLE \citep{sawala2016}. They found that as the galaxies cross the virial radius of the host haloes, their total mass (i.e. mass bound within the stellar half-mass radius, dominated by dark matter) begins to decrease due to tidal effects. Additionally, the stellar half-mass radius increases and the stellar mass (bound within the original stellar half-mass radius) decreases \citep[see][their Fig.~3]{fattahi2018}. They note that this is consistent with the evolutionary tracks derived from \citet{penarrubia2008} \citep[see also][]{errani2015} for dwarf galaxies embedded in dark matter haloes. Using the evolutionary tracks and assuming a stellar mass loss of $50\%$ \citep[e.g. a heavily harassed galaxy from][]{bialas2015}, we would expect the stellar half-mass radius to increase by a factor of $\sim 14\%$. 

To see if the prediction in $R_e$ agrees with our observations, we must quantify the change in $R_e$ from our observations. As such, we estimated the difference in residual $\log_{10}(R_e)$ by taking median values of galaxies within halo-centric distances of 0.1\,Mpc and beyond 1.1\,Mpc for the centre and outskirts, respectively. This produced a value of $\Delta \log_{10}(R_e)=0.036$, which translates to a $\sim 8\%$ increase in $R_e$. Hence, the trend in residual $R_e$ shown in Fig.~\ref{fig:cluster_dist} can potentially be explained by tidal interactions of galaxies with the cluster potential.

Regarding surface brightness, Fig.~\ref{fig:cluster_dist} shows a negative correlation for residual $\bar{\mu}_{e,r'}$ with halo-centric distance for Fornax main galaxies, such that the surface brightnesses of galaxies become fainter as galaxies fall towards the cluster centre. This emerges naturally from the correlation found for $R_e$ given how $\bar{\mu}_{e,r'}$ is defined. Additionally, however, another likely cause is the quenching of star formation from ram pressure stripping, which causes the galaxy to fade over time. Hence, the cumulative effects of tidal interactions and ram pressure stripping could produce the observed trend in surface brightness with halo-centric distance.

In Fornax group galaxies, by contrast, we do not find any significant trends in residual $R_e$ and $\bar{\mu}_{e,r'}$ with group-centric distances. This suggests that the aforementioned cluster--galaxy tidal effects are not as efficient in Fornax group. For this scenario, we can consider the effects of the cluster/group mass on the tidal effects. From Eqn.~(10) of \citet{venhola2019} \citep[adapted from][]{king1962}, we see that 
\begin{equation}
    R_{\text{tidal}} = R_{\text{peri}} \left( \frac{M_{\text{galaxy}}}{M_{\text{cl}}(<R_{\text{peri}}) (3+e)} \right)^{1/3},
    \label{eqn:r_tidal}
\end{equation}
where $R_{\text{tidal}}$ denotes the maximum radius at which material is still bound to the galaxy, $R_{\text{peri}}$ is the pericentre distance to the cluster centre, $M_{\text{galaxy}}$ is the mass of the galaxy, $M_{\text{cl}}(<R_{\text{peri}})$ is the cluster mass within $R_{\text{peri}}$, and $e$ is the orbital eccentricity of the galaxy. So, for a given galaxy (i.e. same $M_{\text{galaxy}}$, $e$, and $R_{\text{peri}}$), Eqn.~(\ref{eqn:r_tidal}) can be reduced to $R_{\text{tidal}} \propto M_{\text{cl}}(<R_{\text{peri}})^{-1/3}$. To estimate $M_{\text{cl}}(<R_{\text{peri}})$, we use the integrated mass profile of NGC~1399 from \citet{paolillo2002} (right panel, Fig. 17) for Fornax main. In the case of Fornax group, given that it is dominated by NGC~1316, we calculate the halo mass profile following Eqn.~(6) from \citet{fabricant1980} and substituting the values\footnote{we used $\mu=0.6$, $\rho \sim r^{-1.53}$, and a constant $kT=0.8$\,keV. This results in a halo mass profile of $M(<r)\approx 4.511\times 10^{10} (r/\text{kpc})\,M_{\odot}$.} from \citet{kim1998}. We find that $R_{\text{tidal}}$ for a given galaxy in the Fornax group is $\sim 10\%$ higher than in Fornax main, for $R_{\rm peri}=1$\,kpc. Thus, galaxies in Fornax group should tend to be more stable against cluster--galaxy interactions than galaxies in Fornax main. 

To compare the efficiency of galaxy--galaxy interactions between Fornax main and Fornax A, we use Eqn.~(11) of \citet{venhola2019} \citep[adapted from Eqn.~(8.54) of][]{binney2008} to estimate the disruption timescale due to high-speed encounters\footnote{One caveat is that Eqn.~(\ref{eqn:t_disrupt}) assumes the impact speed is much greater than the orbital speeds of stars inside the galaxies. This is not the case for galaxies in the Fornax group, and somewhat questionable for the Fornax main. Nevertheless, it allows for a rough comparison between the two environments.}: 
\begin{equation}
    t_d = \frac{0.043}{W}\frac{\sqrt{2}\sigma M_s r_h^2}{GM_p^2n_pa^3}, \label{eqn:t_disrupt}
\end{equation}
where $W$ is related to the mass distribution of the perturber galaxy, $\sigma$ is the cluster velocity dispersion, $M_s$ is the total mass of the galaxy being disrupted, $r_h$ is the half-mass radius of the perturber galaxy, $M_p$ is the total mass of the perturber galaxy, $n_p$ is the number density of perturber galaxies in the cluster, and $a$ is the orbital radius (of stars) within the disrupted galaxy. As with the tidal radius, for a given galaxy in either Fornax main and Fornax group, $t_d \propto \frac{\sigma r_h^2}{M_p^2n_p}$. For the velocity dispersion we adopt values of 318\,km\,s$^{-1}$ and 204\,km\,s$^{-1}$ for Fornax main and Fornax group, respectively, from \citet{maddox2019}. 

We selected galaxies with $\log_{10}(M_*/M_{\odot})>9$ as perturber galaxies. For Fornax main we also limit the perturber galaxies to those within the virial radius, as the probability of galaxies interacting at large halo-centric distances is low. To calculate $M_p$ we first convert the stellar mass to halo mass using the relation from \citet{moster2010}\footnote{We used Eqn.~(2) and values from Table~1 of \citet{moster2010}.} before taking the median. Additionally, we used the conversion factor of $r_h/R_e=3.6$ from \citet{venhola2019} to convert our aperture $R_e$ values. Lastly, we estimated $n_p$ based on the number of perturber galaxies within a spherical volume of a given radius. For Fornax main we used the virial radius ($=700$\,kpc), whereas for Fornax group we used the projected distance to the outermost member perturber galaxy ($\approx 364$\,kpc). Table~\ref{tab:harassment_params} shows the parameter values corresponding to Fornax main and Fornax group. Substituting in the values, we find that for a given galaxy, $t_d$ of Fornax group is $\sim 5\%$ shorter than in Fornax main, implying that galaxy--galaxy tidal interactions are marginally more effective at disrupting galaxies in the Fornax group. 

To summarise, we investigated whether tidal interactions could lead to the observed halo-centric trends (or lack thereof) in $R_e$ and $\bar{\mu}_{e,r'}$. We find the trend in $R_e$ could be explained by tidal interactions, and a combination of tidal interactions and ram pressure stripping could account for the trend in $\bar{\mu}_{e,r'}$ in the Fornax main. Taking the calculations of the effects of galaxy--galaxy and cluster--galaxy interactions into account, the marginal difference in the effectiveness of tidal interactions should result in similar halo-centric trends in Fornax main and Fornax group. However, from Fig.~\ref{fig:cluster_dist} there appears to be little correlation between group-centric distance with $R_e$ and $\bar{\mu}_{e,r'}$. This suggests that tidal interactions are inefficient in the Fornax group, or that the galaxies have not spent enough time in the group environment. For the latter case, \citet{raj2020} used phase-space analysis to show that a significant fraction of the brightest galaxies must have spent several Gyr in the Fornax group. This would suggest that, for whatever reason, tidal interaction is not efficient in transforming the galaxies in the Fornax group. 

\begin{table}
    \caption{Parameters calculated based on the perturber galaxies ($\log_{10}(M_*/M_{\odot})>9$) within Fornax main and Fornax group.}
    \centering
    \begin{tabular}{lrrr}
        \hline
          Environment & $\log_{10}(M_p/M_{\odot})$ & $n_p$ [Mpc$^{-3}$] & $r_h$ [kpc] \\
        \hline
        Fornax main & 11.57 & 20.88 & 9.18 \\
        Fornax group & 11.39 & 44.50 & 10.62 \\
        \hline
    \end{tabular}
    \label{tab:harassment_params}
\end{table}

\subsection{Structures in galaxies}
In this work we find that the structural complexities exhibited in FDS galaxies correlate with their $r'$-band magnitudes, such that the faintest galaxies are well described by a Sérsic profile (not counting a possible nucleus component) and the brightest galaxies require more than one component to account for their morphological structures. To verify that this trend exists in other studies, we compared the simplicity fraction of FDS galaxies to other multi-component decomposition studies, namely \citet{janz2014}, which was based on Stellar content, MAss and Kinematics of Cluster Early-type Dwarfs (SMAKCED) and \citet{salo2015}, which was based on S$^4$G. The SMAKCED survey contains $H$-band images of 121 early-type dwarf galaxies in the Virgo cluster spanning $-16 > M_{r'} > -19$ \citep[assuming a distance modulus of 31.09 for the Virgo cluster,][]{mei2007}. The S$^4$G provides a total of 2352 nearby galaxies ($<40$\,Mpc) for which multi-component decompositions were made based on 3.6\,$\mu$m images. The $r'$-band magnitudes for SMAKCED were obtained from \citet{janz2008}. For S$^4$G galaxies, we estimate the $r'$-band magnitudes based on $3.6$\,$\mu$m and SDSS $r'$-band magnitudes (see Appendix~\ref{sect:rband_s4g}). 

Applying the same limit on axial ratios (see Sect.~\ref{sect:fds_model_complexity}), the SMAKCED sample consisted of 103 ETGs (dwarf ellipticals) from the Virgo cluster with multi-component decomposition models. Similarly, a sample of 479 LTGs ($T \ge 1$) was selected from S$^4$G. For the S$^4$G sample, we apply an additional constraint in determining the complexity of a galaxy model to combat unresolved bulges. This is due to the lower resolution sampling of the Spitzer instrument (0.75\,arcsec/pixel). Additionally, the S$^4$G sample also contains galaxies at greater distances (up to 40\,Mpc), so it is not surprising that less structure may be resolved in the images. The unresolved bulges manifest as nucleus components in decomposition models, hence nucleated S$^4$G galaxies with $M_{r'}<-17$ were considered to be complex. 

\begin{figure}
\centering
\includegraphics[width=\hsize]{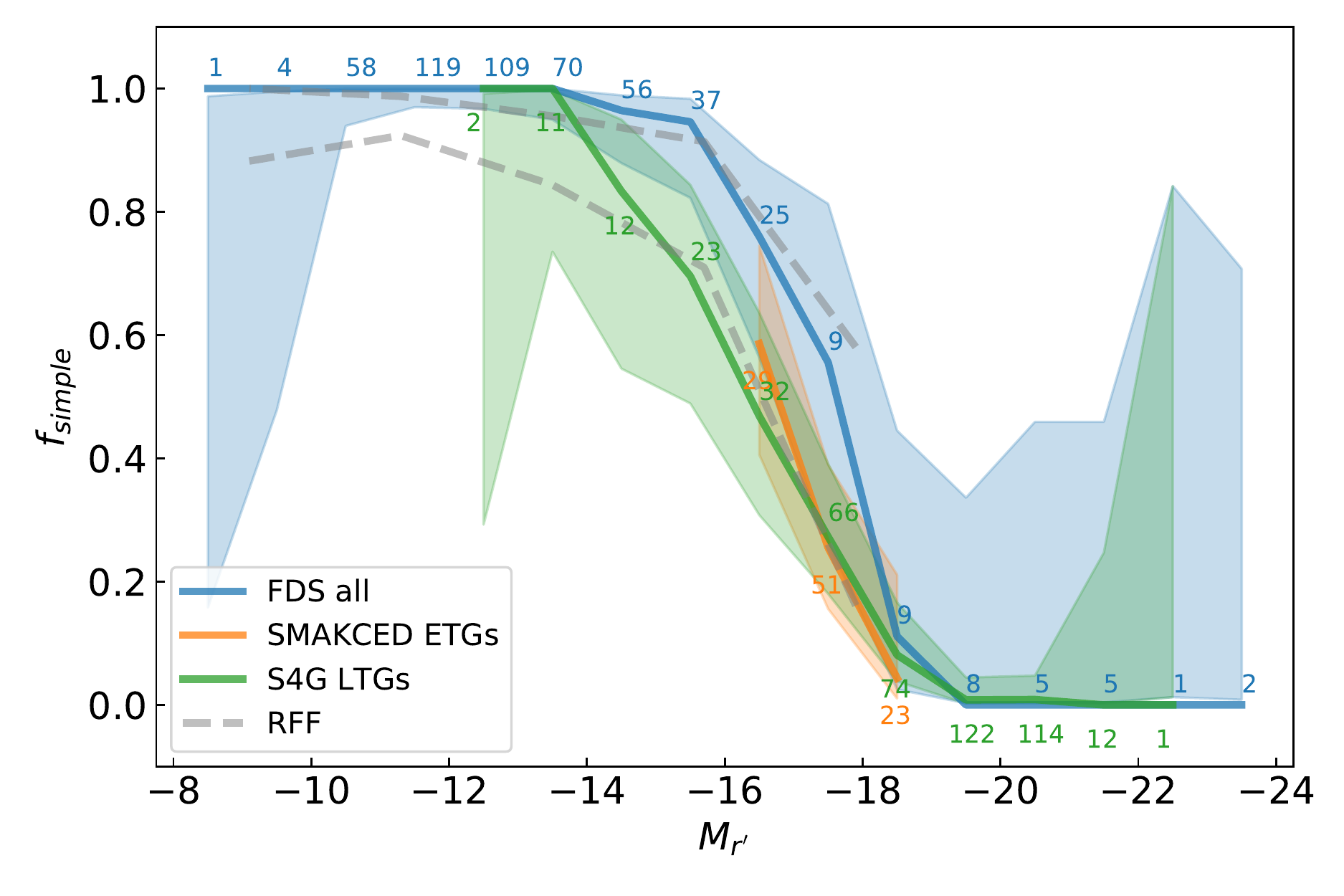}
  \caption{Simplicity fraction ($f_{\text{simple}}$) as a function of $r'$-band magnitude for this work (FDS, \textit{blue}), SMAKCED (\textit{orange}), and S$^4$G (\textit{green}) galaxies. Numbers alongside the solid lines denote the sample size per bin. We also include upper and lower limits (\textit{grey}) based on RFF from FDSDC for comparisons. A bin size of 1\,mag was used for FDS, SMAKCED, and S$^4$G. The shaded regions denote the $95\%$ confidence interval (same as in Fig.~\ref{fig:simple_fraction_fornax}). }
     \label{fig:simple_fraction_mix}
\end{figure}

Fig.~\ref{fig:simple_fraction_mix} shows the distributions of simplicity fraction ($f_{\text{simple}}$) for the FDS, SMAKCED, and S$^4$G samples. Overall, the distributions show a common feature: there is a drop in simplicity fraction around a certain range of magnitudes ($-14 \gtrsim M_{r'} \gtrsim -19$). Accounting for the uncertainties, the $f_{\text{simple}}$ distributions for SMAKCED and S$^4$G are comparable to the FDS sample (within the confidence intervals). Similarly, the majority of FDS $f_{\text{simple}}$ values within the same magnitude range is comparable to the confidence intervals of S$^4$G and SMAKCED.

Additionally, we include the change in $f_{\text{simple}}$ as a function of magnitude based on the RFF given in FDSDC. It is worth noting the subtle differences between the RFF and multi-component models. The RFF shows overall deviation of the galaxy image from a Sérsic model, but the cause of the deviation may not necessarily warrant an additional component. As an example, a high RFF value could indicate that a bar component is not fitted by the Sérsic model, or that there are clumps of star formation within the galaxy, which would still be deemed as a disk component in the multi-component model. In Fig.~\ref{fig:simple_fraction_mix} we use two limiting values of RFF=0.02 and RFF=0.06 as dashed grey lines. The difference in $f_{\text{simple}}$ due to the two RFF limits provides a sense of how subjectivity in choosing the best-fitting decomposition models\footnote{In other words, the number of presumably physical components in the decomposition model.} can affect $f_{\text{simple}}$. Overall, the limits based on RFF also overlap with the confidence intervals of FDS, SMAKCED, and S$^4$G, showing a similar trend in $f_{\text{simple}}$ with absolute magnitude. The observed decrease in simplicity fraction is in line with what was found by \citet{janz2016}, who also compared the SMAKCED sample to S$^4$G and observed the decrease in simplicity fraction with brighter absolute magnitude. In Appendix~\ref{app:downsize} we show that the observed trends in $f_{\text{simple}}$ are unlikely to be attributable to lower S/N or resolution eroding our ability to detect and measure morphological structures in low mass galaxies.

It is important to keep in mind the differences in the datasets. One potential difference is the image depth. In principle, one could expect that deeper data would result in more well defined structures in the galaxy images, which could lead to differences in the decomposition models. For FDS, the $1\sigma$ surface brightness limit within 1\,arcsec$^2$ area reaches 27.8\,mag\,arcsec$^{-2}$ for $r'$-band. S$^4$G was designed to reach stellar mass surface densities of $\sim 1M_{\odot}$pc$^{-2}$, which translates to a $1\sigma$ surface brightness limit of 27\,mag\,arcsec$^{-2}$ in 3.6\,$\mu$m \citep{sheth2010}. However, depending on the galaxy, the S$^4$G typically reached limiting depth of 25--26\,mag\,arcsec$^{-2}$ \citep{salo2015}. The SMAKCED survey aimed to reach comparable depths (S/N of 1 per pixel) for all galaxies at 2$R_e$. This corresponded to a surface brightness of 22.2--23.0\,mag\,arcsec$^{-2}$ in $H$-band \citep{janz2014}. Another important difference is the wavelength of the observations. In general, images at shorter wavelengths are more affected by the structures of dust and star formation regions. Both effects can potentially cause differences in the multi-component models. Given that the samples used in the simplicity fraction comparison are from three different datasets which sample different populations/environments of galaxies, the similar trends in model complexity with increasing brightness suggest that the ability to form/maintain structures in galaxies is likely driven by galaxy mass \citep{janz2016}.

To summarise, we find that ram pressure stripping can explain the difference in the distributions of $g'-r'$ colour and $\Delta (g'-r')$ for galaxies in Fornax main. Furthermore, we find that tidal interaction in Fornax main is efficient enough to match the observed cluster-centric trends in $R_e$ and $\bar{\mu}_{e,r'}$. We speculate that ram pressure stripping can occur in the Fornax group, albeit on longer timescales such that $g'-r'$ colour and $\Delta (g'-r')$ group-centric trends are not significant ($p$-value $>\alpha$). Regarding the overall lack of group-centric trends, it is also worth keeping in mind that the sample size of the Fornax group is smaller than the Fornax main, which likely compounds the lack of significant trends (see Fig.~\ref{fig:test_samplesize}). According to multi-component decomposition models, the aforementioned environmental mechanisms does not appear to significantly alter the structural complexity of Fornax galaxies, or indeed galaxies in other environments. Given the plethora of structural quantities measured for the Fornax galaxies, it would be prudent to quantitatively compare our observations with other clusters, as well as those from cosmological simulations. Indeed, work is being conducted on the quenching of dwarfs between Fornax and Virgo (Janz et al. submitted) as well as on low surface brightness galaxies and comparisons of the luminosity function to IllustrisTNG (Venhola et al. in prep.).


\section{Conclusions}\label{sect:conclusion}

In this paper we studied the structure of galaxies in the Fornax cluster as part of the Fornax Deep Survey. Our main aim was to investigate preprocessing of galaxies from the group to the cluster environment using the Fornax main cluster and the Fornax group as testbeds. We measured the structures of 582 Fornax member galaxies ($5<\log_{10}(M_*/M_{\odot})<11.4$) through 2D structural decompositions with Sérsic+PSF models and non-parametric morphological measures: Concentration, Asymmetry, Smoothness, Gini, and $M_{20}$ (Sect.~\ref{sect:scale_rel} and \ref{sect:pars_vs_dist}). Additionally, we conducted multi-component decompositions to quantify morphological structures in detail (Sect.~\ref{sect:decomp_multicomp} and \ref{sect:fds_model_complexity}). Finally, we produced radial asymmetry profiles (Sect.~\ref{sect:asym_prof}) and inner-to-outer colour difference (Sect.~\ref{sect:col_diff}). In comparing these properties of galaxies between the Fornax main and Fornax group samples, we found the following:

\begin{itemize}
    \item There are significant (KS test $p$-value $<\alpha=0.05$) differences in the distributions of residual (i.e. stellar mass-trend removed) quantities based on Sérsic profiles ($g'-r'$, $r'-i'$, $R_e$, $\bar{\mu}_{e,r'}$, see Fig.~\ref{fig:scale_rel_nomasstrend}), residual non-parametric indices ($A$, $S$, see Fig.~\ref{fig:nonparametric_scalerel_full}), and $\Delta(g'-r')$ (see Fig.~\ref{fig:colour_grad}) between Fornax main and Fornax group. These differences suggest that galaxies in the Fornax group exhibit: i) different stellar populations (e.g. bluer $g'-r'$), ii) smaller effective radii, iii) brighter surface brightnesses, and iv) less smooth light distributions, as compared to their Fornax main counterparts. 
    
    \item Fornax main galaxies beyond $R_{\text{vir}}$ share similar distributions in $g'-r'$ colour with Fornax main galaxies within $R_{\text{vir}}$ and similar $R_e$ and $\bar{\mu}_{e,r'}$ with the Fornax group galaxies (see Fig.~\ref{fig:scale_rel_nomasstrend_3sub}). 
    
    \item Based on projected halo-centric distance (see Fig.~\ref{fig:cluster_dist} and \ref{fig:nonparametric_dist}), there are weak, but significant ($p$-values from Spearman's $\rho$) trends which suggest that as galaxies fall towards the centre of Fornax main, they become: i) more extended ($R_e$), ii) fainter in surface brightness ($\bar{\mu}_{e,r'}$), iii) generally smoother, more symmetric with evenly distributed light profiles ($A$, $S$, $G$, $M_{20}$), and iv) have bluer inner regions/redder outskirts ($\Delta (g'-r')$). We did not find significant trends in Fornax group, which is (at least) partially due to its small sample size.
    
    \item From multi-component decompositions, the complexity of models for our galaxies follow the same trend with $r'$-band magnitude, regardless of whether they reside in Fornax main or Fornax group (see Fig.~\ref{fig:simple_fraction_fornax}), nor whether they are ETGs or LTGs (see Fig.~\ref{fig:simple_fraction_class}). Furthermore, the trend we found for galaxies in Fornax also holds for galaxies in the Virgo cluster and in the field (see Fig.~\ref{fig:simple_fraction_mix}). This suggests that the formation/maintenance of morphological structures in galaxies is likely driven by galaxy mass, rather than environment. 
    
\end{itemize}

The difference in the residual $g'-r'$ colour and $\Delta (g'-r')$ distributions between Fornax main and Fornax group suggests that ram pressure stripping is an important environmental mechanism in Fornax. Similarly, the differences in residual $R_e$ and $\bar{\mu}_{e,r'}$ suggest that tidal interactions are also significant. For Fornax group, the group-centric trends with the largest Spearman's $\rho$ were found for the global $g'-r'$ and $\Delta(g'-r')$. Although the $p$-values suggest they are not significant in Fornax group, it is likely that some gas stripping can occur in galaxy groups, which expedites the transformation in colour when they fall into a cluster. Interestingly, the order-of-magnitude calculations on the tidal radius and disruption timescale suggest similar efficiencies between Fornax main and Fornax group, although the observed group-centric trends were not significant in the group. Overall, the generally weak rank correlations (i.e. Spearman's $\rho$) with projected halo-centric distance suggest that environmental mechanisms in the Fornax region is not particularly efficient, for which the relatively low cluster mass could be a factor. Nevertheless, it is possible that, despite the inefficiency of environmental mechanisms to fully transform galaxies before cluster infall, galaxies in groups can still be partially preprocessed. Ultimately, our work provides perspective on cluster- and group-wide environmental studies, as well as a good basis of comparison for future studies on the structures of galaxy in clusters and groups.

\begin{acknowledgements}
      We acknowledge financial support from the European Union’s Horizon 2020 research and innovation program under the Marie Skłodowska-Curie grant agreement No.~721463 to the SUNDIAL ITN network. HS, EL, and AV are also supported by the Academy of Finland grant No.~297738. GvdV acknowledges funding from the European Research Council (ERC) under the European Union's Horizon 2020 research and innovation programme under grant agreement No.~724857 (Consolidator Grant ArcheoDyn). We acknowledge the usage of the HyperLeda database (http://leda.univ-lyon1.fr). 
\end{acknowledgements}

\bibliographystyle{aa}

\bibliography{references}

\appendix

\section{Excluded FDSDC sources}\label{app:duplicate}
The galaxies excluded from the final sample are listed in \ref{tab:duplicates}. 

\begin{table}
    \caption{Sources from FDSDC which were excluded from our sample.}
    \begin{tabular}{ll}
    \hline
      FDS ID & Remarks \\
    \hline
    FDS9\_0209 & Appears to be stream of another object\\
    FDS10\_0363 & Duplicate of FDS10\_0349\\
    FDS13\_0162 & Duplicate of FDS13\_0165\\
    FDS13\_0496 & Duplicate of FDS13\_0074\\
    FDS16\_0172 & Duplicate of FDS16\_0170\\
    FDS20\_0014 & Duplicate of FDS20\_0008\\
    FDS31\_0318 & Duplicate of FDS31\_0314\\
    FDS31\_0319 & Duplicate of FDS31\_0314\\
    \hline
    \end{tabular}
    \label{tab:duplicates}
\end{table}

\section{Uncertain multi-component decomposition models}\label{app:uncertain}
Table \ref{tab:uncertain_decomp} shows the galaxies which were deemed to have uncertain multi-component decomposition models and were excluded from the analyses. 
\begin{table}
    \caption{Uncertain multi-component decompositions}
    \centering
\begin{tabular}{p{0.2\linewidth}p{0.7\linewidth}}
\hline
FDS ID &                                           Comments \\
\hline
FDS10\_0261b &  Object next to/overlap with another source. \\
FDS11\_0002  &  Unsure if the multi-component model is physically meaningful. \\
FDS11\_0166  &  Galaxy overlaps with bright foreground star and NGC1399. \\
FDS11\_0229  &  Small object within NGC1399, but parameters seem physical. \\
FDS12\_0327  &  Object next to very bright foreground star. \\
FDS16\_0109  &  Object next to very bright foreground star. Unsure of masking. \\
FDS17\_0161  &  Model did not fit central region well. \\
FDS17\_0226  &  Significant part of galaxy masked due to overlap with star. \\
FDS19\_0001  &  Object next to very bright foreground star. Unsure of masking. \\
FDS19\_0435  &  Object next to very bright foreground star. \\
FDS20\_0472  &  Object next to very bright foreground star. \\
FDS21\_0202  &  Object next to very bright foreground star. Unsure of masking. \\
FDS22\_0244  &  Residuals show asymmetric structure, perhaps off-centre? \\
FDS26\_0000  &  Model did not fit outer region well (appears to be Type III disk break). \\
FDS28\_0054  &  Small object embedded within other objects. \\
FDS31\_0208  &  Object next to very bright foreground star. Unsure of masking. \\
FDS31\_0230  &  Model did not fit galaxy well (model is off-centred compared to galaxy) \\
FDS4\_0000   &  Model did not fit outer region well (appears to be Type III disk break). \\
FDS4\_0053   &  Model did not fit galaxy well. \\
FDS5\_0010   &  Abnormally low Sersic index (n$\ll$0.5). \\
FDS6\_0001   &  Unsure if the multi-component model is physically meaningful. \\
FDS7\_0326   &  Model did not fit galaxy well. \\
FDS9\_0357   &  Model did not fit galaxy well. \\
\hline
\end{tabular}

    \label{tab:uncertain_decomp}
    \tablefoot{Short comments on the galaxies which had uncertain multi-component decomposition models. }
\end{table}

\section{Illustrating data processing steps}
\subsection{Sky subtraction}\label{app:skysub}

\begin{figure}
\centering
\includegraphics[width=\hsize]{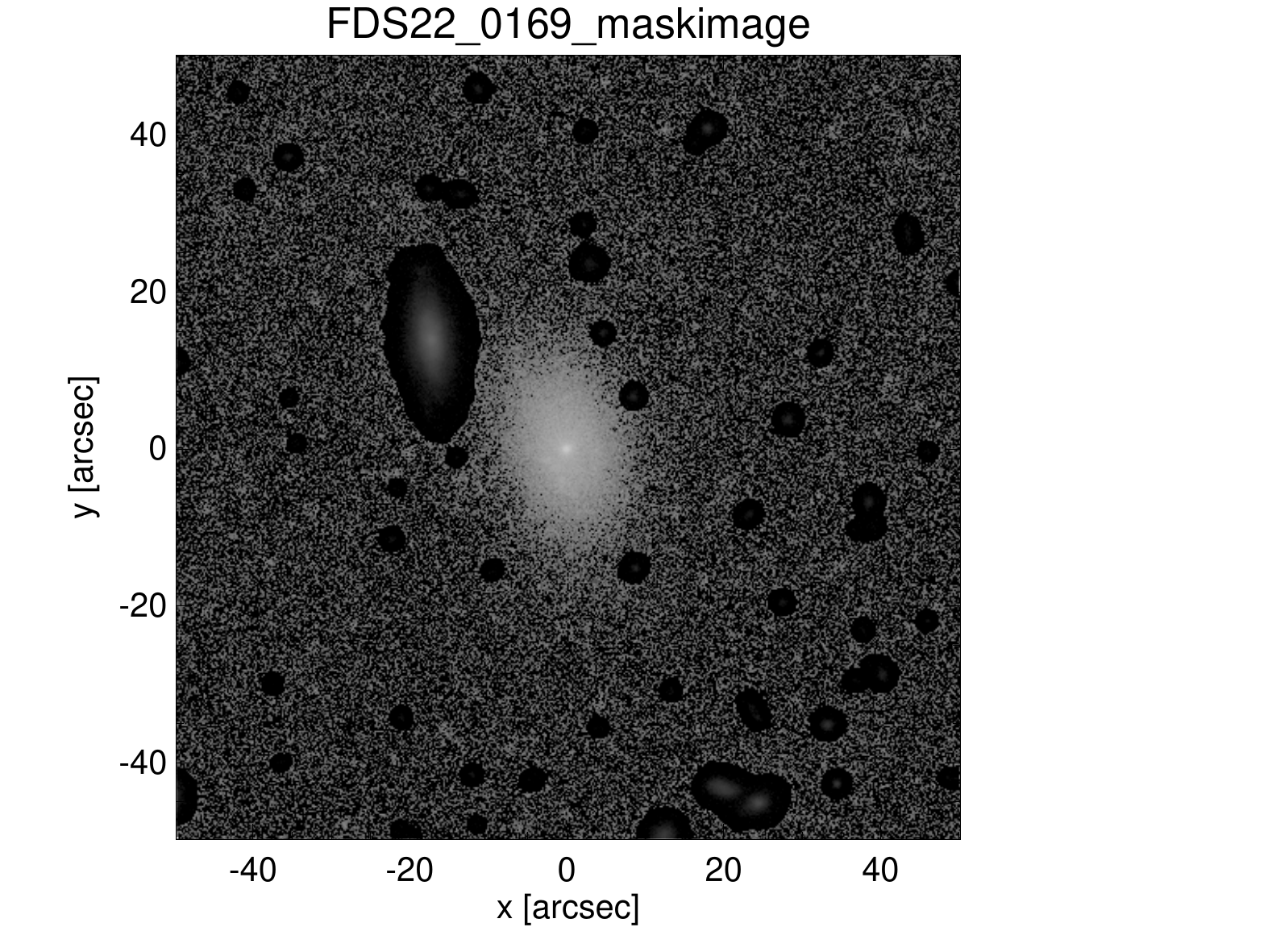}
\caption{Postage stamp image of FDS22\_0169 (FCC~70). The masked regions are shown with +5\,mag in the observed magnitudes.}
\label{fig:maskim}
\end{figure}

\begin{figure*}
\centering
\resizebox{0.9\hsize}{!}{\includegraphics[width=0.75\hsize]{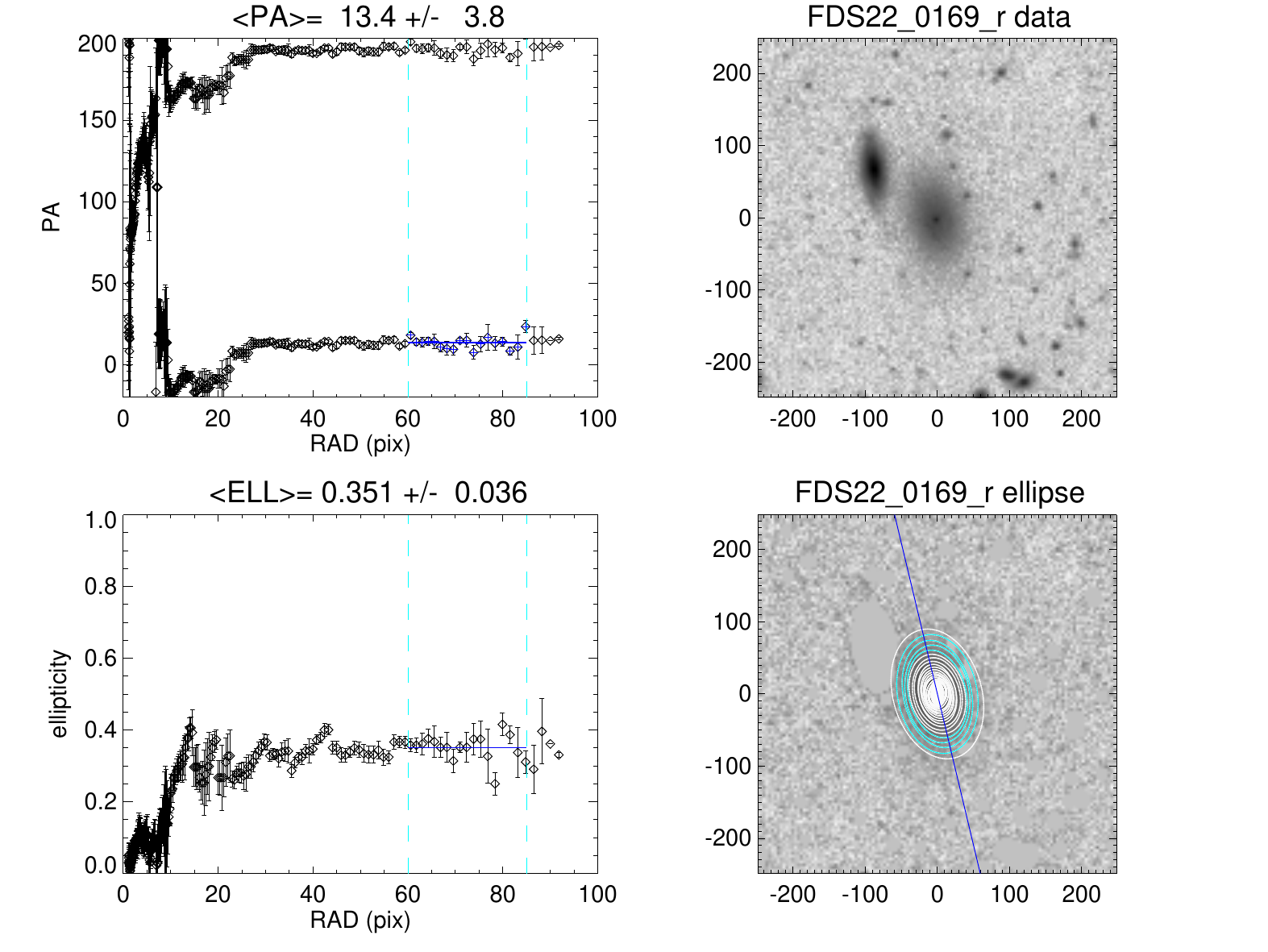}}
\caption{\textit{Upper left}: Position angle profile as a function of semi-major axis. \textit{Lower left}: Ellipticity profile as a function of semi-major axis. \textit{Upper right}: Sky-subtracted $r'$-band image of FDS22\_0169 (for reference). \textit{Lower right}: Masked image overlaid with isophotal ellipses. The dashed \textit{cyan} lines show the region where the mean outer isophotal ellipticity and PA was calculated from. }
\label{fig:ellipse}
\end{figure*}

\begin{figure*}
\centering
\resizebox{0.9\hsize}{!}{\includegraphics[width=\hsize]{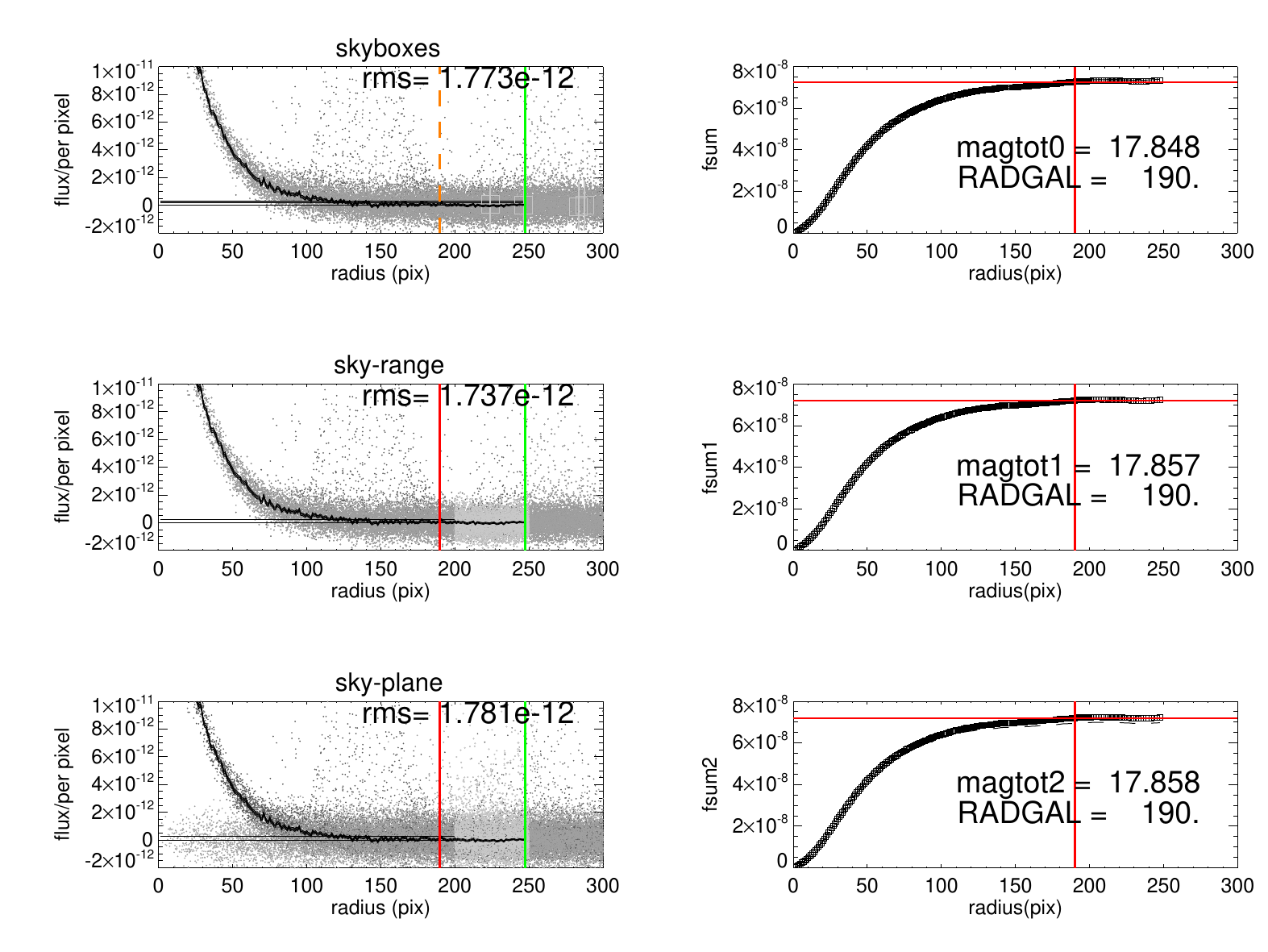}}
\caption{\textit{Left panels}: Azimuthally averaged flux profiles (solid \textit{black} lines) of FDS22\_0169 with different sky subtraction based on the ($r'$-band) cleanimages: skyboxes (\textit{upper}), sky annulus (\textit{centre}), and skyplane (\textit{lower}). The \textit{dark grey} points show the flux of individual pixels (in the AB system \citep{oke1983}; in units of 3631\,Jy $= 3631 \times 10^{-26}$\,W\,m$^{-2}$\,Hz$^{-1}$) in the postage stamp image, while the \textit{light grey} denotes the sky annulus region. The \textit{red} (and \textit{orange}) vertical lines denote the visually estimated maximum extent of the galaxy (i.e. radgal). The \textit{green} vertical line marks the outermost elliptical annulus within the dimensions of the postage stamp image. \textit{Right panels}: The corresponding cumulative flux profiles of FDS22\_0169 with different sky subtraction. The solid \textit{red} lines denotes the same extent of the galaxy, from which the total aperture magnitude is calculated.}
\label{fig:fsumplot1}
\end{figure*}

\begin{figure*}
\centering
\resizebox{0.9\hsize}{!}{\includegraphics[width=\hsize]{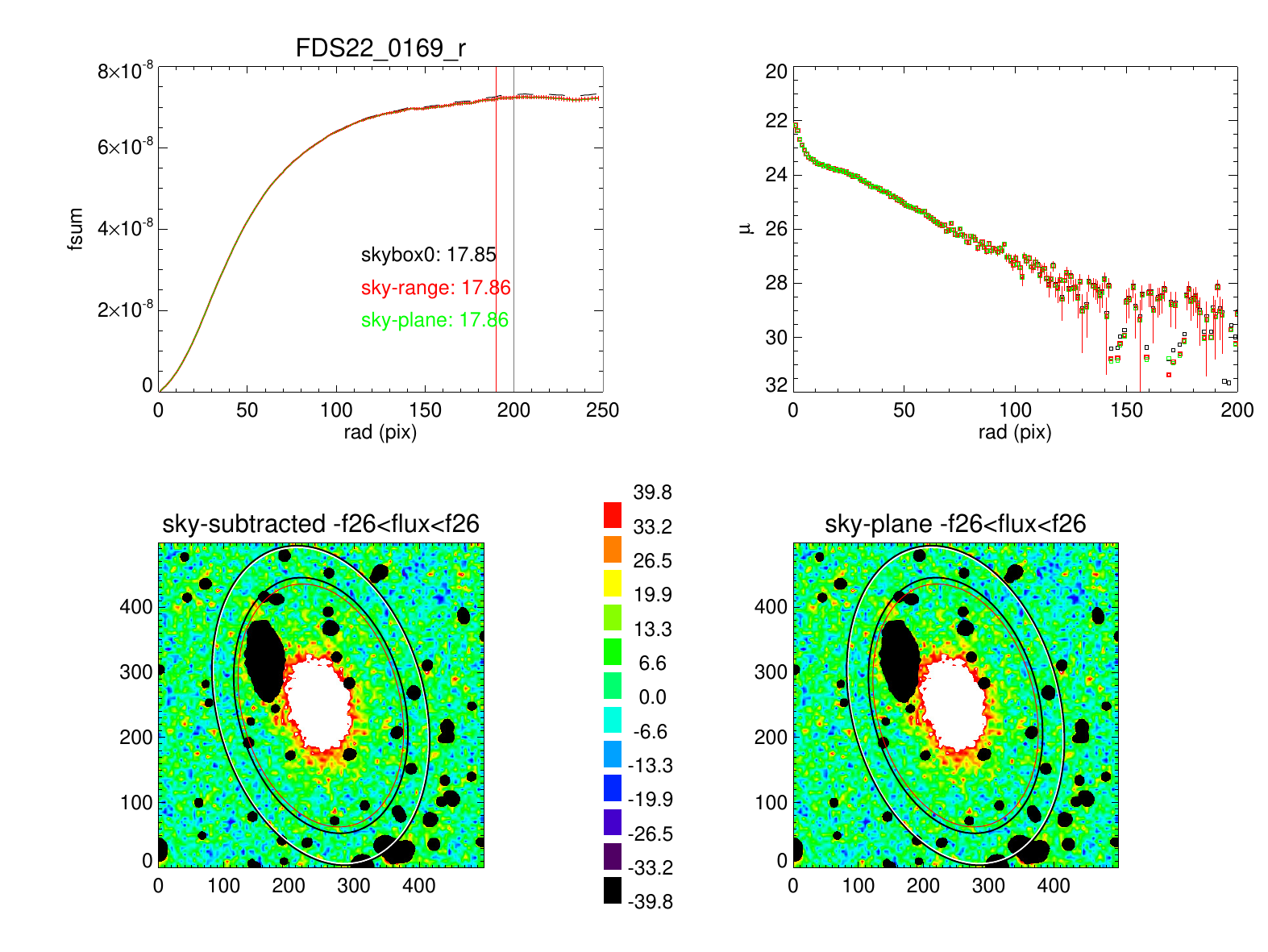}}
\caption{\textit{Upper left}: Cumulative flux profiles of FDS22\_0169 after different sky subtractions. \textit{Upper right}: The azimuthally averaged surface brightness profiles based on the different sky subtraction. The postage stamp images with sky annulus subtraction (\textit{lower left}) and skyplane subtraction (\textit{lower right}). The colours denote flux levels above and below surface brightness of 26 magnitudes in 12 levels (i.e. flux range shown is $\pm f_{26}=39.8\times10^{-12}$). The \textit{red} ellipse denotes the extent of the galaxy, whilst the two \textit{black} ellipses show the sky annulus. The \textit{white} ellipse indicates the outermost elliptical annulus.}
\label{fig:fsumplot2}
\end{figure*}

Here we illustrate the processing steps by following the example of FDS22\_0169 (FCC~70). Figure~\ref{fig:maskim} shows the postage stamp image of FDS22\_0169 with the corresponding masked regions. Figure~\ref{fig:ellipse} shows the isophotal ellipticity and position angle profiles as a function of semi-major axis. The highlighted regions (in cyan) show the range where the mean outer isophotal PA and ellipticity were measured. Figure~\ref{fig:fsumplot1} shows the flux profile for the three sky level estimations used, as well as the cumulative flux as a function of radius. Figure~\ref{fig:fsumplot2} illustrates the different sky subtractions, namely the sky annulus subtraction and the sky plane-fitting subtraction. In this case, there is practically no difference between the sky subtractions (despite a nearby companion). 

\subsection{PSF}\label{app:psf}

\begin{figure*}
\centering
\resizebox{\hsize}{!}{\includegraphics[width=\hsize]{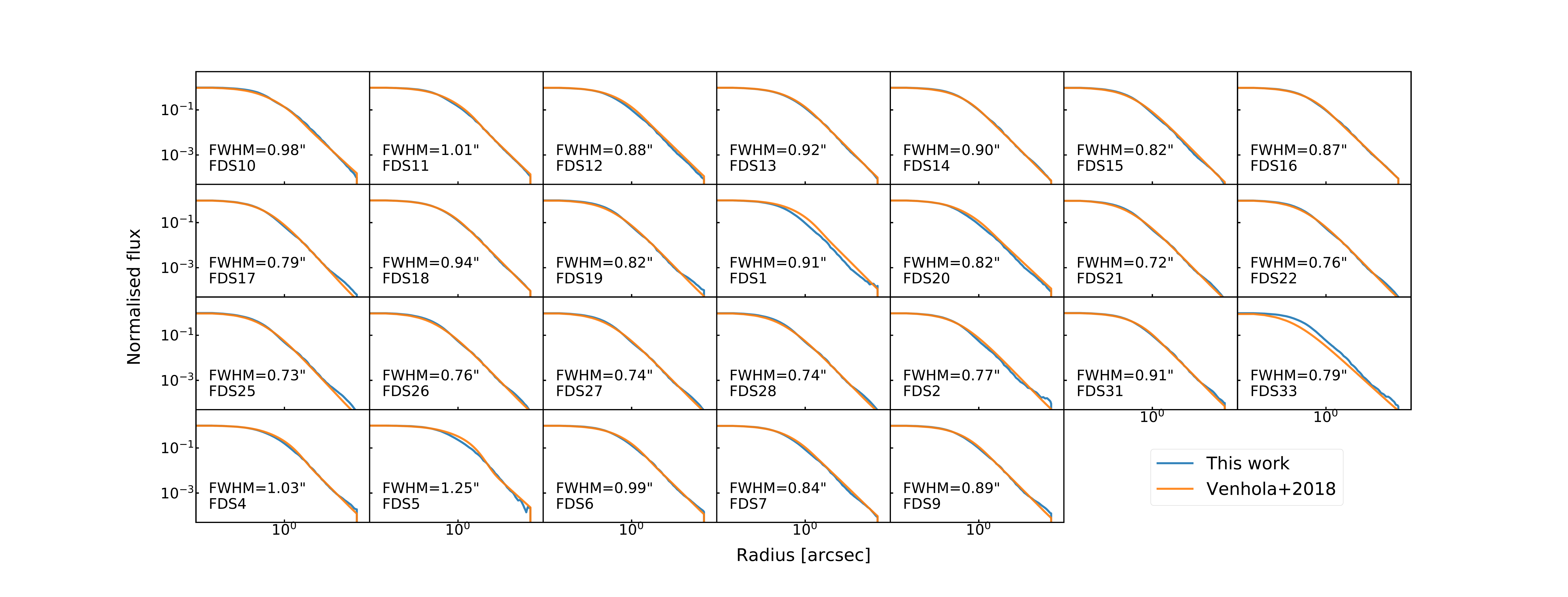}}
\caption{Normalised flux profiles of the PSFs constructed in 26 FDS fields. The profiles were normalised by peak flux. The FWHMs of the PSFs constructed in this work are labelled in each subplot.}
\label{fig:psf_rall}
\end{figure*}

Figure \ref{fig:psf_rall} shows the cumulative profiles of the PSFs constructed in all FDS fields in the $r'$-band from this work and in \citet{venhola2018}. It is worth noting that the PSFs constructed in this work were based on the azimuthally averaged intensity profiles of point sources, whereas the PSFs from \citet{venhola2018} were constructed by fitting analytical functions to the intensity profiles.

\section{Website}\label{app:website}
In order to easily access the various decomposition models constructed as well as images and measurements made during the data processing steps, we present the results and images via a website\footnote{\url{https://www.oulu.fi/astronomy/FDS_DECOMP/main/index.html}}, whilst the decomposition parameters were stored as machine-readable tables. The format of the website closely follows that of the S$^4$G decompositions \citep{salo2015}. As a brief overview, the website contains four levels (in descending order): Main (i.e. the homepage); Index (overview of each galaxy); Decomposition (closer look at Sérsic+PSF and multi-component decompositions in $r'$-band); Other decompositions (e.g. decomposition models for $g'$- and $i'$-bands).

The 'Main' level is the main page from which all other levels/pages can be accessed. The page includes a detailed description of the contents of the website (i.e. what is contained in all other levels) as an 'instructions manual'. Beyond the descriptions section is a list of all galaxies in the sample. Each galaxy ID is presented as a link which takes the user to the 'Index' level pages. The 'Main' page lists the galaxies in order of FDSDC dwarfs, then the additional galaxies. Currently there are also links to other 'Main' level pages which have the galaxies ranked by stellar mass in ascending and descending order.  

\begin{figure*}
\centering
\resizebox{0.9\hsize}{!}{\includegraphics[width=\hsize]{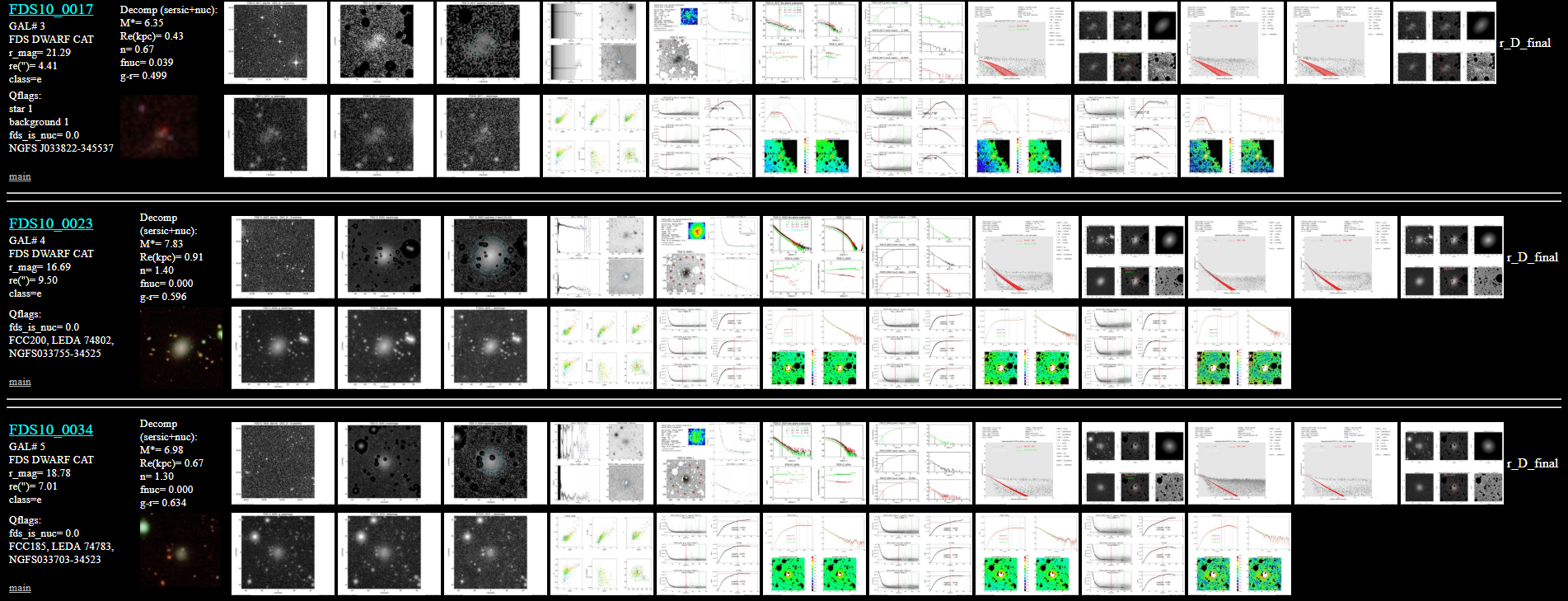}}
\caption{Example screenshot of an 'Index' level page. Each 'Index' page contains up to 100 galaxies. Note the "main" button located below the other catalogue designations, which takes the user back to the 'Main' page.}
\label{fig:web_index}
\end{figure*}

Figure \ref{fig:web_index} shows the structure of an 'Index' level page. Each galaxy contains a header of properties which include aperture photometry, decomposition parameters, quality flags if applicable, and other catalogue designations. The two rows of images contain isophotal measurements made during preprocessing as well as decomposition models. From the top row, counting from the left (we note that $r'$-band data images were used unless specified otherwise):
\begin{enumerate}
    \item Digitized Sky Survey (DSS) image (links to interactive 'map')
    \item $r'$-band data image with mask applied (masked regions appear 5 magnitudes fainter)
    \item $r'$-band data image with elliptical isophotes overplotted (via IRAF \texttt{ellipse})
    \item Ellipticity and position angle profiles (via IRAF \texttt{ellipse})
    \item Initial sky background determination (with contour plot of galaxy centre)
    \item Surface brightness and colour profiles in $g'$, $r'$, and $i'$ bands 
    \item Aperture $R_e$ calculated via cumulative flux profile
    \item 2D surface brightness profile for Sérsic+PSF decomposition 
    \item Model and residual images for Sérsic+PSF decomposition 
    \item 2D surface brightness profile for multi-component decomposition 
    \item Model and residual images for multi-component decomposition 
    \item Azimuthally averaged surface brightness profile for multi-component decomposition 
\end{enumerate}

\begin{figure*}
\centering
\resizebox{0.85\hsize}{!}{\includegraphics[width=\hsize]{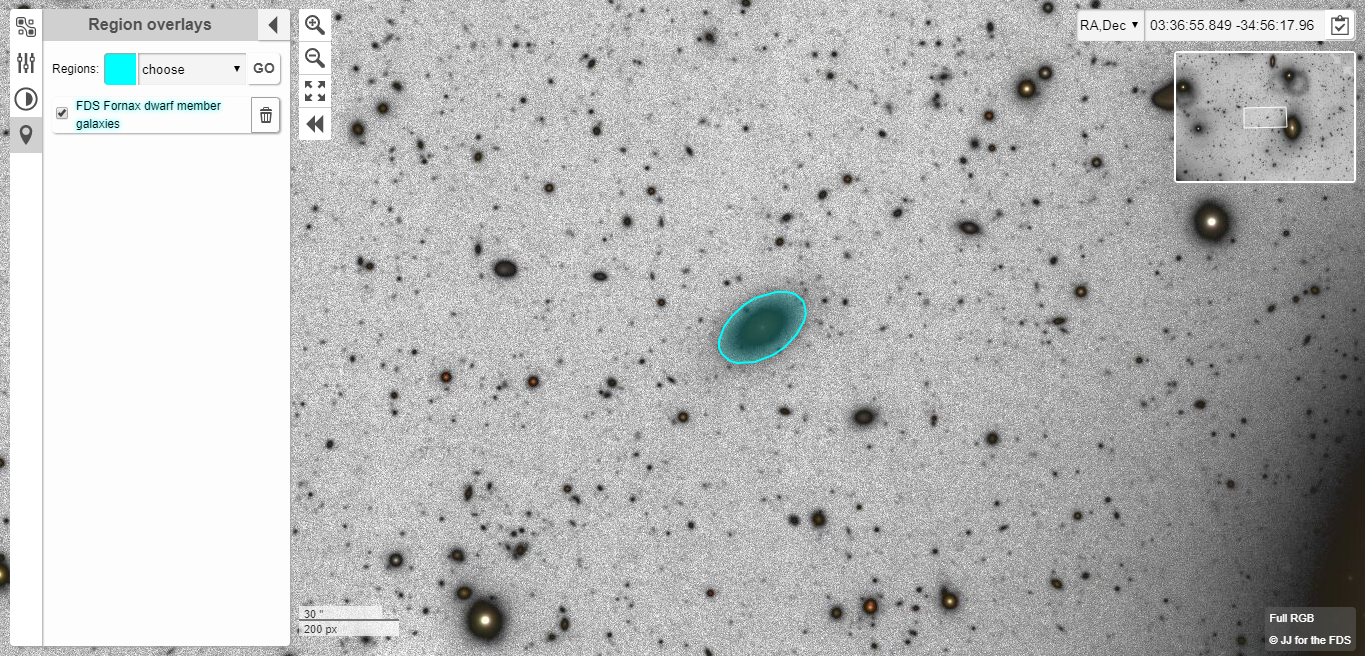}}
\caption{Screenshot of the FDS overview website. The blue region is based on FDSDC parameters. The image was made as a composite of $g'$, $r'$, and $i'$ bands based on FDS mosaics \citep{venhola2018}.}
\label{fig:fds_overview}
\end{figure*}

The link connected to the DSS image leads to an overview of the FDS (created using VisiOmatic \citep{bertin2015}). This is a separate website which allows the user to explore the FDS data in an interactive way. The website consists of a large image composed of all FDS fields which can be navigated via cursor. Coordinates can be queried to look for specific objects within FDS. Additional features include overlaying of the size and location of galaxies based on catalogues (FDSDC \citep{venhola2018}, FCC \citep{ferguson1989}, NGC \citep{sulentic1973}, ESO \citep{lauberts1982}, PGC/HyperLEDA \citep{paturel2003}). Figure~\ref{fig:fds_overview} shows a screenshot of the website with the FDSDC catalogue overlayed to highlight a member galaxy.

\begin{figure*}
\centering
\resizebox{0.8\hsize}{!}{\includegraphics[width=\hsize]{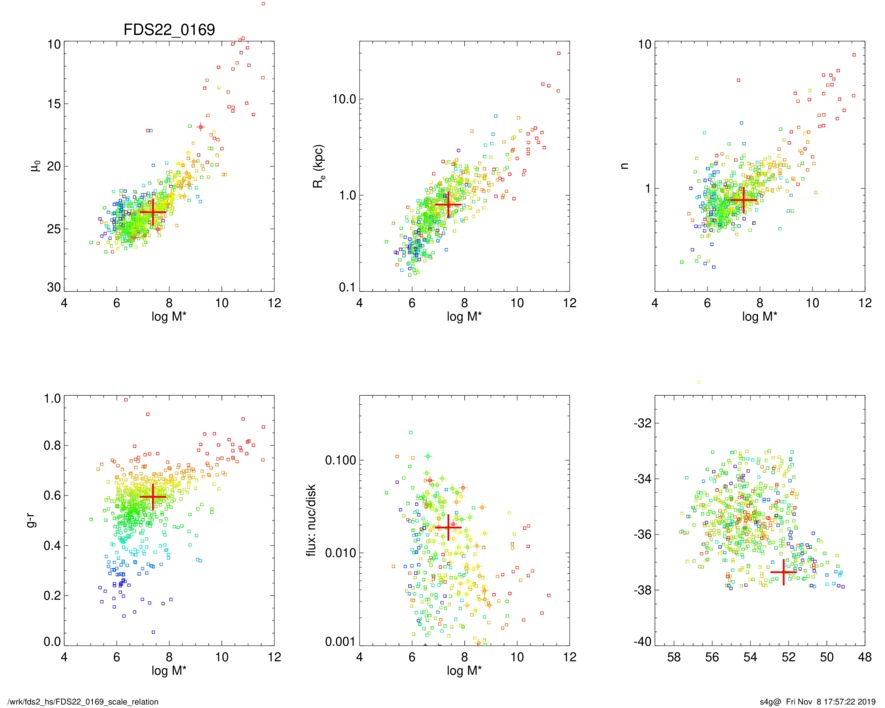}}
\caption{Various scaling relations (based on quantities derived from Sérsic+PSF decompositions) as a function of stellar mass. The \textit{red} cross indicates the location of the galaxy (in this case FDS22\_0169). The colours of data points denote the $g'-r'$ colours of the galaxies. \textit{Upper left}: The central surface brightness. \textit{Upper centre}: Effective radius. \textit{Upper right}: Sérsic index. \textit{Lower left}: $g'-r'$ colour. \textit{Lower centre}: Flux ratio between the PSF and Sérsic components (denoted as nucleus and disk). \textit{Lower right}: Location of the galaxy in Fornax (x-axis: right ascension, y-axis: declination).}
\label{fig:web_scaling}
\end{figure*}

From the second row, counting from the left:
\begin{enumerate}
    \item composite colour image based on $g'$-, $r'$-, and $i'$-band images
    \item $g'$ band data image 
    \item $r'$ band data image 
    \item $i'$ band data image 
    \item Scaling relations based on Sérsic+PSF decompositions (see Fig.~\ref{fig:web_scaling})
    \item $g'$ band sky level determination and aperture magnitude calculation
    \item $g'$ band sky plane comparison plots (cumulative flux profile, surface brightness profiles, and sky-subtracted images)
    \item $r'$ band sky level determination and aperture magnitude calculation
    \item $r'$ band sky plane comparison plots (cumulative flux profile, surface brightness profiles, and sky-subtracted images)
    \item $i'$ band sky level determination and aperture magnitude calculation
    \item $i'$ band sky plane comparison plots (cumulative flux profile, surface brightness profiles, and sky-subtracted images)
\end{enumerate}

Figure \ref{fig:web_scaling} shows scaling relations for a number of parameters as a function of stellar mass for the whole sample of 586 galaxies. The parameters shown are based on Sérsic+PSF decompositions. The plots provide a quick look at how an individual galaxy compares to the whole sample, and where it is located. 

From the 'Index' level pages, the user can view any of the images in more detail by clicking on them. To further inspect the decompositions made for a particular galaxy, the user can click on the hyperlinked galaxy ID name to enter the Decomposition level pages. These pages show (from left to right) the masked data image, ellipticity and position angle profiles, and the preliminary sky level plot with contoured centres as the top row for each galaxy. The two rows below show (from left to right) the two dimensional surface brightness profile, the model and residual images, and the azimuthally averaged surface brightness profile, for the Sérsic+PSF decomposition and multi-component decomposition, respectively. 

As mentioned, the decomposition parameters are stored in ASCII tables which are machine-readable to allow for easy access. The 'Main' page contains links to five separate tables, each containing parameters from different decomposition models:
\begin{itemize}
    \item Sérsic+PSF models (magnitudes as a free parameter in $g'$-, $r'$-, and $i'$-bands)
    \item Sérsic+PSF models (magnitudes, $R_e$, Sérsic $n$ as free parameters in $g'$-, $r'$-, and $i'$-bands)
    \item Multi-component models in $r'$-band
    \item Multi-component models in $g'$-band (magnitudes free)
    \item Multi-component models in $i'$-band (magnitudes free)
    \item Aperture photometry 
\end{itemize}

As a general overview, the multi-component table is structured as: 
\begin{itemize}
    \item Running number (i.e. 1 to 594)
    \item Galaxy ID (e.g. FDS10\_0000)
    \item Decomposition model (N=nucleus, D=disk, B=bulge, bar=bar, bl=barlens)
    \item Number of components in decomposition model
    \item Quality flag based on the trustworthiness of decomposition model (1=uncertain, 0=certain)
    \item Component parameters based on number of possible parameters (parameters filled with "-" for unused GALFIT functions)
\end{itemize}
The Sérsic+PSF parameter tables have the decomposition model, number of components, and quality flag columns omitted, and only component parameters from GALFIT functions "sersic" and "psf" are present. 

\clearpage

\section{Stellar mass vs halo-centric distance}\label{app:dist_vs_mass}
\begin{figure}
\centering
\includegraphics[width=\hsize]{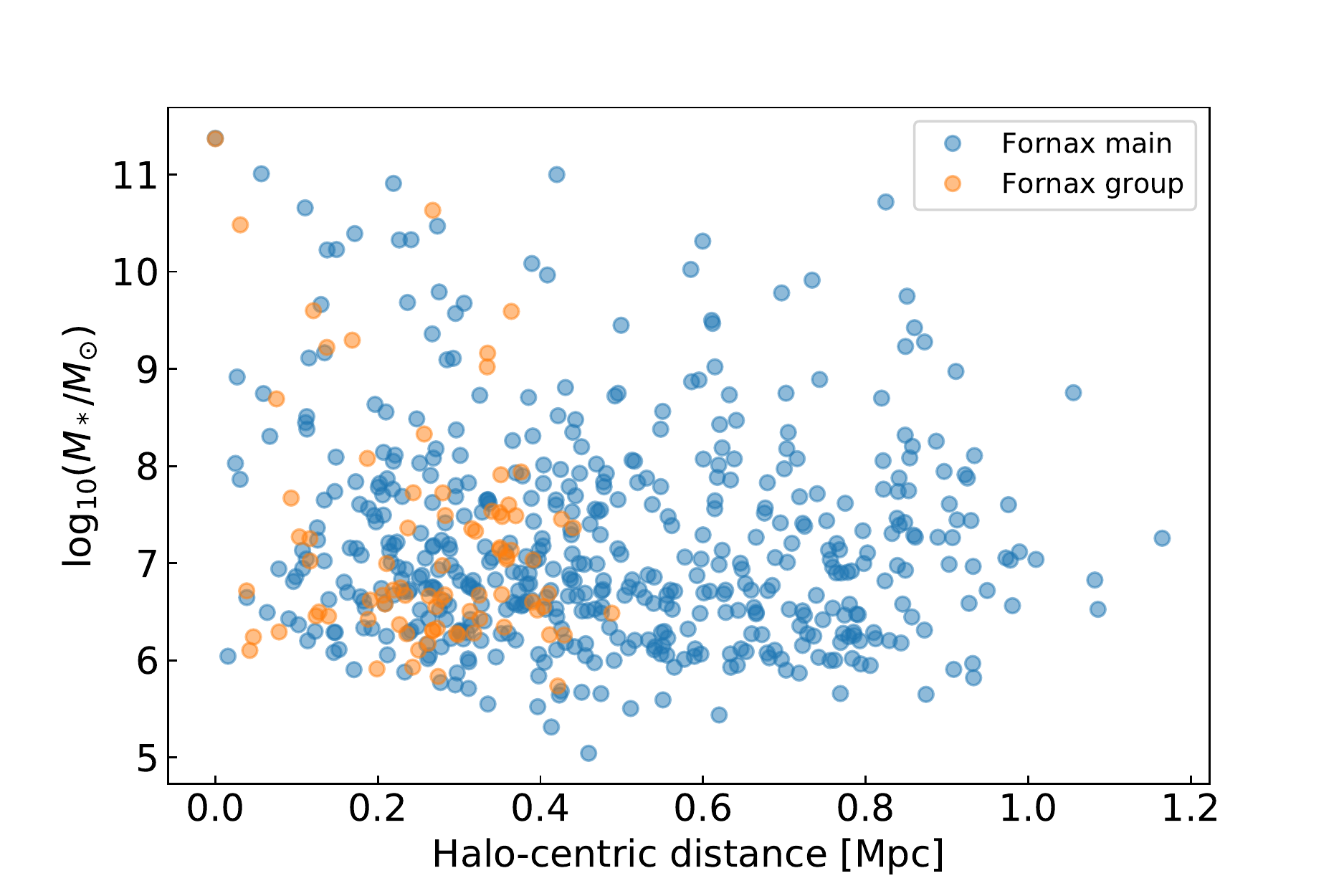}
\caption{Stellar mass as a function of halo-centric distance for our galaxy sample, split between Fornax main (\textit{blue}) and Fornax group (\textit{orange}). }
\label{fig:mass_vs_dist}
\end{figure}

Before comparing the quantities of galaxies as a function of halo-centric distance, we first check if there is a correlation between the stellar mass and the projected halo-centric distance to see if the halo-centric trends could be driven by trends in mass. Figure \ref{fig:mass_vs_dist} shows the stellar mass as a function of halo-centric distance for our sample. We find that using the whole sample, Spearman's $\rho=-0.067$ with a $p$-value$=0.106$, which is not significant. The story does not change when we split the sample into Fornax main ($\rho=-0.075$, $p$-value$=0.093$) and Fornax group ($\rho=-0.013$, $p$-value$=0.909$). Overall, we do not find any significant correlation between the stellar mass and the halo-centric distance for our galaxy sample. 

\section{Difference in magnitudes with NGFS}\label{app:ngfs_mag_offset}
\begin{figure}
\centering
\includegraphics[width=\hsize]{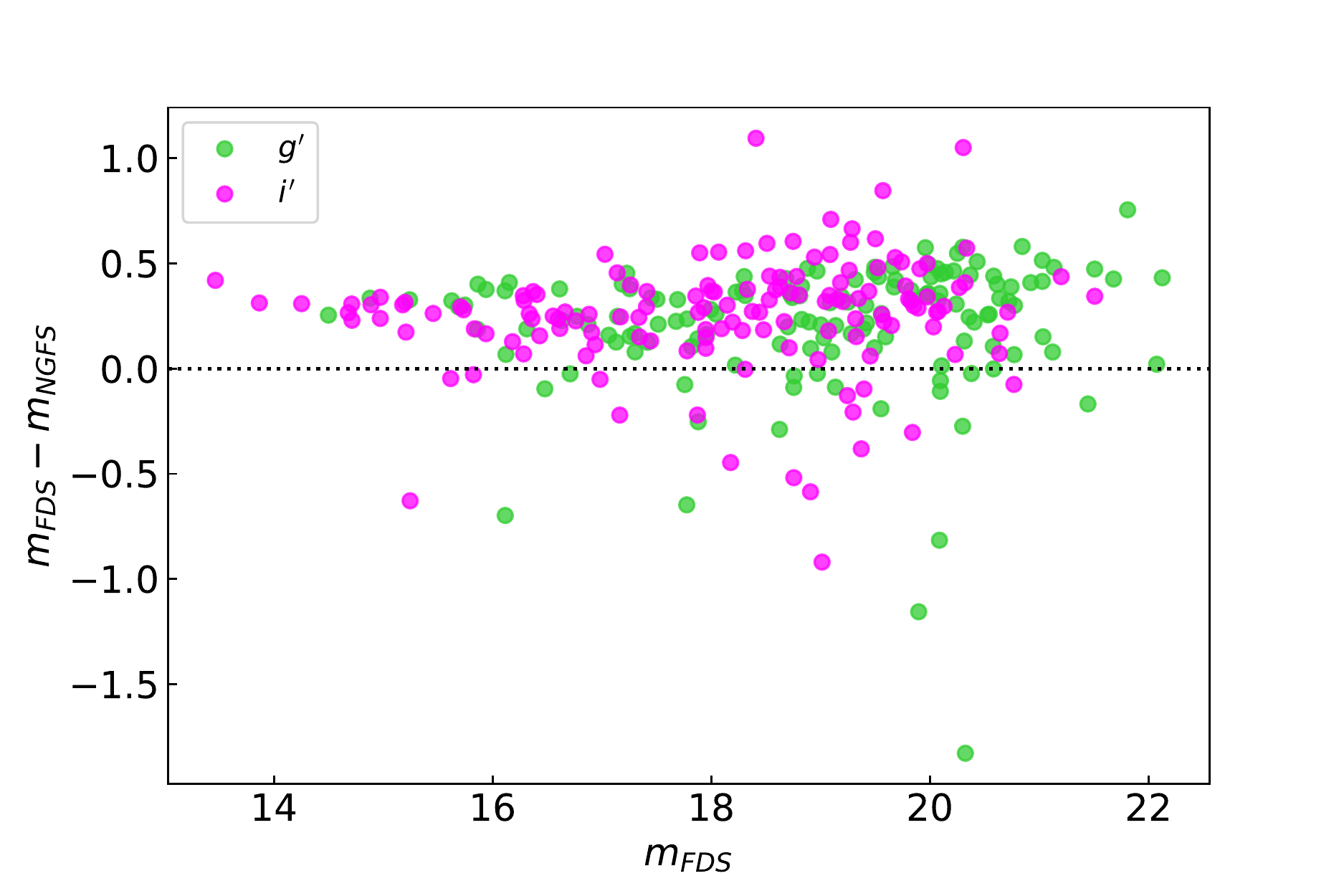}
\caption{Comparison of $g'$ (\textit{green}) and $i'$ (\textit{fuchsia}) apparent magnitudes between this work (FDS) and \citet{eigenthaler2018} (NGFS).}
\label{fig:ngfs_mag_comp}
\end{figure}

In \citet{venhola2018} a comparison of $g'$-band magnitudes and various colour-magnitude relations with \citet{eigenthaler2018} (NGFS) were conducted (see their Figs.~24 and D.3, respectively). They found a systematic offset in the $g'$-band magnitudes as well as in the $u'-g'$ and $u'-i'$ colours. In Fig.~\ref{fig:ngfs_mag_comp} we also show a systematic difference in the $g'$-band as well as in the $i'$-band magnitudes between this work and from \citet{eigenthaler2018}. We find the median of $m_{FDS} - m_{NGFS}$ for $g'$ and $i'$ to be 0.30 and 0.29, respectively. However, the offset in the two bands cancel out when we take the $g'-i'$ colour, hence the similarity in the moving averages in Fig.~\ref{fig:ngfs_comparison}.

\section{$r'$-band magnitudes estimate for S$^4$G}\label{sect:rband_s4g}
In order to estimate the $r'$-band magnitude for galaxies in the S$^4$G sub-sample, we collected SDSS \citep[DR16,][]{ahumada2020} $r'$-band magnitudes and compare them to the 3.6\,$\mu$m magnitudes (AB) from S$^4$G. Due to the difference in coverage between S$^4$G and SDSS, we matched 1287 out of 2352 S$^4$G galaxies (within 5\,arcsec).

\begin{figure}
\centering
\includegraphics[width=\hsize]{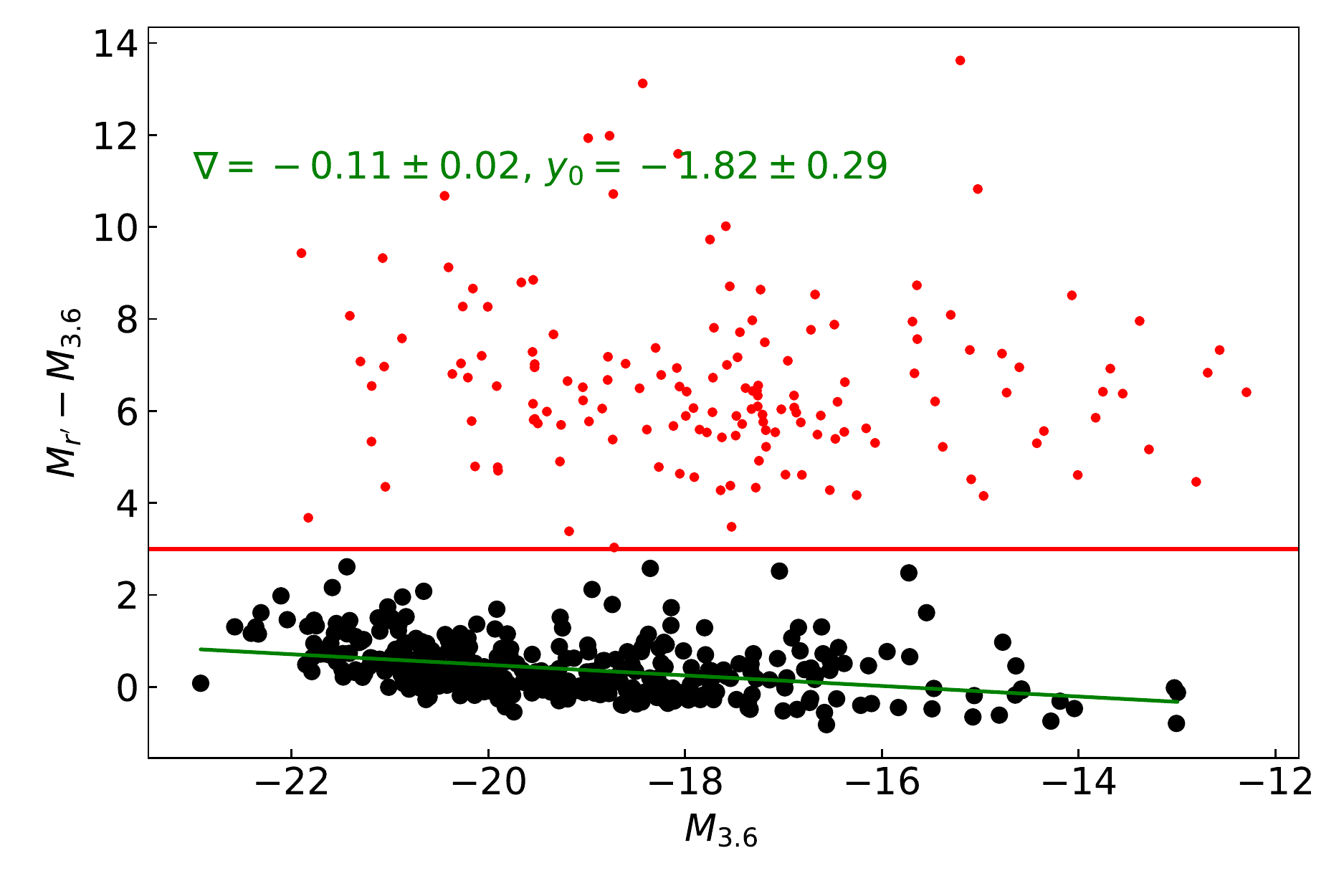}
\caption{$r'-3.6$ colour as a function of $M_{3.6}$ for the S$^4$G sample. The \textit{black} points show the galaxies used to derive a transformation between $r'$-band and 3.6\,$\mu$m magnitudes. The solid \textit{red} line ($M_{r'}-M_{3.6}=3$) denotes the limit where the SDSS $r'$-band magnitudes were untrustworthy (i.e. $M_{r'}-M_{3.6}$ colour too red to be physical). The \textit{green} line denotes the linear fit, with the gradient and intercept annotated in the plot. }
\label{fig:s4g_transform}
\end{figure}

The $r'$-band magnitudes were converted to absolute magnitudes based on the mean redshift-independent distances (taken from NASA/IPAC Extragalactic Database, NED\footnote{\url{https://ned.ipac.caltech.edu/}}) from Pipeline 3 of S$^4$G \citep{munozmateos2015}. Figure~\ref{fig:s4g_transform} shows the $M_{r'}-M_{3.6}$ colour as a function of absolute 3.6\,$\mu$m magnitudes for the S$^4$G sample. From inspection, a limit of $M_{r'}-M_{3.6} \ge 3$ was used to remove SDSS sources which had unphysical colours. These sources were excluded from the fitting process. A linear fit was applied to the remaining sources in order to obtain a relation which transforms $M_{3.6}$ to $M_{r'}$:
\begin{equation}
    M_{r'}=0.89 M_{3.6}-1.82.\label{eqn:s4g_transform}
\end{equation}
For galaxies which had $M_{r'}-M_{3.6} \ge 3$ or no matched SDSS $r'$-band magnitudes, Eqn.~(\ref{eqn:s4g_transform}) was applied to get an estimate of their $r'$-band magnitudes. The typical uncertainty in $M_{3.6}$, due to the scatter in distance measures, is $\sim 0.3$\,mag \citep{munozmateos2015}. Propagating this value along with the uncertainties of the intercept and gradient from Fig.~\ref{fig:s4g_transform}, the uncertainty of the extrapolated $M_{r'}$ is $\sim 0.4$\,mag.

\section{Detection of galaxy structures}\label{app:downsize}

\begin{figure*}
\centering
\resizebox{\hsize}{!}{\includegraphics[width=\hsize]{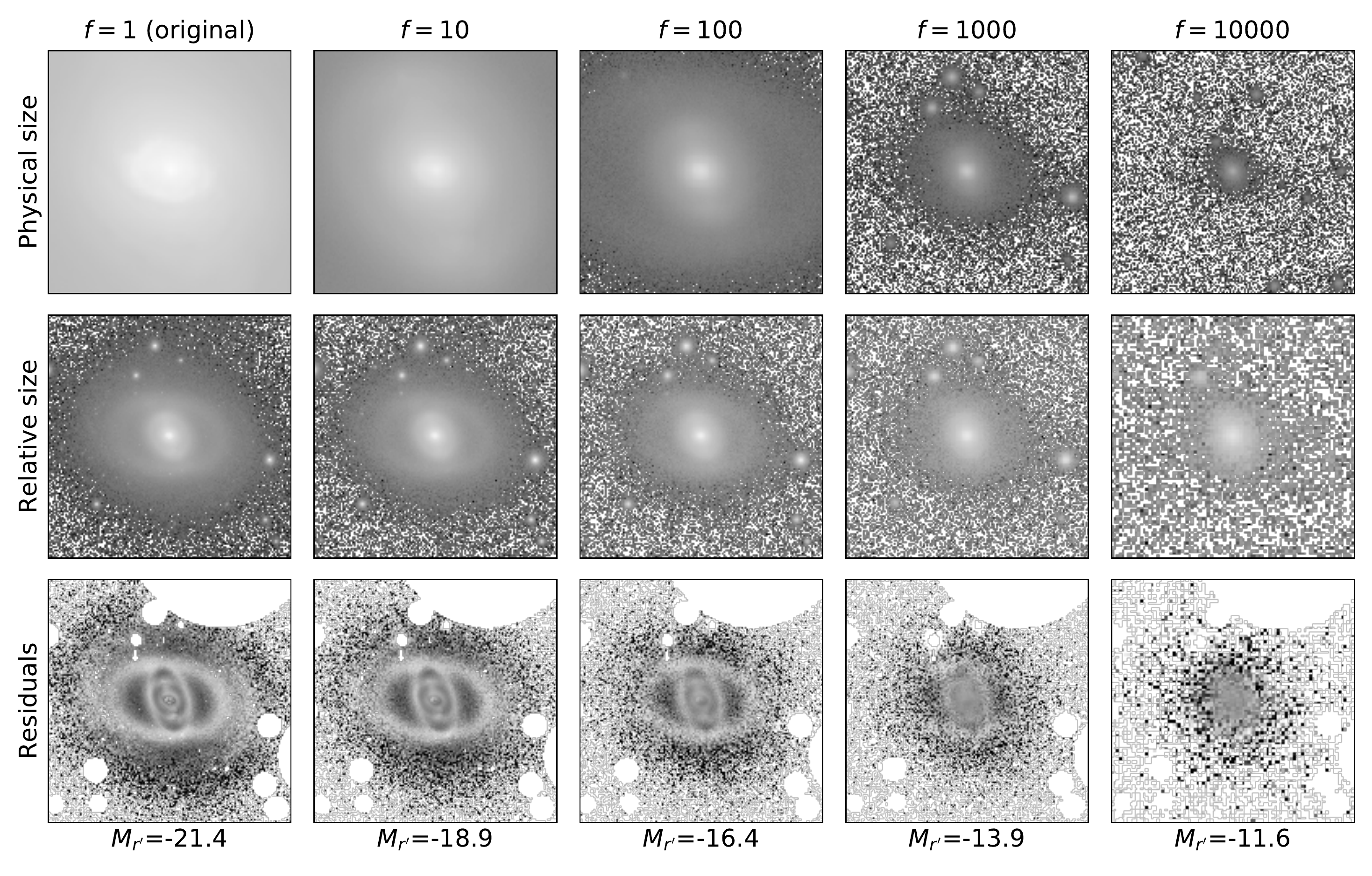}}
\caption{Images of FDS25\_0000 (NGC~1326) with a sequence of downsize factors $f$. The original image has a downsize factor of $f=1$. The \textit{upper row} shows the central $180 \times 180$\,pix of the galaxy at different $f$, presented with a magnitude scale of 30 to 15\,mag\,arcsec$^{-2}$. The \textit{middle row} shows the inner $10R_e \cdot f^{1/3}$ of the galaxies, where the $f^{1/3}$ factor accounts for the reduction in $R_e$ due to downsizing. A term of $-2.5\log_{10}(f)/3$ was added to the magnitude range of 30 to 15\,mag\,arcsec$^{-2}$ in order to account for the dimming of surface brightness due to downsizing. The \textit{lower row} shows the same region as the row above, where a magnitude range of -1 to 1 mag\,arcsec$^{-2}$ was used to display the residual images. The absolute $r'$-band magnitudes ($M_{r'}$) were calculated based on the corresponding multi-component decompositions.}
\label{fig:downsize_test}
\end{figure*}

To make sure that the observed trends from Fig.~\ref{fig:simple_fraction_mix} are not due to low S/N or poor spatial resolution for the fainter galaxies, here we present an experiment to mimic similar effects for a bright galaxy. In principle, we would like to take a bright, massive galaxy and 'downsize' it to a faint, small galaxy. This can be achieved by reducing the galaxy's total flux/stellar mass by a factor $f$ such that
\begin{equation}
    F_0 = f \cdot F, \label{eqn:downsize_flux}
\end{equation}
where $F$ and $F_0$ are the downsized and original total galaxy flux, respectively. Given that the total flux of the galaxy has been changed, the galaxy size must also be reduced accordingly. We used the scaling relation found in Fig.~\ref{fig:scale_rel_nomasstrend} and estimated that $M_* \propto R_e^{1/3}$. Hence, the size of the galaxy is reduced following 
\begin{equation}
    R_{e,0} \approx f^{1/3} \cdot R_{e}, \label{eqn:downsize_re}
\end{equation}
where $R_e$ and $R_{e,0}$ denote the downsized and original effective radius, respectively. Combining Eqn.~(\ref{eqn:downsize_flux}) and (\ref{eqn:downsize_re}), we get the following relation for the surface brightness of the galaxy
\begin{equation}
    \mathit{SB}_0 \sim F_0/R_{e,0}^2 = f^{1/3} \cdot \mathit{SB}, 
\end{equation}
where, $\mathit{SB}$ and $\mathit{SB}_0$ are the downsized and original surface brightness (in flux units), respectively. Therefore, to downsize a galaxy image the original galaxy image size and the pixel flux values must be reduced by a factor of $f^{1/3}$. This ensures that the total flux of the galaxy is reduced by a factor of $f$. Additionally, the seeing and the noise due to the sky must be preserved in the downsized galaxy image, so we convolve the image with the PSF and add Gaussian noise with the same standard deviation as measured within the sky annulus from the original image (see Sect.~\ref{sect:skysubtraction}) to the downsized image. 

As an example, we chose FDS25\_0000 (NGC~1326), which is a bright galaxy with $M_{r'}\sim -21.4$\,mag. In Fig.~\ref{fig:downsize_test} we show the galaxy images with a range of downsize factors $1\le f \le 10000$ ($f=1$ denotes the original galaxy image without downsizing). The top row shows the images at a fixed physical size (180\,pix $\times$ 180\,pix). The section shows only the very central region of the galaxy for $f=1$, whilst the $f=10000$ panel shows the whole galaxy and much of the sky. Additionally, the middle panel of Fig.~\ref{fig:downsize_test} shows the equivalent inner $10R_e$ (i.e. $10R_{e,0} \times f^{-1/3}$) of the galaxy images. Although there is a degradation in the amount of visible structures with increasing $f$, one can clearly still see structures up to $f=1000$. The same can also be seen in the corresponding residual images, displayed in the lower row of Fig.~\ref{fig:downsize_test}. 

In Table~\ref{tab:downsize_parameters} we present the flux fractions and relative sizes of each component in the decomposition models for $f=1$--10000. Overall, we find that the decomposition parameters remain rather similar across different values of $f$. The largest difference in parameters occurs for $f=10000$, where the barlens and disk components become smaller and larger, respectively, in size. In addition, we find that the barlens component becomes relatively brighter at the expense of the disk component for $f=10000$. 

To quantify the amount of observable structures in the downsized galaxies, we also calculate the RFF values using Sérsic+PSF decompositions. The choice of Sérsic+PSF models allows us to compare how well a 'simple' model would describe each downsized galaxy, as opposed to the multi-component model. From Table~\ref{tab:downsize_rffs} we see a clear decrease in RFF with increasing $f$. At $f=10000$, RFF $< 0.01$, which suggests that the Sérsic+PSF model can more or less account for the galaxy structures in the downsized image, and hence the downsized galaxy would likely to be considered a simple galaxy in Fig.~\ref{fig:simple_fraction_mix}. Combining the decomposition parameters, RFF, and residual images from the downsizing experiment, we find that structures can conservatively still be seen for $f=1000$. Referring back to Fig.~\ref{fig:simple_fraction_mix}, this experiment shows that if faint galaxies (e.g. $M_{r'} \sim -13.9$) exhibit morphological structures, they would be observable and hence fitted in the multi-component decompositions. This translates to a conservative lower stellar mass limit of $\log_{10}(M_*/M_{\odot}) \sim 7$ for our galaxies. In the case of $f=10000$, the lower limit would change to $\log_{10}(M_*/M_{\odot}) \sim 6$. 

\begin{table}[!ht]
    \caption{Multi-component decomposition parameters as a function of downsizing factor $f$.}
    \centering
    \begin{tabular}{lrrrrrrr}

\multicolumn{1}{c}{} & \multicolumn{4}{c}{Flux fraction} & \multicolumn{3}{c}{Rel. size}\\
\cmidrule(r){2-5}\cmidrule(l){6-8}
{$f$} &  {nuc} &  {bl} &  {bar} &  {disk} &  {bl} &  {bar} &  {disk} \\

\hline
1     &      0.01 &      0.23 &      0.19 &      0.57 &      1.00 &      1.00 &      1.00 \\
10    &     <0.01 &    0.24 &      0.19 &      0.57 &      1.04 &      1.01 &      1.00 \\
100   &     <0.01 &    0.24 &      0.19 &      0.57 &      1.03 &      1.02 &      1.02 \\
1000  &     0.01 &      0.23 &      0.19 &      0.56 &      1.04 &      1.02 &      1.10 \\
10000 &     0.01 &      0.28 &      0.20 &      0.51 &      0.78 &      1.08 &      1.20 \\

\hline
\end{tabular}

    \tablefoot{Relative decomposition parameters for each value of the downsizing factor $f$ for FDS25\_0000 (NGC~1326). The same components were used for each decomposition model. For the relative sizes (e.g. scale lengths), the size parameters were normalised by the $f=1$ size parameters multiplied by a factor of $f^{-1/3}$. This allows for a comparison of the effects of downsizing on the decompositions.}
    \label{tab:downsize_parameters}
\end{table}

\begin{table}[!ht]
    \caption{RFF values for different downsizing factors $f$. }
    \centering
    \begin{tabular}{lr}
    \hline
    $f$ &       RFF \\
    \hline
    1     &  0.16 \\
    10    &  0.14 \\
    100   &  0.09 \\
    1000  &  0.04 \\
    10000 &  <0.01 \\
    \hline
    \end{tabular}
    \label{tab:downsize_rffs}
    \tablefoot{The RFF were calculated based on Sérsic+PSF models, in order to provide insight as to how well 'simple' models can account for the (downsized) galaxies and their structures. A higher RFF suggests less of the structures are accounted for in the simple model. For reference, we used RFF=0.06 and 0.02 in Fig.~\ref{fig:simple_fraction_mix} to illustrate the effects of subjectivity when considering whether a galaxy is simple or not.}
\end{table}

\section{Table of decompositions} \label{app:multicomp_table}
Table~\ref{tab:multicomp_summary} shows an excerpt of the table of decomposition models made for galaxies in our sample. The full table can be found online.

\begin{sidewaystable*}
\caption{Summary of multi-component decomposition models. }
\begin{tabular}{lllllrlrrrl}
\hline
FDS ID &  FCC ID &       Alias &    RA\,[deg] &    Dec.\,[deg] & $\log_{10}(M_*/M_{\odot})$ & Multi-comp model &  $B/T$ & $Bar/T$ & $R_e$\,[arcsec] & S$^4$G overlap \\
\hline

FDS11\_0003 &  FCC213 &     NGC1399 &  54.6209 &  -35.4504 &                  11.37 &               BD & 0.39 &   - &       297.01 &         Yes \\
FDS26\_0001 &  FCC021 &     NGC1316 &  50.6823 &  -37.1931 &                  11.37 &               BD & 0.66 &   - &       281.50 &          No \\
FDS11\_0166        &  FCC219 &     NGC1404 &  54.7164 &  -35.5933 &                  11.01 &               BD & 0.60 &   - &       125.12 &          No \\
FDS17\_0365        &  FCC121 &     NGC1365 &  53.4015 &  -36.1408 &                  11.00 &          NblbarD &  - &  0.15 &       167.79 &         Yes \\
FDS11\_0006        &  FCC167 &     NGC1380 &  54.1150 &  -34.9760 &                  10.91 &         BblbarDD & 0.09 &  0.10 &       210.57 &         Yes \\
FDS14\_0133        &  FCC088 &     NGC1350 &  52.7835 &  -33.6284 &                  10.72 &         BblbarDD & 0.06 &  0.07 &       132.72 &         Yes \\
FDS11\_0001        &  FCC184 &     NGC1387 &  54.2376 &  -35.5066 &                  10.66 &            BbarD & 0.31 &  0.06 &        49.82 &         Yes \\
FDS25\_0000        &  FCC029 &     NGC1326 &  50.9848 &  -36.4644 &                  10.63 &          NblbarD &  - &  0.19 &        63.82 &         Yes \\
FDS26\_0254        &  FCC022 &     NGC1317 &  50.6889 &  -37.1051 &                  10.48 &         BbarbarD & 0.29 &  0.08 &        55.69 &          No \\
FDS6\_0001         &  FCC276 &     NGC1427 &  55.5810 &  -35.3925 &                  10.47 &              BDD & 0.30 &   - &        99.26 &         Yes \\
FDS11\_0002        &  FCC161 &     NGC1379 &  54.0165 &  -35.4412 &                  10.39 &              BDD & 0.48 &   - &        51.44 &         Yes \\
FDS16\_0001        &  FCC147 &     NGC1374 &  53.8191 &  -35.2263 &                  10.33 &              BDD & 0.24 &   - &        60.67 &         Yes \\
FDS12\_0003        &  FCC179 &     NGC1386 &  54.1925 &  -35.9993 &                  10.33 &               BD & 0.20 &   - &        38.09 &         Yes \\
FDS19\_0000        &  FCC083 &     NGC1351 &  52.6457 &  -34.8539 &                  10.31 &              BDD & 0.61 &   - &        99.22 &         Yes \\
FDS11\_0004        &  FCC170 &     NGC1381 &  54.1318 &  -35.2952 &                  10.23 &              BZZ & 0.40 &   - &        47.38 &         Yes \\
FDS11\_0000        &  FCC193 &     NGC1389 &  54.2989 &  -35.7461 &                  10.22 &            BbarD & 0.74 &  0.09 &        75.14 &         Yes \\
FDS6\_0000         &  FCC290 &     NGC1436 &  55.9045 &  -35.8531 &                  10.08 &           BbarDD & 0.04 &  0.05 &        54.29 &         Yes \\
FDS1\_0000         &  FCC312 &  ESO358-063 &  56.5797 &  -34.9418 &                  10.02 &               DD &  - &   - &       172.84 &         Yes \\
FDS15\_0002        &  FCC153 &      IC0335 &  53.8794 &  -34.4470 &                   9.97 &              NZZ &  - &   - &        32.04 &         Yes \\
FDS13\_0000        &  FCC249 &     NGC1419 &  55.1754 &  -37.5108 &                   9.91 &               BD & 0.73 &   - &       144.44 &         Yes \\
FDS10\_0000        &  FCC177 &    NGC1380A &  54.1978 &  -34.7397 &                   9.79 &           BbarZZ & 0.07 &  0.09 &        43.26 &         Yes \\
FDS7\_0737         &  FCC310 &     NGC1460 &  56.5571 &  -36.6964 &                   9.78 &          NblbarD &  - &  0.19 &        35.28 &         Yes \\
FDS20\_0000        &  FCC047 &     NGC1336 &  51.6339 &  -35.7136 &                   9.75 &              NBD & 0.51 &   - &        43.03 &         Yes \\
FDS16\_0000        &  FCC148 &     NGC1375 &  53.8200 &  -35.2656 &                   9.68 &             BblD & 0.14 &   - &        71.52 &         Yes \\
FDS12\_0002        &  FCC176 &     NGC1369 &  54.1886 &  -36.2562 &                   9.68 &          NblbarD &  - &  0.05 &        30.53 &          No \\
FDS11\_0005        &  FCC190 &    NGC1380B &  54.2873 &  -35.1951 &                   9.66 &          BblbarD & 0.09 &  0.03 &        37.68 &         Yes \\
FDS28\_0420        &  FCC013 &     NGC1310 &  50.2646 &  -37.0979 &                   9.60 &             barD &  - &  0.02 &        27.02 &         Yes \\
FDS22\_0000        &  FCC062 &     NGC1341 &  51.9933 &  -37.1495 &                   9.59 &             barD &  - &  0.12 &        19.60 &         Yes \\
FDS6\_0002         &  FCC277 &     NGC1428 &  55.5949 &  -35.1541 &                   9.57 &               BD & 0.80 &   - &        37.62 &         Yes \\
FDS4\_0001         &  FCC255 &  ESO358-050 &  55.2650 &  -33.7791 &                   9.50 &           NbarDD &  - &  0.05 &        30.39 &         Yes \\
FDS7\_0736         &  FCC308 &    NGC1437B &  56.4785 &  -36.3569 &                   9.47 &                D &  - &   - &        37.12 &         Yes \\
FDS7\_0000         &  FCC301 &  ESO358-059 &  56.2649 &  -35.9727 &                   9.45 &            barDD &  - &  0.17 &        20.41 &         Yes \\
FDS19\_0001        &  FCC055 &  ESO358-006 &  51.8251 &  -34.5265 &                   9.42 &               ND &  - &   - &        13.73 &         Yes \\
FDS16\_0002        &  FCC143 &     NGC1373 &  53.7467 &  -35.1711 &                   9.36 &            BbarD & 0.51 &  0.04 &        17.35 &         Yes \\
FDS26\_0003        &  FCC033 &    NGC1316C &  51.2432 &  -37.0096 &                   9.30 &            NbarD &  - &  0.17 &        17.96 &         Yes \\
FDS2\_0000         &  FCC335 &  ESO359-002 &  57.6530 &  -35.9094 &                   9.28 &              NDD &  - &   - &        38.46 &         Yes \\
\end{tabular}

\tablefoot{We include identifications of our galaxy sample as outlined in Sect.~\ref{sect:sample}, as well as any FCC and alias i.e. designations based on cross-match with HyperLeda \citep[][\url{http://leda.univ-lyon1.fr/}]{makarov2014}, if one exists. We also include parameters from decompositions such as the multi-component models, stellar mass, bulge-to-total, bar-to-total, effective radius, and whether the galaxies overlap with the S$^4$G. The effective radius was calculated from the outermost disk component from the multi-component model. The galaxies are ordered by their stellar masses, in descending order. The full table can be found online.}
\label{tab:multicomp_summary}
\end{sidewaystable*}

\end{document}